\documentclass[twoside,openright,a4paper,11pt]{report}
\usepackage{amsmath}
\usepackage{amsthm}
\usepackage{pmat}
\usepackage[pdftex]{graphicx} 
\usepackage{algorithm}
\usepackage{algorithmic}
\usepackage{fancyhdr}
\usepackage{cite}
\usepackage{geometry}
\usepackage{subfig}
\usepackage{float}
\usepackage{amsfonts}
\usepackage{amssymb}
\usepackage{multirow}
\usepackage{psfig}
\usepackage{epsfig}
\usepackage{epstopdf}
\usepackage{flafter}
\usepackage{array}
\renewcommand{\baselinestretch}{1.5}
\usepackage{layout}
\usepackage{geometry}
\usepackage{relsize}

\renewcommand{\headrulewidth}{0pt}
\renewcommand{\footrulewidth}{0pt}
\numberwithin{equation}{section}
\usepackage[colorlinks,linktocpage=true]{hyperref}
\hypersetup{
hypertexnames=false,
pdftitle={User-friendly Algorithms for Cyber-Physical Smart Grid system and Cognitive applications},
pdfauthor={Swaroop Ranjan Mishra},
pdfkeywords={music, nonlinear time series analysis, correlation dimesnion, lyapunov exponent},
pdfstartview={FitH},
colorlinks=true,
urlcolor=blue,
linkcolor=black,
anchorcolor=black,
citecolor=black,
filecolor=black,
}%

\usepackage[nottoc]{tocbibind}

\usepackage{amsmath}
\usepackage{fancyhdr}
\renewcommand{\headrulewidth}{0.4pt}
\renewcommand{\footrulewidth}{0.4pt}

\begin{document}
\pagenumbering{roman}
\fancyfoot[CO]{\thepage}

\titlepage
\begin{center}

	\vspace{4mm}
	\noindent {\bf{\Large{DEVELOPMENT OF USER-FRIENDLY SMART GRID ARCHITECTURE}}}\\
	\vspace{4mm}
\end{center}
\vspace{40mm}
\begin{center}
by\\
\vspace{5mm}
{\bf{\Large Swaroop Ranjan Mishra}}\\
{\bf{\Large (14104174)}}\\
\end{center}
\vspace{40mm}
\begin{figure}[h]
\centering
\includegraphics[width=0.25\textwidth]{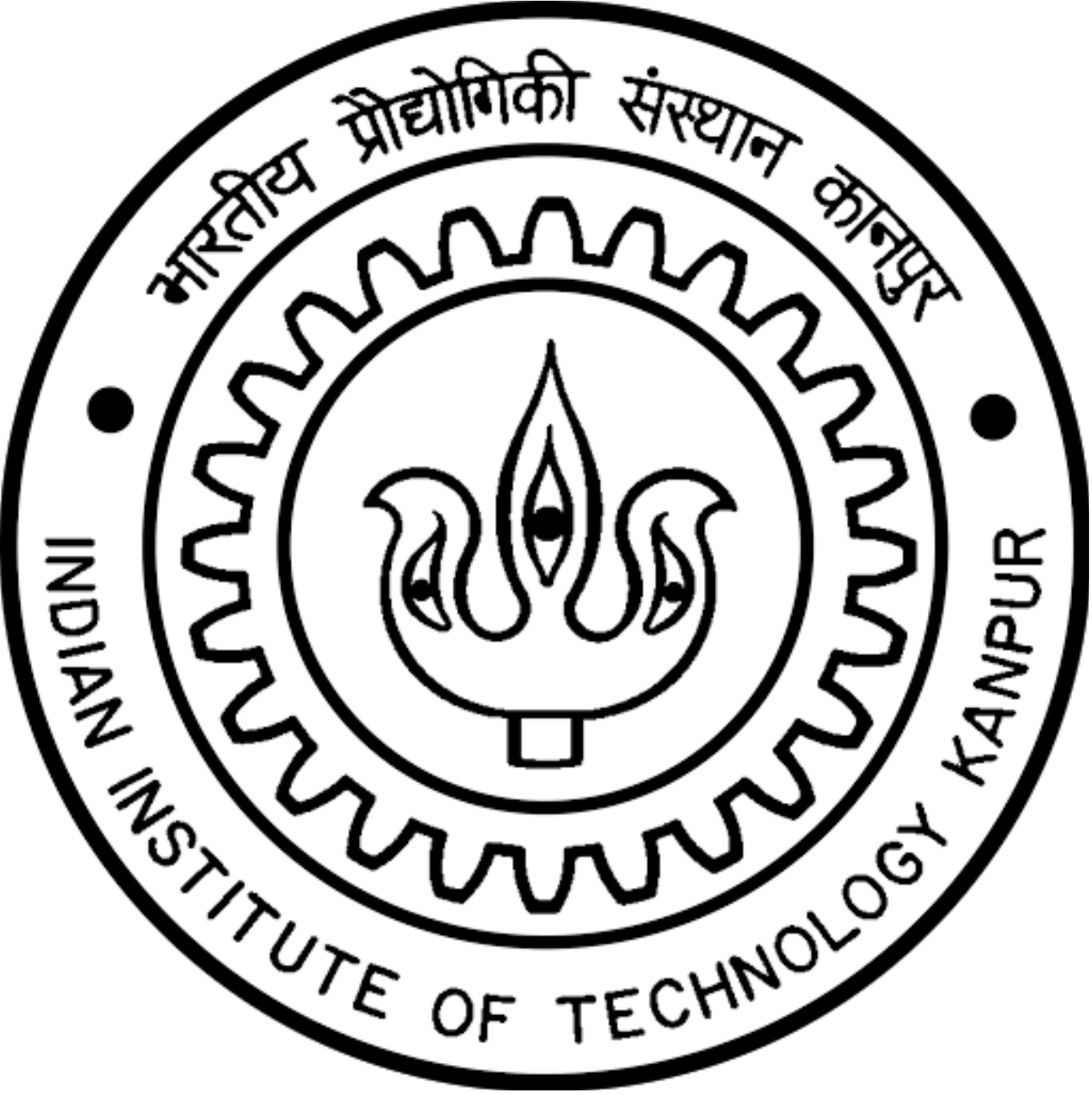}
\end{figure}
\begin{center}
\vspace{3mm}
{\bf {\large {\sc DEPARTMENT OF ELECTRICAL ENGINEERING}}}\\
\vspace{2mm}
{\bf {\large {\sc INDIAN INSTITUTE OF TECHNOLOGY KANPUR}}}\\
\vspace{3mm}
{\textbf{ May, 2016}}\\
\end{center}

\newpage
\titlepage
\begin{center}

	\vspace{4mm}
	\noindent {\bf{\Large{DEVELOPMENT OF USER-FRIENDLY SMART GRID ARCHITECTURE}}}\\
	\vspace{4mm}

\end{center}
\vspace{20mm}
\begin{center}
A Thesis Submitted\\
in Partial Fulfilment of the Requirements
\\
for the Degree of
\\
\vspace{25mm}
{\bf {\large MASTER OF TECHNOLOGY}}
\\
by
\\
{\bf{\Large Swaroop Ranjan Mishra}}\\
{\bf{\Large (14104174)}}\\
\end{center}
\vspace{15mm}
\begin{figure}[h]
\centering
\includegraphics[width=0.25\textwidth]{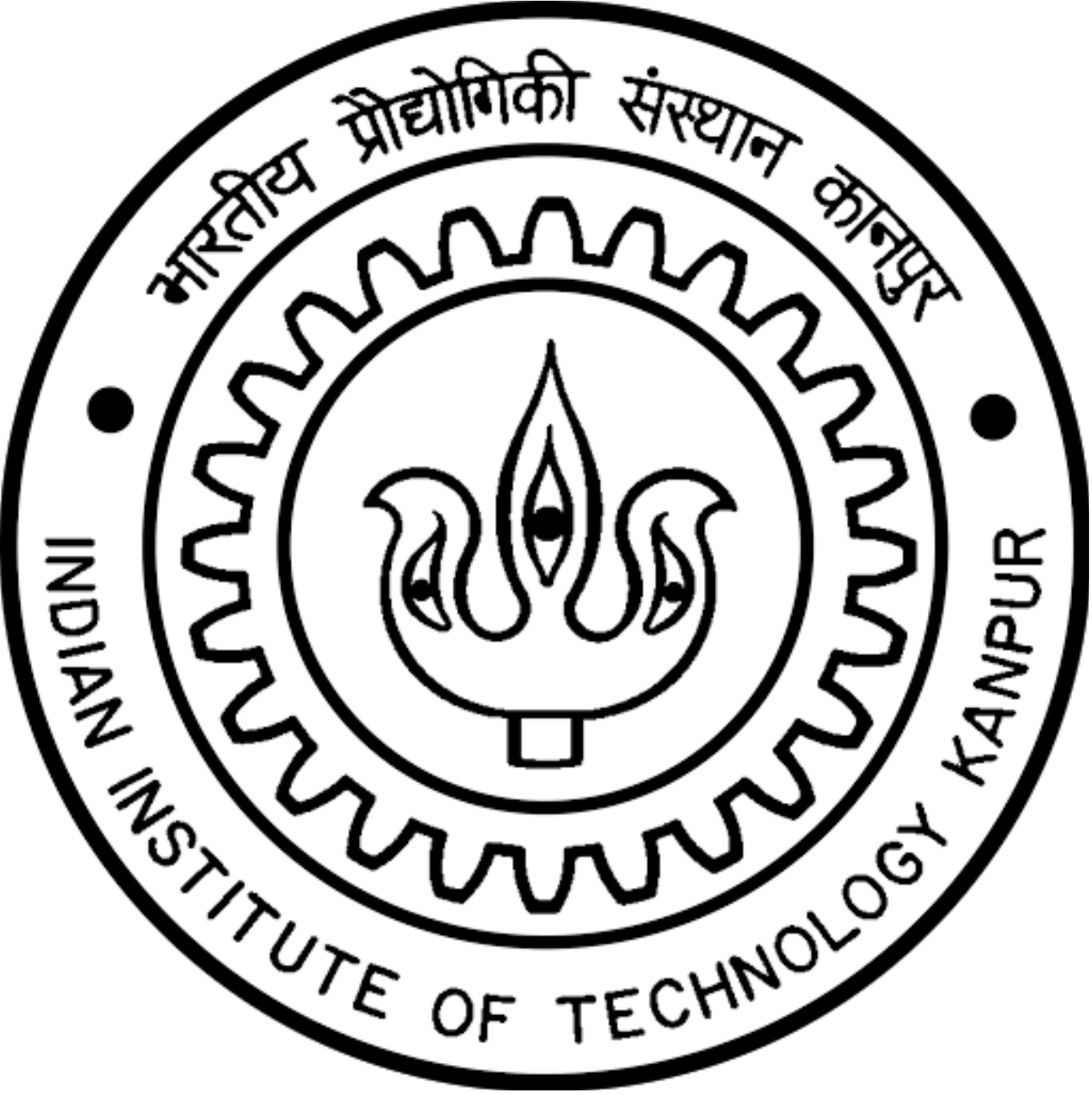}

\end{figure}
\begin{center}
\vspace{3mm}
{\bf {\large {\sc DEPARTMENT OF ELECTRICAL ENGINEERING}}}\\
\vspace{2mm}
{\bf {\large {\sc INDIAN INSTITUTE OF TECHNOLOGY KANPUR}}}\\
\vspace{3mm}
{\textbf{ May, 2016}}\\
\end{center}

\newpage
\titlepage

\begin{center}
\noindent{  }\\
\vspace{90mm}
\noindent {\textit{\LARGE {Dedicated to the service of Almighty}}}\\
\end{center}
\newpage

\vspace {1.0in}
\begin{center}
\begin{large}
{\bf CERTIFICATE}
\end{large}
\end{center}
\vskip 0.2in
It is certified that the work contained in this 
thesis entitled ``{\bf{\textit{Development of user-friendly smart grid architecture}}}'', by {\textbf{Mr. Swaroop Ranjan Mishra (Roll No. 14104174)}}, 
has been carried out under my supervision and this work has not been submitted elsewhere for a degree.
\vskip 1in

\begin{flushright}
May, 2016\\
IIT Kanpur
\end{flushright}
\begin{flushleft}
    {\bf Dr. Laxmidhar Behera,} \\ 
    Professor,\\
    Department of Electrical Engineering, \\
    Indian Institute of Technology, Kanpur\\
    Kanpur, 208016.
\end{flushleft}		
\newpage
%
%

\begin{center}
\begin{large}
{\it{\bf ACKNOWLEDGEMENTS} }
\end{large}
\end{center} 

I would like to express my sincere gratitude to my thesis guide, Dr. Laxmidhar Behera, for his invaluable guidance. I learnt how to work independently under him. I was afforded a lot of freedom by Dr. Behera, and this helped me gain confidence in my own abilities to think independently and contribute. I shall always remain grateful to him.

I have learnt a lot through detailed discussions with Dr. Pawan Goyal. I shall always remain grateful to him for his time and patience.

I believe the final outcome of my research was possible because of the excellent environment. I sincerely thank all my friends and seniors - Raj, Srinath, Soumya, Chinmay, Chandryee, Shubham, Prakhar, Ayush, Aquib, Aritra, Ajay Pratap, Sonal Dixit, Niladri Das, Samrat Dutta, Anima Majumdar, Ranjith Nair, Vipul Arora, Uday Majumdar, Meher Pritam, Anuj, Komal, Tharun, Vibhu, Sunil, Ravi, Radhe Shyam and Tushar. 

Special mention must be made of Meher Preetam, he has been in every sense a source of constant help and support specially in making various important decisions. The hour long discussions were not only thoroughly enjoyable but also intellectually stimulating. I shall always remain thankful to him.

\vskip 0.5in
\begin{flushright}
Swaroop Ranjan Mishra\\
May, 2016
\end{flushright}

\newpage
 \textbf{
\begin{center}
\vspace{10mm}
ABSTRACT
\end{center}
}

As systems like smart grid continue to become complex on a daily basis, emerging issues demand complex solutions that can deal with parameters in multiple domains of engineering. The complex solutions further demand a friendly interface for the users to express their requirements. Cyber-Physical systems deals with the study of techniques that are committed to modeling, simulating and solving the problems that emerge from a multi-disciplinary outlook towards futuristic systems. This thesis is mainly concerned with the development of user-friendly cyber-physical frameworks that can tackle various issues faced by the utilities and users of smart grid through a suitable choice of smart microgrid architecture. \\
   
The first contribution in this thesis aims to find the MIMO state feedback controller with the most stablilizing routing configuration between sensors and actuators while considering two challenges- (i) inclusion of general constraints with respect to sensor communication and actuators;(ii) consideration of individual capacities of these elements. 
The major contribution of this work is the development of a generalized algorithm to find optimal combination of sensors and controllers to be connected so as to make the system highly stable. The proposed algorithm minimizes a suitable cost or enhances reliability while guaranteeing stability using Lyapunov stability theory and linear matrix inequalities (LMI). The efficacy of the algorithm has been demonstrated through application on a 4-bus cyber physical smart microgrid system.\\

The second work deals with the experimental validation of proposed algorithm for various range of applications including smart grid. A communication test-bed consisting of wireless nodes has been designed and fabricated to function as the cyber system for the cyber-physical smart microgrid. The developed cyber system has been coupled with a virtual grid running on a server to test our algorithms for optimal sensor-controller connection design.\\

The third contribution deals with inclusion of robust and adaptive formulations in the existing framework to improve stability under various criticial conditions like load variation, delay variation and loss of communcation node. Optimization frameworks have been developed based on a specially designed sensor-controller connection design algorithm for no-delay and non-negligible delay cases to find controllers pertinent to the mentioned situations.\\

In the fourth work, we developed algorithms to achieve peak load shaving using the existing smart grid framework under physical and communication constraints. Electric devices are modelled as real-time scheduling tasks with timing parameters and are scheduled with real time algorithms like Earliest Deadline First, Least Laxity First and Dynamic Rate Priority to balance the peak load by switching the electric devices. The developed generalized algorithm finds optimal combination of sensors and controllers which can make the system highly stable and achieve peak load shaving at the same time as per real-time user requirements.\\

The fifth contribution in the thesis aims at improving the interaction of non-technical masses with the complex technological solutions through an intelligent user-friendly platform. A cognitive network has been designed based on experiential analysis of learning and its relation to emotion in humans. To help users self-analyze and grow personally, a reverse methodology is used to track user attributes based on the tuning process selected by the user. To demonstrate the user-friendly feature of this network in adding cognition, it has been used for summary finding applications (as a part concept teaching). For any given type of story and characters, the network finds the most appropriate moral from a set of morals. Unlike existing networks which require apriori skills to use and implement own ideas, this network simply requires the skill to express human emotion. 


\tableofcontents
\listoftables 
\listoffigures
\newpage
\pagenumbering{arabic}
\fancyhead[RO]{\thepage}
\fancyhead[LO]{\slshape \leftmark}
\fancyfoot[CO]{}
\renewcommand{\headrulewidth}{0.5pt}
\chapter{Introduction\label{intro}}
\subsection{Motivation}
Development of user-friendly algorithms is very much important to help people stay more human and stop end up machines working with other machines. The degree of necessity of user-friendly algorithms is directly proportional to the complexity of the system. Because of the existence of challenges of many types at the junction of cyber and physical systems, we decided to develop user-friendly algorithms for Cyber Physical Smart Grid Systems (CPS).

Cyber Physical Systems (CPS) integrate the dynamics of the physical processes with those of the software and communication provide abstractions and modelling, design and analysis techniques for the integrated whole. The dynamics among computers, networking, and physical system interact in ways that require fundamentally new design technologies- the technology depends on various disciplines such as embedded system, computer, communications etc. Many futuristic systems as smart grid are progressing towards incorporating more and more advanced sensors and actuators but without an appropriate model of the system and scheme of control, it would be quite cumbersome to manage and utilize the vast amounts of data available. Modelling these systems as a CPS with both physical and cyber input output signals, internal dynamics, local sensing and actutation would provide the basic foundation to further build upon and support the ideas of advanced control in these systems. 

This thesis is mainly concerned with the development of user-friendly cyber-physical frameworks that can tackle various issues faced by the utilities and users of smart grid through a suitable choice of smart microgrid architecture. First we focused on achieving stability of the cyber-physical smart grid system by designing a proper cyber system under various constraints present in both physical and cyber system. Then we developed communication hardware modules and tested our proposed architecture  there. We focused on the cyber aspect more in the hardware development so that this can be useful for other Cyber Physical Systems. While testing we used virtual grid modelled in the form of state space equations. Then we added robust and adaptive formulations in our architecture to deal with various critical conditions like delay, communication node failure, load variation. Further we made our architecture user-centric by addition of real time peak load shaving concept in it. After realizing the importance of intelligent algorithms in every sector (including daily life activities) and the amount of complexity involved in their use (tuning for example), we used our experience in user-friendly architecture design to develop a cognitive network. We used our network for text summarization application due to its huge research interest due to the natural presence of time and space constraints in various domains. Specifically, the main contributions of this thesis are:
\begin{itemize}
\item[1.] A Generalized Novel Framework for Optimal Sensor-Controller Connection Design to Guarantee a Stable Cyber Physical Smart Grid. This framework can be applied to systems having any number of sensors, controllers of variable capacity. This also can handle any type of constraints present in cyber as well as physical system in a user-friendly way. 
\item[2.] Cyber Architecture Development in the form of hardware communication modules for Experimental Validation of our proposed algorithm. Because of the presence of physical system independency in these modules, this cyber architecture can be applied to any cyber physical systems apart from smart grid.
\item[3.] Robust and Adaptive Formulation with Sensor-Controller Connection Design Algorithm. This formulation is capable to keep our system robust and adapt to the situation under various critical conditions like  node failure, load variation and delay.
\item[4.] Peak Load Shaving through Real Time Modeling for Cyber Physical Smart Grid Stability. This formulation is capable of handling any number of tasks with different deadlines under any types of constraints like delay, scability requirements, communication limitations. This can also satisfy different types of user requirements like stability, cost, reliabilty.
\item[5.] User-Friendly Cognitive Network Design using an Emotion based Algrotihm. For any type of story and characters, the network can find moral(s) from a set of morals. This network is meant to develop a user-friendly intelligent platform in future which the common public can use without having any knowledge about mathematical tools. To help users self-analyze and grow personally, reverse concept can be used to track user attributes based on the tuning process selected by the user. Further this can be extended to develop an emotional language which can help us get rid of this world of divisive language.
\end{itemize}
\section{Literature Survey}
\subsection{Literature in Cyber Physical Smart Grid Stability}
Cyber Physical Systems (CPS) consist of physical systems to be controlled, sensors, communication network and controllers. Physical system can be any system where the input and output has to be controlled. Sensors are used for the measurements of control and state variables. Computation units perform computation over measured quantities which are transmitted using a communication network (if sensor and controllers are not at one place). Controllers receive these information to actuate control signals for the physical system. CPS has a wide range of applications in areas such as smart grid \cite{rp:cps_app, korukonda2016improving}, mobile robotics, and unmanned aerial vehicles (UAVs) \cite{rp:UAV_CPS}. Communication networks in CPS brings in network uncertainties such as packet loss and delays which needs special attention for developing a reliable CPS \cite{rp:communication_failure}. Also, due to cyber physical coupling present in practical systems, design of communication network can have perceptible effect on physical system control\\

The communication between sensors and controllers can be of any type - wired, optical fiber or even wireless but, it should meet the requirement. Issues such as bandwidth introduces the necessity of cost minimization in designing a communication protocol for any CPS. Effective communication protocol from sensors to controller for a state-estimation problem has been derived in \cite{rp:state_estimate}. The work in \cite{rp:distributed_control} presents communication topology for distribution control with an assumption of no delay. In \cite{rp:communication_book,rp:communication_survey} readers can find more information on sensor - controller communication design. H. Li et al. \cite{rp:hli} have suggested that the voltage control can be done locally which may not require extensive communication. However, this argument is not valid when the smart grid consists of multiple distribution generations (DGs).\\

Addressing this voltage stability problem, Li et al. \cite{rp:multicast_paper} studied the effect of varying origins (sensors) and destinations (controllers) on improving smart grid stability as opposed to finding pathways between fixed set of origins and destinations. The major drawback of this work lies in patronizing a greedy scheme for routing which ultimately results in a sub-optimal solution. In this greedy approach, the system with the entire communication link is not studied for stability, instead the stability analysis is done part-wise. This part-wise analysis also may lead to sub-optimal solutions. In addition, the approach is computationally intensive, time consuming and requires more memory.\\

\subsection{Literature in Hybrid Cyber-Physical Voltage Control}
 Cyber-physical systems (CPS) \cite{lee2008cyber}\cite{kim2012cyber}\cite{khaitan2015design} are engineered systems that are built from, and depend upon, the seamless integration of communication elements, computational algorithms and physical components. CPS frameworks will imbue the current embedded system technologies with a whole lot of capabilities in various domains of reliability, scalability, resiliency, security, safety, adaptability and utility. Cyber-physical systems have already begun to broach many unforseen applications \cite{nist} in various areas like robotics \cite{fink2012robust}, health care \cite{lee2012challenges}, social networking \cite{yu2011geo},  smart grid technologies \cite{khaitan2013cyber}, data centers \cite{parolini2010cyber} and environmental monitoring \cite{lim2010national}.\\

The present smart microgrid scenario offers a platform for active deployment of CPS technologies due to inter-disciplinary interactions among power electronics, distributed generation, communication and information technology. Microgrid concepts offer a fair approach to  rethink our new age distribution system problems inundated widely with a large number of distributed generators(DGs) \cite{lasseter2011smart}. Microgrids work cooperatively with the existing grid as well as in stand alone mode when used in remote areas. Especially, when the microgrid becomes islanded, DGs can play significant role in maintaining essential parameters like voltage \cite{li2010adaptive} and frequency \cite{liu2014decentralized}.\\
\begin{figure}[t]
\centering
\includegraphics[width=0.9\textwidth]{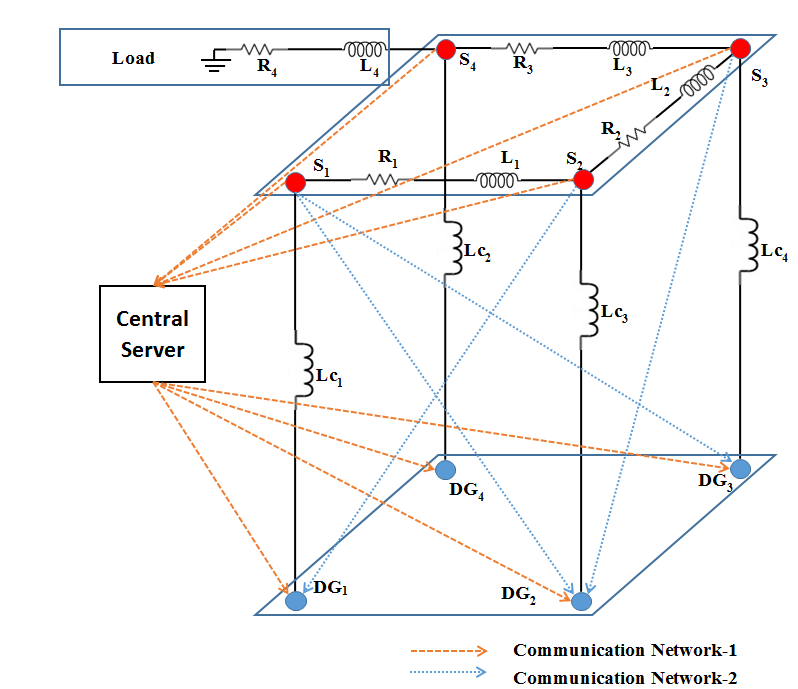}
\caption{The Cyber-Physical Micro Grid}
\label{Fig:fig1}
\end{figure}\\
  Thus, the cyber-physical approach towards solving the problem of voltage control would include models that would take inputs from many fields like power, communication and computation. \\ 
The coordinated voltage control problem in an islanded microgrid system has been presented in figure \ref{Fig:fig1}.
The diagram depicts a 4 bus system in which a single load is being fed by four distributed generators present at four different buses. The voltage at each bus is sensed by a dedicated sensor and controlled by a DG. A conventional approach to voltage control would follow a localized paradigm with each bus sensor sensing the respective voltages, relaying it to the DG on the bus, which, in turn, responds to the discrepancies in the bus voltage. A cyber-physical solution for the same problem would be to equip every DG with the information from sensors on all buses providing scope for each DG to devise better control strategies. Practical problems to such a system would arise both from the domains of control and communication. Controllers designed for such problems should be able to control voltage efficiently in scenarios like variation in loads and generation, while tackling communication constraints like delay in communication lines and unavailability of sufficient bandwidth to connect all sensors to all controllers. Providing joint and optimized solutions to such a complex engineering problem would be the objective of this work.\\

Augmenting the microgrid system with communication for better voltage and frequency stability has been under consideration among leading research circles for quite some time \cite{korukonda2018hybrid}, especially in the area of droop control. The usage of decentralized/distributed \cite{nasirian2014distributed}\cite{anand2013distributed}\cite{bidram2013distributed} control framework with communication has gained importance over a centralized scheme \cite{mehrizi2012constrained} keeping in mind, the reliability and scalability of the system. For example,  \cite{shafiee2014distributed} proposes a distributed secondary control technique which enhances the working of traditional droop control in an islanded microgrid by adding an ubiquitous communication framework.  Liang et al. \cite{liang2013stability} proposed a hybrid control scheme which can work both in the presence and absence of a centralized communication scheme between various DGs in the network. Kahrobaeian et al. \cite{kahrobaeian2015networked} improved upon it to make the scheme robust to communication delays. There have also been certain works like \cite{majumder2012power}\cite{nguyen2010distributed} which came up with communication routing algorithms and protocols to connect DGs and enhance power sharing in various microgrid configurations. Although these works have established that communication system structure affects overall system stability, they have failed to mathematically model the relation between control parameters and communication parameters. \\
The work in \cite{li2012multicast} managed to capture a co-relation between the interconnecting communication structure between controllers and sensors on the basis of voltage stability in a microgrid. This work uses a greedy based algorithm to route the connections between sensors and controllers which results in suboptimal solutions. The authors in \cite{mishra2015generalized} came up with a generalized algorithm to find more optimal configurations. Yet, in these works, the communication constraints are fixed. Also, the notion of stability arrived at in this work is inadequate when applied to varying physical and communication parameters over an extended range. The current work \cite{korukonda2017handling} proposes a constraint based sensor-controller connection design(CBSCD) algorithm which has the ability to take a larger number of connections into purview while arriving at the final topology. The algorithm considers a least cost criteria which  searches for the cheapest set of communication constraints that can provide maximum stability. Furthermore, optimization frameworks have been developed that use the theory of hybrid systems and common Lyapunov function(CLF) \cite{lunze2009handbook} to find controllers that can switch seamlessly between various parameter ranges while also retaining their individual stabilities in their respective bounds. The CLF based method is more suitable for a decentralized control which adopts a full communication structure \cite{ferreira2012analysis} while the CBSCD is most likely to be used for the sparse communication \cite{simpson2015secondary} scenarios.    
\subsection{Literature in Real time Peak Load Shaving}
During same time period if simultaneous requests are coming from many users, it can create problems for the public. Problem can be serious like complete disruption of power. It also can be small to produce temporal power shortage and result in load shedding. It will affect every sector of society along with consumers and utility. Companies can take this as nice opportunity and may increase electricity price to a large extent. That's why peak load shaving is very much important and has been in the center of moder day energy research.\\ 

Various optimization algorithms have also been used to find the best schedule to achieve peak load shaving. Genetic algorithm has been used to deal with combinatorial nature of the problem \cite{rp:cheng2004dynamic}. In this approch search operation to find the optimal schedule continues till it reaches convergence.\\

Along with optimization various analogies and daily life motivations have been applied to tackle this peak load shaving problem. It has been modelled as level packing problem where each load is modelled as a rectangle. Height of the rectange corresponds to the power
consumption and utilization decides its width. Obviously the aim is to minimize peak power in every group \cite{rp:facchinetti2010reducing}.\\

In peak load shaving \cite{mishra2019enabling}, startegies are being developed to move load far from the peak load time period. Energy demand has been modelled as a cost-and-benefit function 
\cite{rp:li2011optimal}. Energy demand of all household applicances has been modelled and control has been performed at the appliance level to reduce peak load. This further reduces variation in demand and uncertainities attached in it.\\ 

Complex systems demand robustness in this peak load shaving formulation for consistent performance. Robustness in this approach i.e. the ability to perform consistently in presence of uncertainities like delay has been focused in literature \cite{rp:facchinetti2011}.\\

Different types of time models have been used in literature. Time slots can be as large as 20 minutes \cite{rp:derin2010scheduling}. The complexity of the problem gets affected depending on the way we select time slots.\\

Beacuse of the involvement of tasks of diverse types and the amount of scheduling latency involved, various approaches have been developed by differentiating tasks and designing separate strategies for each of them. Different techniques have been developed for pre-emptive and non premptive tasks. Peak power requirement is checked without violating any of the constraints present inherently in the system. Feasible combinations are searched, grouped and algorithms have been proposed to find optimal scheduling similar to the way we developed sensor-controller connection design algorithm \cite{rp:lee2012energy}. 
\subsection{Literature in Emotion Recognition and Cognitive Network}
Relations between various disciplines have played a very important role in our process of innovation (Application range varies over a big region e.g. starting from ad-hoc based local innovations (many such are found in Indian market) to path-breaking inventions. Because of presence of so many constraints in our system starting from daily life processes to things we learn in class whose reasons we dont know, we hardly gain confidence to think, relate, form concepts and use it for interdisciplinary applications. So, where our models fail (e.g. physics based models), we treat physical variables as abstract data and operate by developing better algorithms to improve storage speed, computational time etc. But stable and balanced operation of various processes happening in this universe strongly helps us conclude the existence of relation between each and every entity of this world. So prediction of various phenomena is possible by finding generalized conclusions from analysis of few phenomena. Many breakthroughs in diverse areas like science, engineering, design etc. have been made by using relations intelligently. Revolutionary concepts can be developed by suitable applications of correlations. To help everybody beyond deep thinkers use correlation, a user-friendly interface with inherent intelligent structure is needed. The central focus of this work is to design a user-friendly intelligent platform to allow people enjoy relating various phenomena they deal with and develop innovative solutions to different real life problems just by providing instructions \cite{mishra2021natural}. These phenomena can be as simple as daily life processes (taking in to account common public as users).\\

As mathematics is the only technical language we have, an innovator has to follow mathematical modeling most of the times to have a theoretical proof  of his vision. Because of regular use, our thinking process is also affected by the mathematical literature. Through the process of evolution, we have developed this divisive language. Innovation has become tougher with the growth of this divisive language as we have to learn a lot of skills to implement our vision. This has also separated us from enjoying an emotional life and it has led us towards an era of machines working with other machines. So, the interest has been to develop something natural and user-friendly which can enable people to get their works done (starting from implementing own vision to finishing a job).\\

As our world is simply combination of various inputs from our sensory organs and naturally we are more sensitive to emotions than any other thing, the focus in this paper has been towards enabling common public get their works done simply by being more emotional. This indirectly requires the user to be a better human.\\
 
Intelligence has to play a major role in achieving this. But the major problem associated with machine learning algorithms (in the context of its use by a common public) is tuning. Intelligence of machine learning algorithms depend on intelligence of user (mathematical intelligence mostly) because the user has to set various parameters by analyzing the response. Only with experience and knowledge about various mathematical tools, user can develop a better machine learning algorithm. So, special care has been taken to ensure easy tuning of the proposed network.\\ 

Generation of emotion at the very fundamental algorithm level instead of the peripheral level in machines is necessary to enable machines behave considerably like humans. So, the algorithm is enabled to perform  various tasks with very less computation to ensure the involvement of emotion at the very fundamental level.\\ 

Algorithms operating on machines directly focus to get the desired objective whereas humans have to focus on many additional things along with the desired objectives. This is because of the inherent presence of so many elements in human body and desired objectives may not have components corresponding to all these elements (So, machines are far ahead in precision compared to humans and are very much suitable for repetitive works). Multitasking index has been used in this network to take into account these additional things.\\ 

Because of the presence of inherent relations between these additional parameters (associated with different elements of the human body), additional objectives are automatically achieved in different formats. Because of relation between these additional objectives and desired objectives, there exist reaction on various parameters (considering the bidirectional nature i.e. output is also treated same as input and it also gets updated). This leads them to additionally learn apart from the primary learning due to the back-propagated error between actual output and desired output. After understanding the improvement in overall stability because of these additional processes, this concept has been put in our algorithm by adding 10\%\ of the allocated multitasking index. This also helps in development of emotion with lesser emotion meter inputs (naturally also these additional learning processes affect our emotion). \\

In human related activities, relation of one parameter to others are well known (based on our experience, we can establish the relation at least), so every activity we do (including forcible daily life activities) has some inherent contribution factor attached which helps in instantaneous calculation of variation in error to achieve desired outputs. So this concept of contribution factor based real time calculations has been applied (by reducing the time spent in back-propagating error to each layer to update the weights).\\

To supervise and fasten the update process, brain is given the opportunity to experience its role with addition of huge number of extra elements through the process called Dream . By getting chance to experience the process which is actually impossible to achieve (or very difficult to achieve), this increases confidence and other emotional parameters of a human being and that person's belief on its own to achieve desired objectives (which were earlier a dream) increases drastically (similarly it can decrease in the opposite context). Accordingly “contribution factors” vary leading to fasten our speed towards the achievement of desired objectives. So, this concept has been used for updating confidence values.\\

Using the feature of high computational speed of Electro-Magnetic waves, various Neural Network (NN) schemes have been developed to replace actual system models without bothering about the generalized logic to explain a range of phenomenon. These NN schemes have been evolved using various mathematical tools (functions etc). But problem is the need to go to the new domain, mathematics. We can't implement our common sense based ideas unless we convert those to suitable mathematical forms. Because of the difficulty in establishing the relation between various parameters of the network to the error curve, tuning has become a tough task for people beyond data experts. This user-friendly interface based cognitive network has been  developed to solve this major problem. Various parameters of the network are changed to produce the desired moral.\\

After realizing the importance of emotion recognition through thousands of applications,
different types of human features have been analyzed to track emotion, but there exist issues of many types in the process of automatic emotion recognition. This has been illustrated below with various references.\\
\begin{enumerate}
\item Emotion recognition system using multimodal emotional markers\cite{rp:av_h}\cite{rp:mh_c} embedded in voice, face and words spoken etc. has been an area of great interest because of its numerous potential applications \cite{rp:aff_sens}\cite{rp:aff_det}\cite{rp:emo_rob}\cite{rp:mit_ac} in industry as well as academia \cite{rp:emo_reco}\cite{rp:affect_hci}.  Interdisciplinary literature of many related areas (like psychology, neuroscience) have been used in shaping various components of this system\cite{rp:em_r}\cite{rp:eu_r}. But the analysis has been challenging because of different types of issues attached in the process of emotion recognition \cite{rp:univ_cult}\cite{rp:pa_mi}\cite{rp:ss_ed}\cite{rp:em_s}\cite{rp:mh_c}. Cultural differences, inconsistency of clues in a culture, input- related issues like inconsistencies in the expression of emotion due to social norm, deceptive purposes, natural ambiguity, output- related issues based on the nature of the representation of various emotional states, time course and its implications, influence of environmental conditions (like lighting, audio recorder) on the input features, nonlinear behaviour of humans in revealing inner emotional states, nonlinear dynamics\cite{rp:nd_v} etc. \cite{rp:emo_reco}\cite{rp:rev_ds}\cite{rp:ch_r}\cite{rp:ea_d} are few examples.
\item Similarly body expression has been used as an important modality for affective communication and for creating affectively aware technology because of various scientific, technological, and social reasons. Affective body expression perception and recognition analysis has been challenging again because of similar issues mentioned in the previous paragraph (like data collecting, labeling, modeling, and setting benchmarks for comparing automatic recognition systems). The need to use more types of body expressions and create systems that are able to recognize emotion independently of the action (that the person is doing) is very important for ubiquitous deployment \cite{rp:aff_body}.
\item Many other parameters have also been used to recognize emotion. Speech data has been analyzed as an objective predictor of depression and suicidality\cite{rp:rev_ds}. Phonetic syllables have been used for continuous emotion recognition\cite{rp:ce_rc}. Preferences for upright, emotionally valenced biological motion can be used to predict emotion recognition abilities. This can help in development of interfaces that tailor interactions based on the emotional processing abilities of the user\cite{rp:aff_comp}.
\item The level of physical activity has also been used to monitor the evolution of some physical and mental disorders, e.g., obesity or depression. This has been applied for patient-specific treatments. using a virtual agent whose emotional state evolves depending on input data \cite{rp:lin_em}. Other sources like EEG signals \cite{rp:emo_pi}, have also been used. Research is also going on to track emotional states and motivate changes in people (e.g. physical behavior \cite{rp:lin_em}), society \cite{rp:ps_hi}\cite{rp:ss_ed}.
\end{enumerate}
Existance of many types of issues in the process of automatic emotion recognition immediately demands a technique which can be easily enabled to adapt to different issues. So, we analyzed various classifiers which are used in this process.\\
 
Limitation of scientific techniques (like Neural Networks (NN) \cite{rp:emo_reco}\cite{rp:sa_e}\cite{rp:er_f}, optimization algorithms 14\cite{rp:in_pso}
, Hidden Markov Model \cite{rp:err_hmm}\cite{rp:two_lv}\cite{rp:pv_r}\cite{rp:av_h}, SVM \cite{rp:emo_pi}, statistical methods \cite{rp:sm_hmm}, combination of classifiers \cite{rp:fe_ce}) to allow users enable adaptiveness in their dynamics (e.g. dynamics of NN) in accordance with different issues discussed so far has been a major drawback of this study. The absence of user-friendliness is the issue \cite{rp:aff_comp}\cite{rp:pa_mi}\cite{rp:ch_r}\cite{rp:as_r} as already discussed in the beginning of this section.\\

In the search of a user-friendly classifier to process various features, we found some related works. People have developed algorithms and tools by putting emotion in the learning process with inspiration from the biological features of the emotional brain and utilized it for several applications like facial detection, and emotion recognition\cite{rp:pr_nn}\cite{rp:at_e}. That’s adaptive for a particular context and user friendly for people in that particular area. But it’s not the user-friendliness we have been talking because user (specifically a general public without much knowledge about mathematical literature) can’t change various parameters of the network (and implement his common sense based ideas). This has already been discussed in the beginning of this section.\\

With the motivation of a user-friendly cognitive network, we analyzed various human processes in comparison with existing intelligent techniques (NN etc.) as discussed in the beginning of this section. To develop a cognitive network using the comparison, we started searching for the best-fit applications.\\

Several recent works have shown that many of the publicly available large scale datasets contains artifacts\cite{mishra2020dqi,mishra2020our} that provides shortcuts for models to exploit on, resulting in overestimation of model performance \cite{mishra2020we,Mishra2020DQIAG,anonymous2020realtime} and unfair evaluation \cite{mishra2021robust}.
To find best-fit applications which demands this user-friendliness to the highest extent, we went through various relevant literatures. Text summarization and Story generation (being two complementary processes) have been of great research interest because of the natural presence of time and space constraints in various domains[44]. Mobile learning \cite{rp:e_ts}, Biomedical domain\cite{rp:t_sb}[38], Game design\cite{rp:s_sg}\cite{rp:f_sg}, Linguistic Reasoning \cite{mishra2020towards,mitra2019exploring, varshney2020s} are few fast growing areas in this context. Absence of user-friendly, generalized, customized and less memory usage based techniques have been the major issues \cite{rp:e_ts}\cite{rp:a_sc}\cite{rp:a_tsml}\cite{rp:sr_p}. Several techniques have been developed to solve few of the mentioned issues in text summarization\cite{rp:e_ts}\cite{rp:t_ra}\cite{rp:s_ts} as well as Story generation \cite{rp:c_sg}\cite{rp:a_de}\cite{rp:c_l}\cite{rp:s_g}\cite{rp:sr_p}, but to our knowledge there doesn't exist a technique with all 4 features. So we used our network to perform this 2 operations (though we develop the network in general for many other applications as mentioned in Section \ref{Sec:unique}). To add user-centricity in the analysis (and also for simplicity), we have used user-inputs manually in our network.
\section{Thesis Organization}
In Chapter 2 we discuss the detailed procedures involved in optimal senser-controller connection design to guarntee a stable cyber physical smart grid. In Chapter 3, we discuss  the process of cyber architecture development in the form of hardware communication module for experimental validation of proposed algorithm . In Chapter 4, we take up the robust and adaptive formulation with senser-controller connection design algorithm. In Chapter 5, we develop a peak load shaving formualation through real time modeling for cyber physical smart grid stability. In Chapter 6, we discuss the detailed procedures involved in the development of user-friendly cognitive network using an emotion based algrotihm. In Chapter 7, we conclude by giving a summary of the work done in this thesis and by listing the directions of future research.

 \chapter{Optimal Sensor-Controller Connection Design to Guarntee a Stable Cyber Physical Smart Grid\label{micromodel}}
\section{Introduction}
Here, we have proposed a generalized connection design algorithm which can be applied to any CPS application including the smart-grid system. The proposed algorithm ensures optimal cost while providing maximal stability margin. In the proposed algorithm, all those connections which can simultaneously co-exist without violating any constraints are found out. These connections are then grouped depending on further objectives to find out valid connection sets in the system. The algorithm thus proposed has been studied to obtain maximum reliability as well as minimum cost as per the necessity. In addition, DGs of different capacities which act as controllers have been incorporated into the system and has been studied in this chapter.\\
The remainder of this chapter has been organized as follows. Section \ref{Sec:2MIMO} deals with MIMO system model, where the system model, controller structure and the optimization problem are discussed. The cyber physical smart grid system has been introduced in \ref{Sec:2application}. The proposed connection design algorithm has been presented in Section \ref{Sec:2MIMO_proposed_algo}. Simulation results are presented in \ref{Sec:2results} followed by concluding remarks in Section \ref{Sec:2sum}.

\section{System Model} \label{Sec:2MIMO}
\begin{figure}[!ht]
\centering
\includegraphics[width=0.9\textwidth, height=7cm, keepaspectratio]{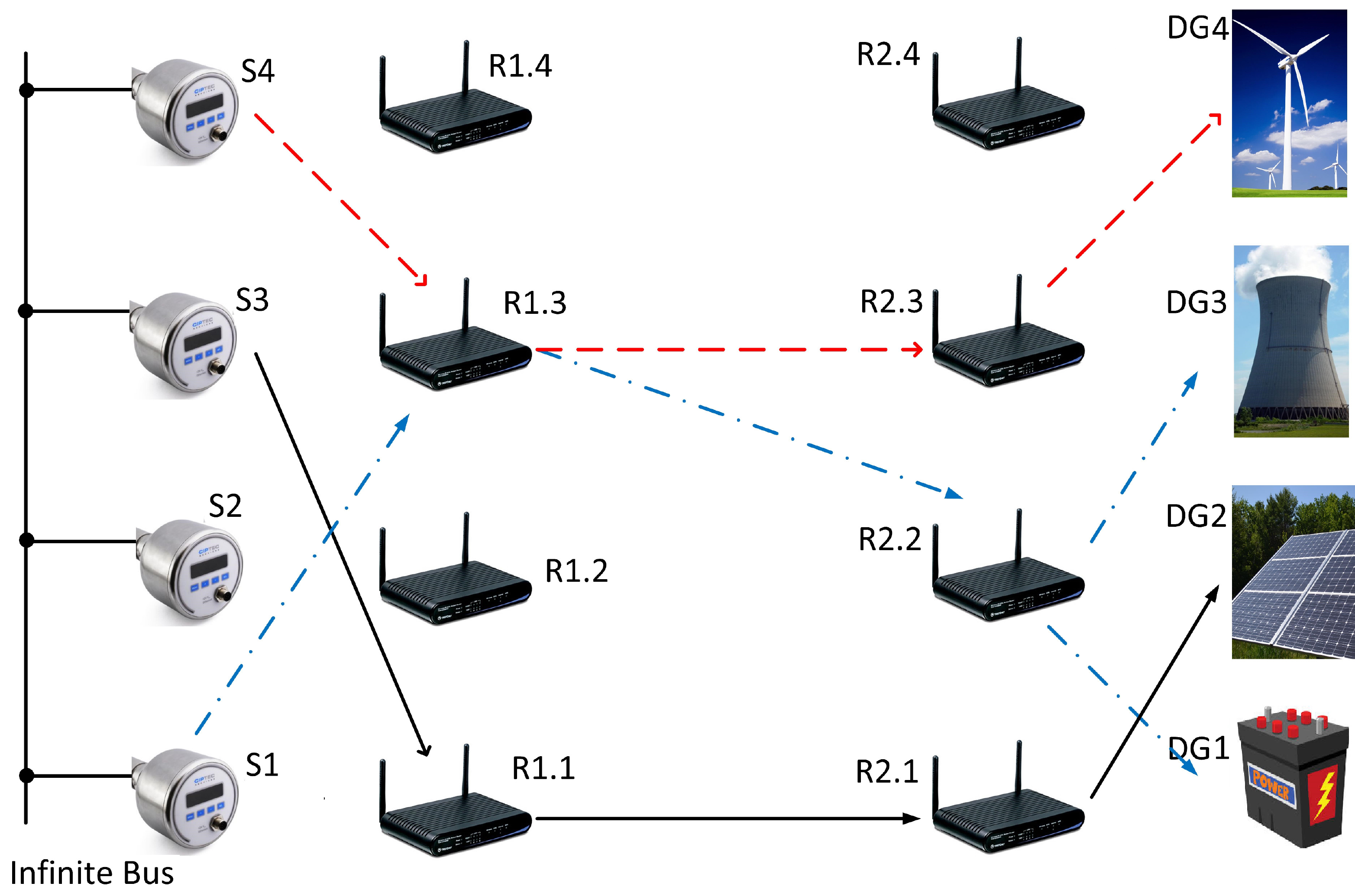}
\caption{An illustration of communications in CPES}
\label{fig:figure11}
\end{figure}

Let's refer Figure \ref{fig:figure11} which schematically depicts the voltage control of a four-bus distribution system operating in islanded mode. This implies that the infinite bus is disconnected as shown in this figure. This figure shows that the system consists of four sensors measuring the voltage at the points of common coupling (PCCs) (the point where the DGs are integrated into network), four DGs of varying capacities and two array of relay nodes for communication between sensors and controllers. The objective is to design a proper communication link between the controllers and sensors such that installed DGs can stabilize the voltage at a pre-defined reference value. 
This system has four inputs and four outputs. Such systems can be modeled as multi-input multi-output (MIMO) system. In this chapter we are concerned only with linear models.
\subsection{MIMO System Model} \label{Sec:MIMO_model}
The generic model of a linear MIMO system is given as:
\begin{equation}
\left.
\begin{aligned}
&\dot{x}(t)=Ax(t)+Bu(t)~~\\
&y(t)=Cx(t)\label{Eq:sys11}
\end{aligned}
\right\}
\end{equation}
where $x \in R^n, A \in R^{n \times n}, B \in R^{n \times m}, u \in R^m, y \in R^p$ and $C \in R^{p \times n}$\\
This model takes into account that there are $p$ sensors and $m$ control inputs (DGs in case of the smart grid system).
\subsection{Controller Structure} \label{Sec:MIMO_Control}
The control law is given as:
\begin{align}
u(t) = Ky(t)
\label{Eq:ut11}
\end{align}
where the matrix $K$ gives the structure of connections between the control input vector $u(t)$, i.e. DGs and sensor vector $y(t)$. This matrix $K$ has the following structure:
\begin{equation}
\nonumber
  k_{ij}\begin{cases}
    &=0 \text{, if no connection between sensor $j$ - controller $i$}\\
    &\neq 0 \text{, otherwise}.
  \end{cases}
\end{equation}
For instance, there are four controllers and four sensors in the MIMO system with reference to Figure \ref{fig:figure11}. Let the controller 1 (DG1) is connected to sensors 1 and 4 while sensors 2 and 4 are connected to controllers 2 and 3 (i.e. DG 2 and DG 3) respectively. Then, the connection matrix $K$ is given as
\begin{equation}
K=\left(
\begin{tabular}{cccc}
$K_{11}$ & 0 & 0 & $K_{14}$ \\
0 & $K_{22}$ & 0 & 0 \\
0 & 0 & 0 & $K_{34}$ \\
0 & 0 & $0 $ & 0
\end{tabular}
\right)\label{Eq:kmatrix11}
\end{equation}
Combining equations \eqref{Eq:ut11} and \eqref{Eq:sys11}, the closed loop system dynamics is obtained as
\begin{equation}
\dot{x}=Ax(t)+BKCx(t)=\bar{A} x(t) \label{Eq:eq411} \\
\end{equation}
where $\bar{A} \triangleq A+BKC$ represents the modified system matrix
\subsection{Optimization Problem} \label{Sec:MIMO_optimization}
Considering the linear system model as in equation \eqref{Eq:eq411}, and given a Lyapunov function $V(x)=x^TPx$, it is well known that equilibrium point $x^e$ goes to zero, if the following two inequalities hold simultaneously for all $x \neq 0$
\begin{eqnarray}
V(x)>0 \label{Eq:v11}\\
\dot{V}(x)<0 \label{Eq:v11.}
\end{eqnarray}
The rate derivative of the Lyapunov function $V$ for the system model \eqref{Eq:eq411} is obtained as:
\begin{eqnarray}
\dot{V}(x)=\dot{x}^TPx+x^TP\dot{x}=x^T(\bar{A}^TP +P\bar{A})x \label{Eq:Lyap11}
\end{eqnarray}
Given that the matrix $P$ is positive definite, equation \eqref{Eq:Lyap} implies that $\dot{V}$ will be negative define if the following condition holds true:
\begin{equation}
\bar{A} ^TP+P\bar{A} ~<~0 \label{Eq:Lyap211}
\end{equation}
Usually the system matrix $A$ is unstable. The matrix $P$ is computed by selecting a positive value of $\beta$ such that $\left( A - \beta I\right)$ is stable and the following condition holds true:
\begin{equation}
(A-\beta I)P+P(A-\beta I)^T=-I \label{Eq:ABPI11}
\end{equation}
It should be noted that for different value of $\beta$, the controller solution will be different. By replacing $\bar{A}$ by $A+BKC$ in \eqref{Eq:Lyap211}, and introducing a design parameter $\gamma$, the optimization problem in the form of linear matrix inequality is obtained as:
\begin{equation}
A^TP +\left(BKC\right)^TP+PA+PBKC+\gamma I<0 \label{Eq:unknown11}
\end{equation}
One can note that larger the $\gamma$ value, larger is the stability margin. As one optimizes equation \eqref{Eq:unknown11}, some constraints arise. First constraint is that $K_{ij}=0$ when the sensor $j$ is not connected to the controller $i$. We also would like to see that controller gains should be bounded, i.e. the second constraint is $||K||_2<\rho$ where $\rho$ is a user defined scalar value.
\section{Cyber Physical Smart Grid System} \label{Sec:2application}
This work adopts an IEEE 4 bus distribution feeder \cite{rp:ieee_4bus} as test system which is illustrated in Fig. \ref{fig:figure11}. As the system operates in islanded mode, the grid is disconnected from distribution feeder and the loads are catered by four DGs. The connection between the voltage sensors placed at PCCs and the controllers, a communication parameter has been identified as a prominent parameter in affecting the stability (physical parameter) of bus voltages. Entire power system has been modeled as linear model as proposed in \cite{rp:multicast_paper} which shall be the primary point of discussion in this work. The Cyber-physical system model can be better described by splitting the system into two parts: the physical system and communication network.
\subsection{Physical Model of the Grid in Island Mode}
\begin{figure}[h!]
\centering
\includegraphics[width=0.9\textwidth, height=7cm]{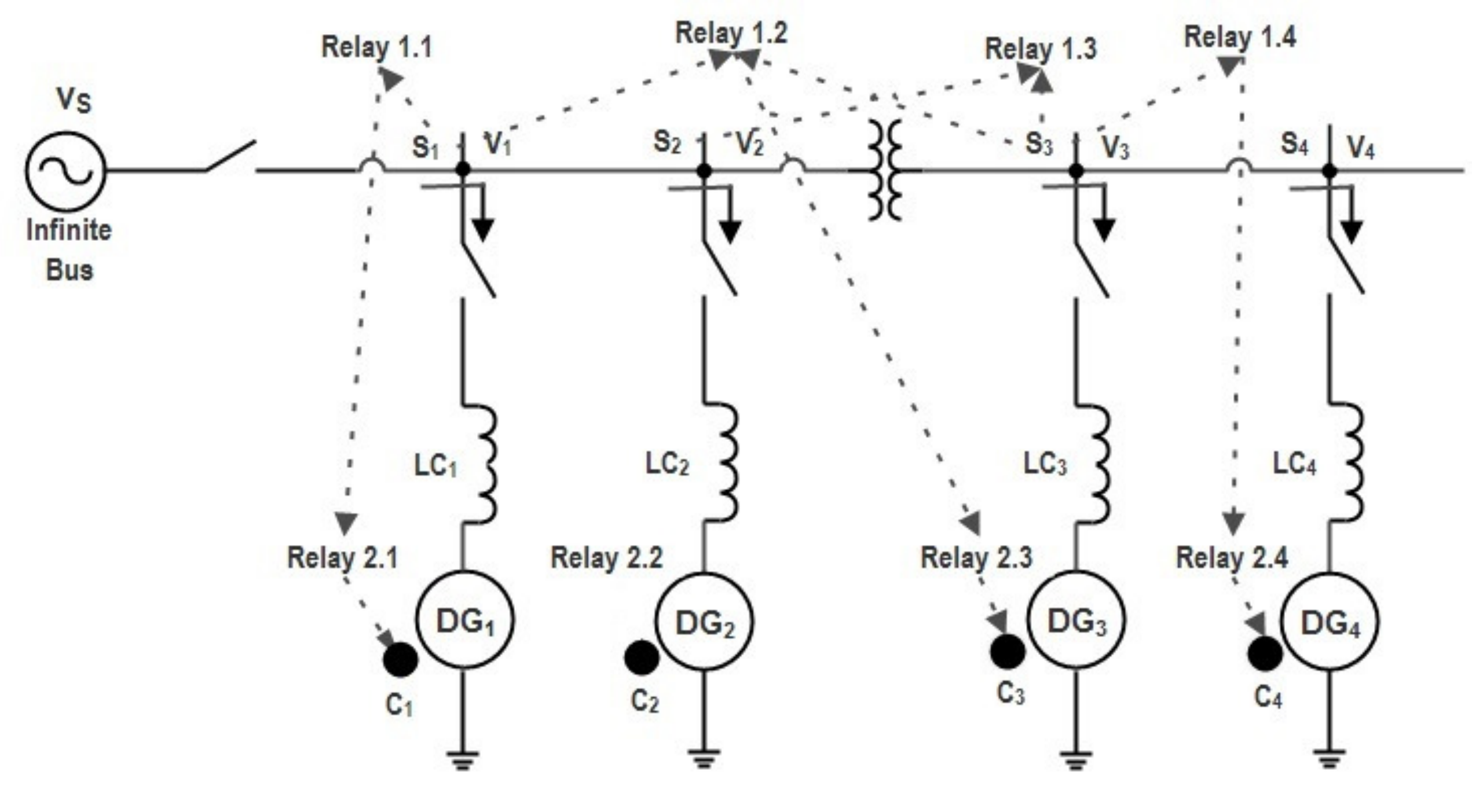}
\caption{Model of multiple DG sets connecting to the power network}
\label{fig: model of multiple DG set11}
\end{figure}

The model of a power network operating in island mode making source voltage $V_s~=~0$ i.e., the feeder is disconnected from the infinite bus as shown in Fig. \ref{fig: model of multiple DG set11}. The DGs are connected to the network at the PCCs through Power Electronic (PE) interfaces. An inverter and a DC link capacitor forms the PE interface. A coupling inductor separates inverter from the rest of the system.\\
The nodal voltage equations using Laplace transform for the above four bus system are given as:
\begin{equation} 
Y(s)v_t(s)=\frac{1}{s}Lv_c(s)
\label{Eq:yvsl11}
\end{equation}
where $Y(s)$ is the admittance matrix, $v_t \in R^p$ represents the vector of output voltages at PCC which is also known as sensor output vector, $v_c \in R^m$ represents the vector of DG output voltages which is also known as input vector in the current formulation, and $L \in R^m$ represent the coupling inductors between DGs and the feeder network. 
The exact expression for the admittance matrix for the four bus system \cite{rp:multicast_paper} is given as: 
\begin{equation} \label{Eq:Ybus11}
Y(s)=\left[ Y_1~~ Y_2~~ Y_3~~ Y_4\right]+\frac{1}{Ls}I\\
\end{equation}
where $Y_i$ represents $i^{th}$ column vector of $Y(s)$. Each column vector is represented as follows:
\begin{equation}
\nonumber
Y_1(s)=\left(
\begin{tabular}{c}
$\small{\frac{1}{0.175+0.0005s}}$\\
$\small{\frac{-1}{0.175+0.0005s}}$\\
$\small{0}$ \\
$\small{0}$ \\
\end{tabular}
\right)
\end{equation}
\begin{equation}
\nonumber
Y_2(s)=\left(
\begin{tabular}{c}
$\small{\frac{-1}{0.175+0.0005s}}$ \\
$\small{\frac{1}{0.175+0.0005s}+\frac{1}{0.1667+0.0004s}}$ \\
$\small{\frac{-1}{0.1667+0.0004s}}$\\
$\small{0}$\\
\end{tabular}
\right)
\end{equation}
\begin{equation}
\nonumber
Y_3(s)=\left(
\begin{tabular}{c}
$\small{0}$\\
$\small{\frac{-1}{0.1667+0.0004s}}$ \\
$\small{\frac{1}{0.1667+0.0004s}+\frac{1}{0.2187+0.0006s}}$\\
$\small{\frac{-1}{0.2187+0.0006s}}$\\
\end{tabular}
\right)
\end{equation}
\begin{equation}
\nonumber
Y_4(s)=\left(
\begin{tabular}{c}
$\small{0}$\\
$\small{0}$ \\
$\small{\frac{-1}{0.2187+0.0006s}}$\\
$\small{\frac{1}{0.2187+0.0006s}+\frac{1}{12.3413+0.014s}}$\\
\end{tabular}
\right)
\end{equation}
The Laplace transform model as given in equation \eqref{Eq:yvsl11} can be converted to linear state-space representation in time-domain \eqref{Eq:sys11} where system and control matrices are represented as:
\begin{align}
A&=10^{3} \left[
\begin{tabular}{cccc}
0.1759 & 0.1768 & 0.5110 &1.0360 \\
-0.3500 & -0.0000 & -0.0000 & -0.0000 \\
-0.5442 & -0.4748 & -0.4088 & -0.8288 \\
-0.1197 & -0.5546 & -0.9688 & -1.0775\\
\end{tabular}
\right]\\
B&=10^3\left[
\begin{tabular}{cccc}
0.0008 & 0.3342 & 0.5251 & -1.0360 \\
-0.3500 & -0.0000 & -0.0000 & -0.0000 \\
-0.0693 & -0.0661 & -0.4201 & -0.8288 \\
-0.4349 & -0.4142 & -0.1087 & -1.0775\\
\end{tabular}
\right]
\end{align}
The state vector $v_t$ represents voltages at PCCs. Since these voltages also represent output $y$, the output matrix $C=I$ where $I \in R^{n\times n}$ is the identity matrix. Obviously, $n=p=4$ in the considered model of the four-bus system. It is to be noted that the perturbations as noise are not considered in this model. The objective of the voltage control in the feeder network is to maintain all PCC voltages to a pre-defined reference value $V_{ref}$. Formulating the problem as that of the state regulation, the state vector $x$ is set as $x~=~v_t-V_{ref}$.
\subsection{Communication Network Structure} \label{Sec:communication network structure}
The communication network considered in this paper consists of three types of nodes: sensors, relays and controllers. It has been assumed that the controllers and sensors are all equipped with communication interfaces, either wired or wireless. A group of additional relay nodes are present which make up the connection between sensors and controllers as shown in Figures \ref{fig:figure11} \& \ref{fig: model of multiple DG set11}.
The following two assumptions have been made to make the present analysis easier as also done in \cite{rp:multicast_paper}:
\begin{enumerate}
\item The data that flows from sensors to controllers are continuous i.e. there is no delay, no packet loss and no network uncertainties. This kind of assumption can be done when the communication speed is very fast as compared to the physical system dynamics.
\item For any link between any two nodes say $n_1$ and $n_2$, the transmission of data from one sensor requires one unit of bandwidth. The bandwidth denoted by $bw_{n_1,n_2}$ is assumed to be an integer. \\
\begin{equation}
\sum_{n_s=1}^{N_s} \Delta (n_s,n_1,n_2) \leq bw_{n_1,n_2} \\
\end{equation}
where $\Delta(n_s,n_1,n_2)=1$ if the data from sensor $n_s$ passes through link between $n_1$ and $n_2$, otherwise, $\Delta (n_s,n_1,n_2)=0$.\\
\end{enumerate}
The communication from sensors to controllers is established via two arrays of relay nodes, thus forming a $4\times4$ array of nodes. It is assumed that the bandwidth constraint equals 1. This would mean that any node or link would accept and transmit data from only one sensor. The data of a particular node can be transmitted only to its next hop neighbors. The controllers are allowed to receive data from multiple sensors.
\subsection{Connection Design for Decentralized Control}
\begin{equation} \label{Eq:uky11}
u(t)=Ky(t)\\
\end{equation}
A decentralized output feedback control as given in \eqref{Eq:uky11} is considered. It is shown that the connection matrix $K$ used in decentralized control plays a very important role in merging both the cyber aspect that is the communication topology and the physical aspect which is the feedback gain for the controller. This is because the structure of $K$ matrix decides the communication topology while the value of elements in $K$ matrix actually generate the control command required for maintaining a physical variable like bus voltage in the power system. The generalized connection design process has been elaborated in the following section.
\section{A Generalized Sensor-Controller Connection Design} \label{Sec:2MIMO_proposed_algo} 
\begin{figure}[!ht]
\begin{minipage}[b]{.5\linewidth}
\centering
\includegraphics[width=0.9\textwidth, height = 5cm]{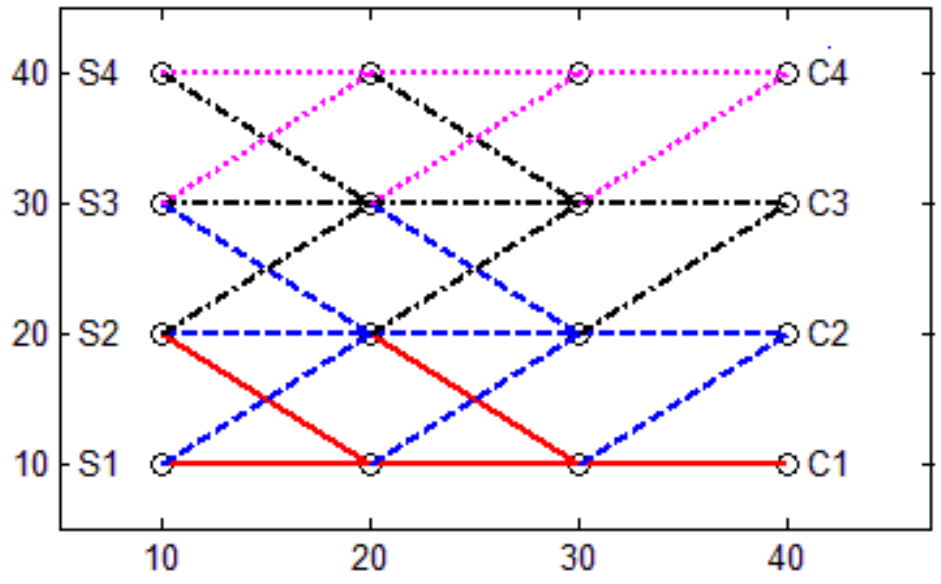}
\caption{Possible connections}\label{Fig:Network_Connections11}
\end{minipage}%
\begin{minipage}[b]{.5\linewidth}
\centering
\includegraphics[width=0.9\textwidth, height =5cm]{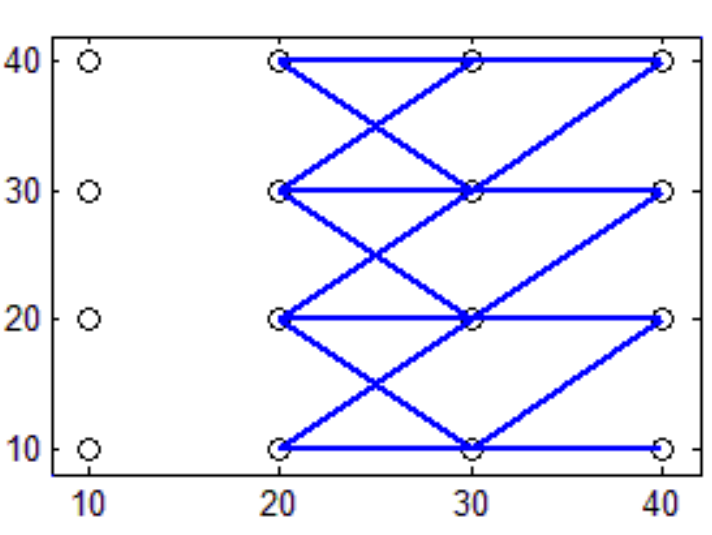}
\caption{Step 5 output}\label{Fig:step511}
\end{minipage}
\end{figure}

\begin{figure}[!ht]
\centering
\includegraphics[width=0.9\textwidth, height = 5cm]{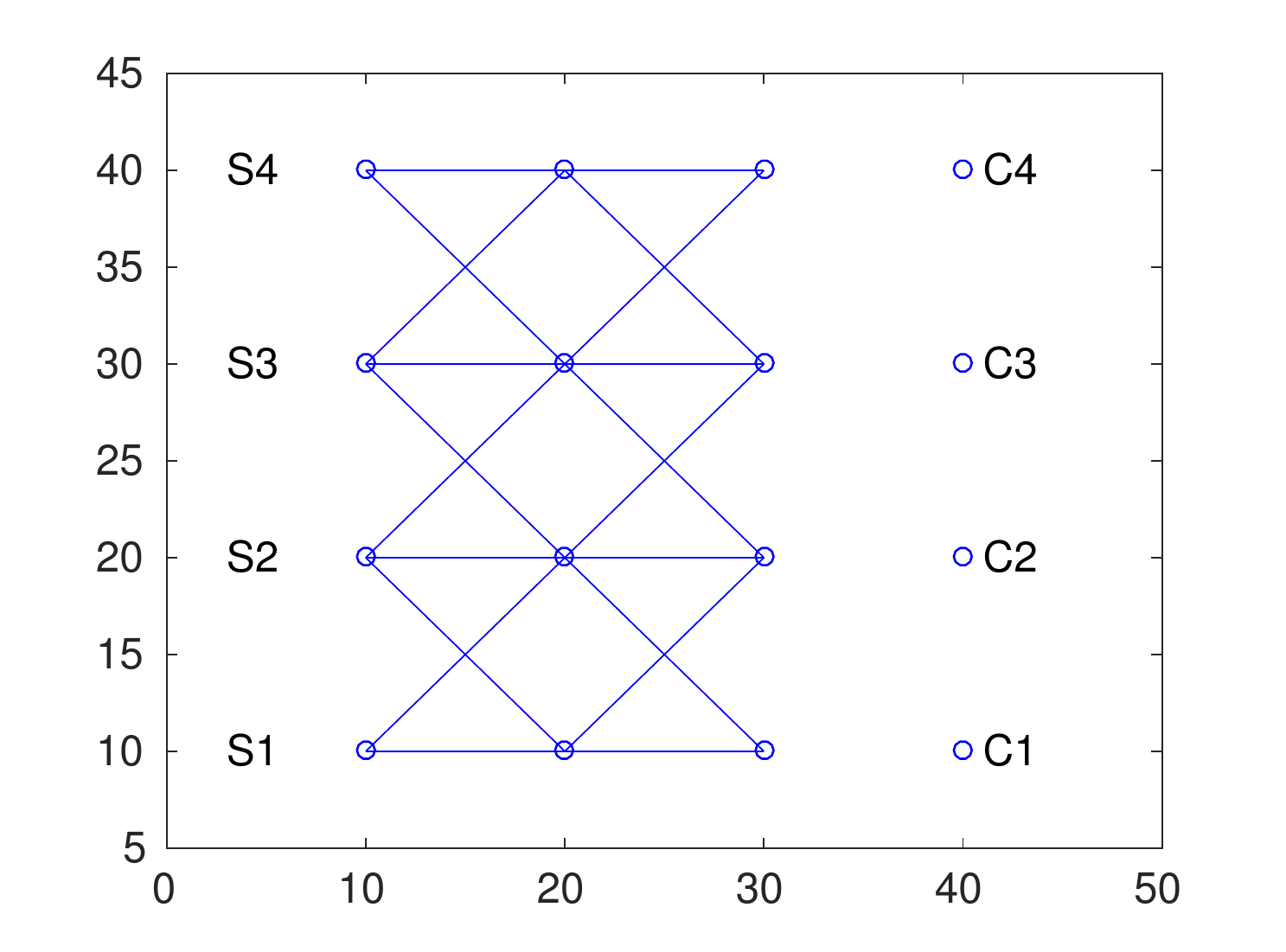}
\caption{Grouping of simultaneously existing connections}
\label{Fig:step611}
\end{figure}

In this section, a novel generalized framework for optimal sensor-controller connection design is presented. The process of finding $K$ as required for equation \eqref{Eq:unknown11} is the root problem to be solved. This would mean to finding both the structure as well as the values present in the  matrix $K$. 
An algorithm is proposed to set out an intelligent search in contrast to exhaustive search for finding the matrix $K$ which provides maximum stability to the system and demonstrated as follows: \\
\textbf{Step 1:} Enter Specifications \\ 
Enter the number of sensors, controllers and intermediate relays.\\
\textbf{Step 2:} Define connection constraints imposed on the system.\\
For the application considered the connection constraints are defined based on the assumptions made in section \ref{Sec:communication network structure}.\\ 
\textbf{Step 3:} Find all possible connections with connection constraints.\\
Starting from the controller layer, successively map possible connections to previous layers till sensor layer respecting the connection constraints.\\ 
Executing this results in a set of $47$ connections as shown in the Figure \ref{Fig:Network_Connections11} below
\\
\textbf{Step 4:} Define the system's physical resource constraints.\\
For the application considered, physical resource constraints are defined as bandwidth constraint.\\
\textbf{Step 5:} Find the set of connections which can exist together without violating physical resources constraint.\\
\emph{for} $i=g\footnote{represents the index of final layer, in this case $g = 4$}:-1:1$\\
$\left\{\right.$\\
$~~~$\emph{if}$~(i\neq h\footnote{index of layer where physical resource constraints are present. For the application considered, physical resource constraints are at sensors, hence h=1})$\\
$~~~$Add connections to its previous layer considering connection constraints.\\
$\left.\right\}$\\
From this, we get $18$ connections as shown in Figure \ref{Fig:step511}.\\
\textbf{Step 6:} Form groups from the results of step 5 in such a way that every group starts from the layer nearest to $h$ as shown in Figure \ref{Fig:step611} (in this case the nearest layer is that of relay 1).\\
\\
\textbf{Step 7:} Determine the secondary objective and subgroup them accordingly\footnote{two cases have been studied in this paper. Other parameters can also be included}.\\
Case 1 : Reliability (this means every controller should be connected to maximum possible sensors. For this application it is $2$).\\
Case 2 : Cost (a different design has been made where controller 2 \& 3 have higher rating compared to controller 1 \& 4, so threshold value of atleast two connections has been defined for controller 2 \& 3).\\
Note that each sub-group is named after its relay index and all the figures presented below in this section pertain to case-1.
\begin{figure}[!ht]
\centering
\includegraphics[width=0.9\textwidth, height = 7cm]{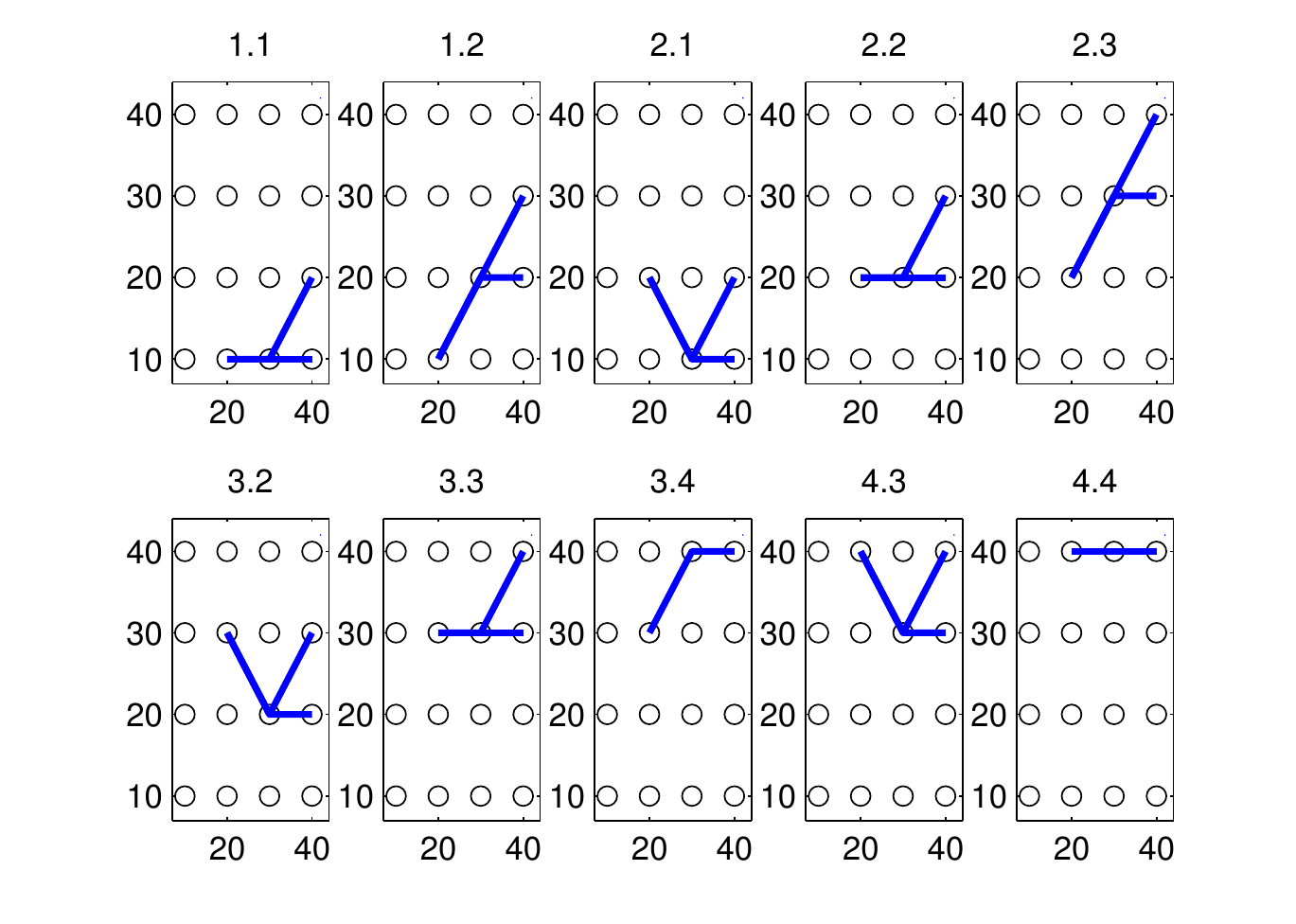}
\caption{Sub groups for case 1}
\label{Fig:step7.111}
\end{figure}\\
$~$\textbf{Step 8:} Find all sets of complete co-existing sub-groups and place them in set $v$.\\
$~$\textbf{Step 8.1:} Define $d$ divisions each representing a group obtained in step 6. From Figure \ref{Fig:step7.111} the divisions are tabulated in Table \ref{Tab:Division11}.\\
\begin{table}
\centering
\begin{minipage}[b]{.3\linewidth}
\centering
\caption{Divisions} \label{Tab:Division11}
\begin{tabular}{|c|c|} \hline
d1 & 1.1  1.2\\ \hline
d2 & 2.1  2.2  2.3\\ \hline
d3 & 3.2  3.3  3.4\\ \hline
d4 & 4.3  4.4\\ \hline
\end{tabular}
\end{minipage}\hfill
\begin{minipage}[b]{.37\linewidth}
\centering
\caption{Co-existing subgroups set $v$} \label{Tab: subgroups set11}
\begin{tabular}{|c|c|} \hline
v1 & 1.1, 2.2, 3.3, 4.4\\ \hline
v2 & 1.1, 2.2, 3.4, 4.3\\ \hline
v3 & 1.1, 2.3, 3.2, 4.4\\ \hline
v4 & 1.2, 2.1, 3.3, 4.4\\ \hline
v5 & 1.2, 2.1, 3.4, 4.3\\ \hline
\end{tabular}
\end{minipage}\hfill
\begin{minipage}[b]{.33\linewidth}
\centering
\caption{Sensors set $w$} \label{Tab: sensor set11}
\begin{tabular}{|c|c|} \hline
w1 & s1, s2, s3, s4\\ \hline
w2 & s1, s2, s4, s3\\ \hline
w3 & s1, s3, s2, s4\\ \hline
w4 & s2, s1, s4, s3\\ \hline
w5 & s2, s1, s3, s4\\ \hline
\end{tabular}
\end{minipage}\hfill
\end{table}
\textbf{Step 8.2:} \emph{for }$i = 1:d$\\
$\left\{\right.$\\
$~~~$\textbf{Step 8.2.1:} List the available sub groups.\\
$~~~$\textbf{Step 8.2.2:} Connect these subgroups to previous division subgroups without violating physical resource constraints (in each subgroup name, the number after the decimal point has to be different while connecting each division to its previous division as illustrated iteration-wise in Figure \ref{Fig:Group of sub groups11}).\\
$\left.\right\}$
\begin{figure}[!ht]
\begin{minipage}[b]{.6\linewidth}
\centering
\includegraphics[width=0.9\textwidth, height = 5cm]{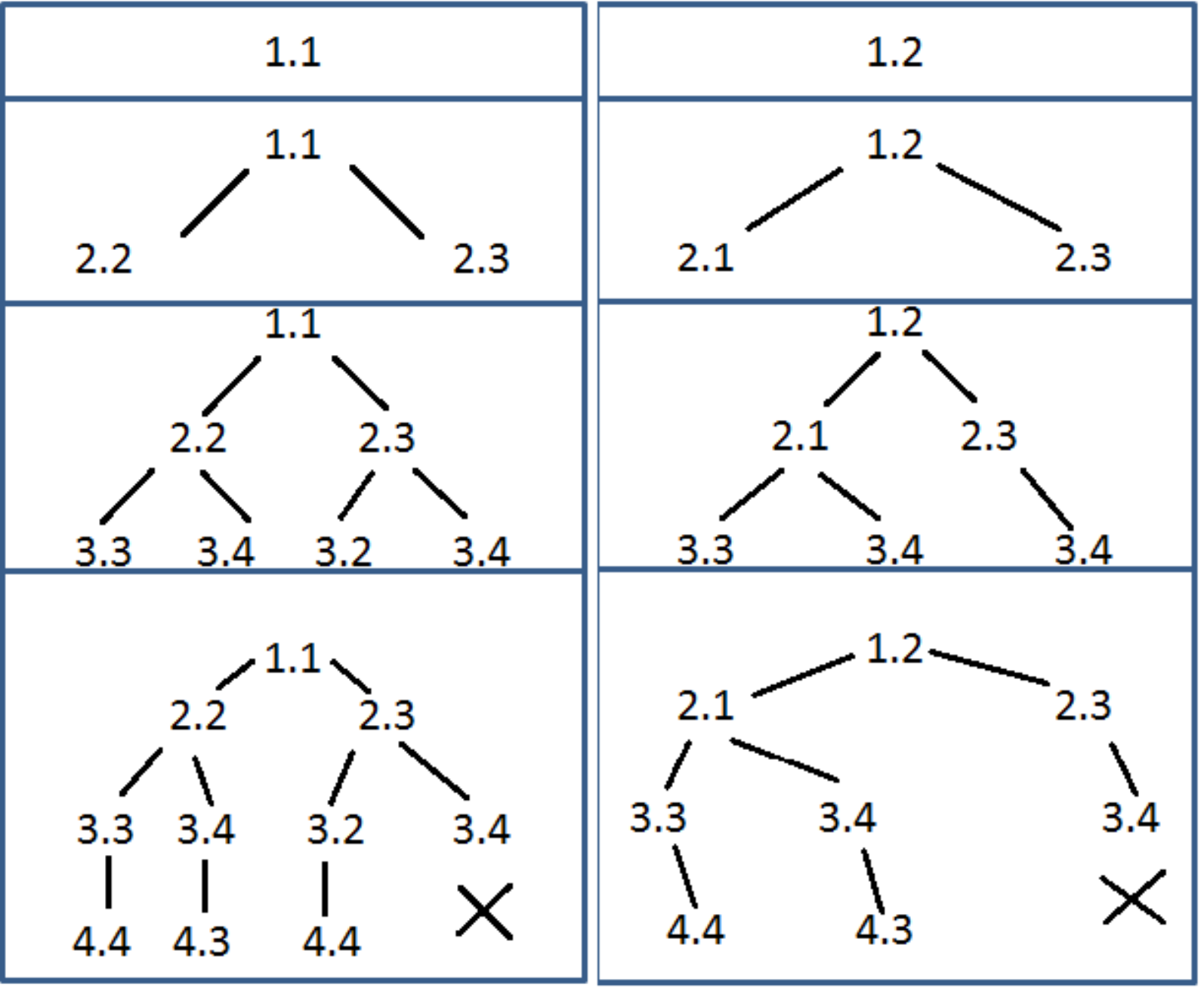}
\caption{Group of sub groups}\label{Fig:Group of sub groups11}
\end{minipage}%
\begin{minipage}[b]{.4\linewidth}
\centering
\includegraphics[width=0.9\textwidth, height = 5cm, keepaspectratio]{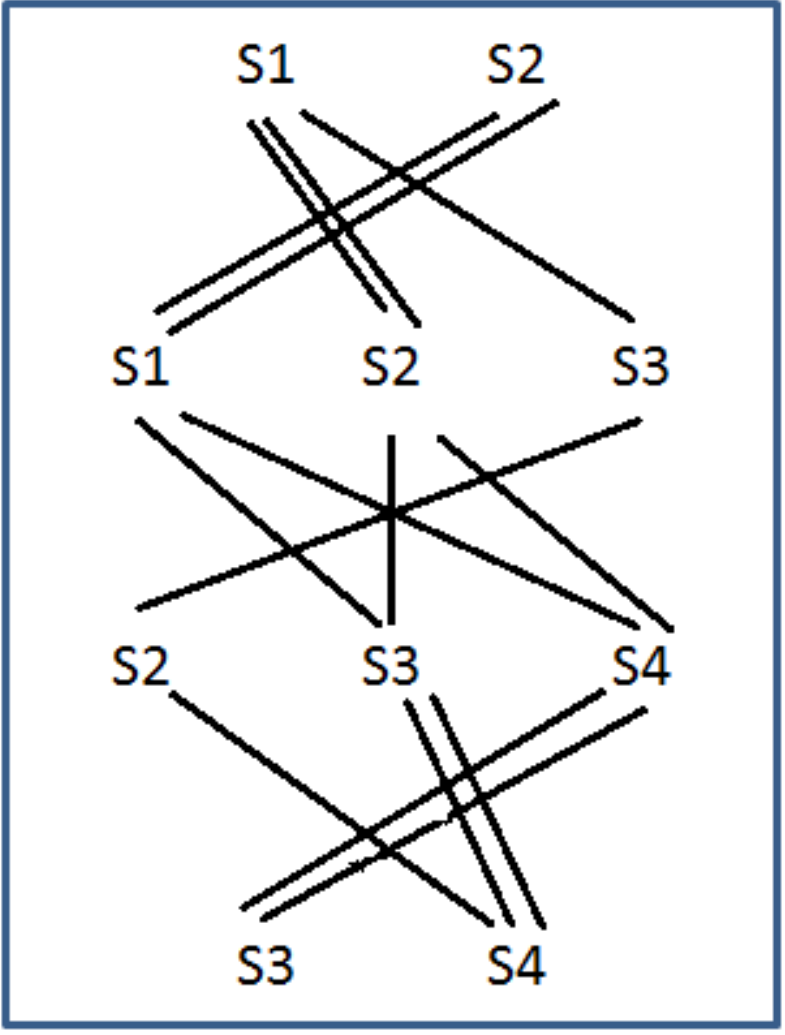}
\caption{Sensors set}\label{Fig:sensors replacement11}
\end{minipage}
\end{figure}\\
%
%
\textbf{Step 8.3:} Write each set of subgroups into set $v$ found at the end of step 8.2 if they are complete i.e. if they have $d$ number of subgroups in their set as shown in Table \ref{Tab: subgroups set11}.\\
\textbf{Step 9:} Find the available complete coexisting elements set $w$ in the layer where physical resource constraints are present to connect to each of the element of set $v$.\\ 
\emph{for} $i = 1:d$ \\
$\left\{\right.$\\
$~~$\textbf{Step 9.1:} Find the number of elements available in the layer where physical resource constraints are present to connect to each of the complete co-existing subgroups set as per the connection constraint.\\
$~~$\textbf{Step 9.2:} Connect them to the elements found in the previous iteration without violating physical resource constraints (here sensor number should be different from rest of the sensor numbers selected in previous iterations).\\
$\left.\right\}$\\
The elements of set $w$ have been shown in Figure \ref{Fig:sensors replacement11} and tabulated in Table \ref{Tab: sensor set11}.\\
\textbf{Step 10:} Find the complete set of connections by connecting each member of set $v$ to that of set $w$.\\
From Tables \ref{Tab: subgroups set11} \& \ref{Tab: sensor set11} we obtain $25$ sets of connections (for case-1) that can simultaneously exist. Each connection set consists of $7$ paths between sensors and controllers.\\ 
\textbf{Step 11:} Solve the matrix equation in \eqref{Eq:ABPI11} for all complete set found in Step 10.\\
\textbf{Step 12:} Solve optimization problem in equation \eqref{Eq:unknown11} using LMI and obtain $\gamma$.\\
\textbf{Step 13:} List the connection sets with maximum $\gamma$.\\
\textbf{Step 14:} From step 13 choose the connection with shortest path and obtain the corresponding $K$.\\
\section{Simulation results} \label{Sec:2results}
In this section, simulation studies have been performed for three different scenarios:
\begin{enumerate}
\item Greedy algorithm based connection design when all DG capacities are equal 
\item Optimal connection design based on proposed algorithm when all DG capacities are equal 
\item Optimal connection design based on proposed algorithm when DG capacities are different
\end{enumerate} 
All simulations have been carried out using MATLAB and YALMIP toolbox \cite{rp:yalmip}. In all the scenarios, $\rho=5$ i.e. the 2-norm of the feedback gain matrix $K$ is constrained within 5 and the value of $\beta$ is chosen as 5000.
\subsection*{Scenario 1}
\begin{figure}[!ht]
\begin{minipage}[b]{.49\linewidth}
\centering
\includegraphics[width= 0.9\textwidth, height=3.5cm, keepaspectratio]{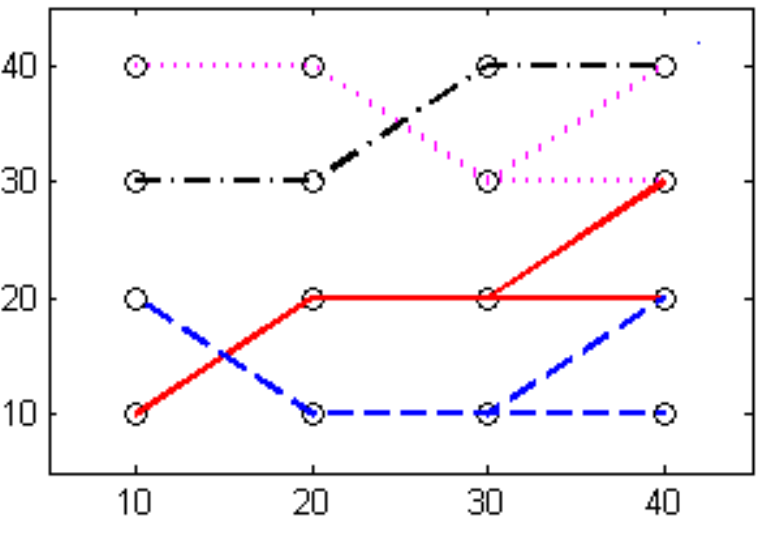}
\caption{Greedy algorithm results}\label{Fig:Scenario1_plot11}
\end{minipage}%
\begin{minipage}[b]{.51\linewidth}
\centering
\includegraphics[width= 0.9\textwidth, height=3.5cm, keepaspectratio]{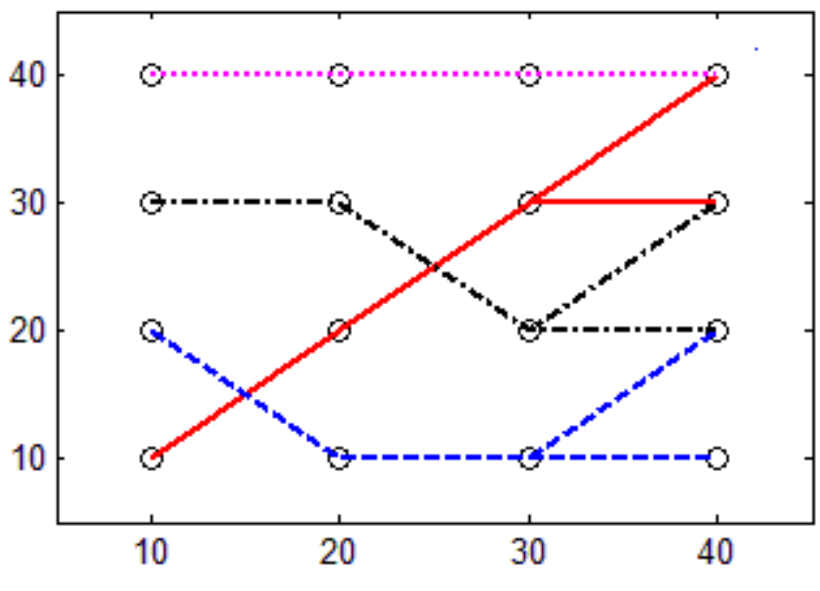}
\caption{Final topology for scenario 2}\label{Fig:Scenario2_plots_reduced11}
\end{minipage}
\end{figure}

The greedy algorithm proposed in \cite{rp:multicast_paper} has been implemented to find a routing mode that stabilizes the power system.The set of connections and the trend of $\gamma$ have been reproduced with the chosen $\beta$. The results obtained for this case are $\gamma$ = 30.7836, maximum eigenvalue value: -213.2269 with $K$ matrix in \eqref{Eq:K_scenario17} and the final topology is presented in Figure \ref{Fig:Scenario1_plot11}\\ 
\begin{equation}
K=\left(
\begin{tabular}{cccc}
0 & 1.3568 & 0 & 0 \\
-0.9309 & -0.7023 & 0 & 0 \\
-0.4902 & 0 & 0 & -2.1480 \\
0 & 0 & 0.2385 & 0.2955\\
\end{tabular}
\right)\label{Eq:K_scenario17}
\end{equation}
\subsection*{Scenario 2}
The proposed algorithm mentioned in Section \ref{Sec:MIMO_proposed_algo} has been applied to find the set of connections between sensors and controllers. This scenario comes under case 1 of the proposed algorithm where reliability has been defined as the secondary objective. The resultant configuration was found to come with $\gamma$ increased to 74.3584 and maximum eigenvalue reduced to -431.7517. This suggests that the proposed algorithm finds a better and stable solution when compared to the greedy approach. The final topology obtained is shown in Figure \ref{Fig:Scenario2_plots_reduced11} and $K$ matrix as in \eqref{Eq:K_scenario27}
\begin{equation}
K=\left(
\begin{tabular}{cccc}
0 & 1.3877 & 0 & 0 \\
0 & -0.7842 & -1.3757 & 0 \\
0.2672 & 0 & 1.6004 & -0\\
1.4442 & 0 & 0 & 0.8173\\
\end{tabular}
\right)\label{Eq:K_scenario27}
\end{equation}
\subsection*{Scenario 3}
\begin{figure}[!ht]
\begin{minipage}[b]{.49\linewidth}
\centering
\includegraphics[width= 0.9\textwidth, height=3.5cm, keepaspectratio]{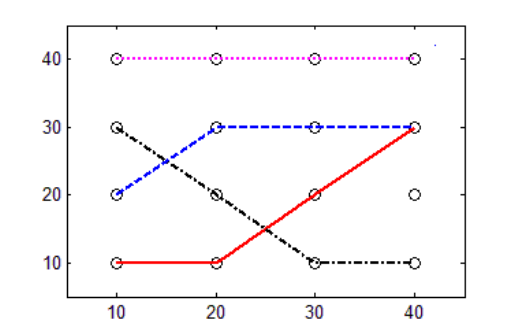}
\caption{Intermediate topology}\label{Fig:Scenario31_plots1}
\end{minipage}%
\begin{minipage}[b]{.51\linewidth}
\centering
\includegraphics[width= 0.9\textwidth, height=3.5cm, keepaspectratio]{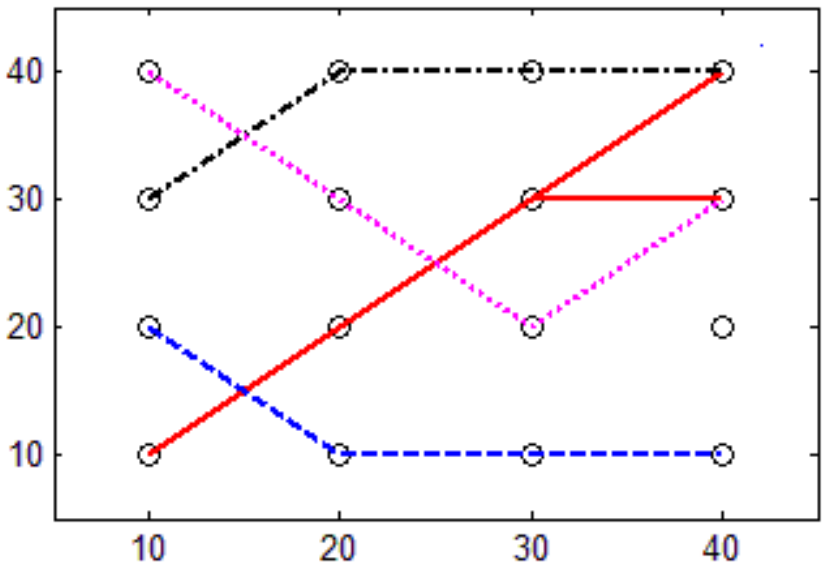}
\caption{Final topology for scenario 3}\label{Fig:Scenario32_plots1}
\end{minipage}
\end{figure}

In practical systems, different generators can have different capacities and hence communication resources need to be allocated accordingly. This scenario deals with finding the connections when the secondary objective is to minimize cost when DGs are of different capacity. It has been assumed that cost is proportional to number of connections between sensors and DGs. Here, controllers 2 \& 3 have higher capacity compared to controllers 1 \& 4. A threshold value has been set such that the controllers 2 \& 3 each receive atleast two inputs or no input as shown in Figure \ref{Fig:Scenario31_plots1}.\\
\\
It has been observed that there are less no.of connections which is advantageous when cost minimization is the main objective. Set of connections found are much economical but, they are not stabilizing the system as $\gamma$ is -47.1822. Hence, with little compromise on cost the next most economical connection set has been adopted as per the algorithm increasing $\gamma$ to 19.9634 and reducing eigenvalue to -127.297 thus, making system stable. The final topology is shown in Figure \ref{Fig:Scenario32_plots1} and $K$ matrix is mentioned in \eqref{Eq:K_scenario37}.
\begin{equation}
K=\left(
\begin{tabular}{cccc}
0 & 1.0332 & 0 & 0 \\
0 & 0 & 0 & 0 \\
-0.2090 & 0 & 0 & -2.0210\\
0.2409 & 0 & -0.0137 & 0\\
\end{tabular}
\right)\label{Eq:K_scenario37}
\end{equation}
\section{Summary}\label{Sec:2sum}
This chapter presents a generalized novel framework for optimal sensor-controller connection design in decentralized control of MIMO systems. The salient features of this algorithm are:
\begin{enumerate}
\item It can be applied to system containing any number of sensors, controllers and relays
\item It can deal with any type of connection constraints present in the system 
\item Any sort of physical resource constraints present in the real world system like bandwidth in this paper can be taken into account while finding the optimal connection set
\item It also provides scope to add secondary objectives like reliability and cost along with stability, the primary objective while finding the connections
\item The effect of variable capacities present in any of the system elements like controllers in scenario 3 can also be taken into account
\end{enumerate}
The proposed approach has been applied to study the effect of communication network design on stability for voltage control problem in cyber physical smart grid system. Simulation studies have been performed for three different scenarios: (i) Greedy algorithm based connection design when all DG capacities are equal (ii) Optimal connection design based on proposed algorithm when all DG capacities are equal and (iii) Optimal connection design based on proposed algorithm when DG capacities are different. It has been found that the proposed algorithm provides the connection set which gives maximum stability to the power system unlike the greedy scheme in scenario 1. Also, when analyzing a system with variable DG capacities in scenario 3 it has been found that stability is achieved with lesser number of connections to satisfy secondary objective which is cost. Currently, we are working on developing a wholesome design of the microgrid system that can take into account network uncertainties like delay and packet loss as well as nonlinearity of grid dynamics into account.
 
 \chapter{Cyber Architecture Development for Experimental Validation of Control Algorithms\label{AFSMC}}
\section{Introduction}
This section deals with the development of communication hardware required for networking the sensors and controllers as discussed in the previous sections. A communication test-bed consisting of wireless nodes has been designed and fabricated to function as the cyber system for the cyber-physical smart microgrid. As of now, the developed cyber system has been coupled with a virtual grid running on a server to test the developed algorithms for optimal sensor-controller connections as shown in the Figure \ref{Fig:virtual_grid}.
\section{Description}
\subsection{Communication Protocol}
The Wi-Fi mode of communication has been chosen for the purpose. The Message Queuing Telemetry Transport (MQTT) communication protocol has been adopted owing to the following features:
\begin{enumerate}
\item Publish-Subscribe as opposed to Request-Response
\item Light weight, designed for low bandwidth applications
\item Aimed at low complexity and low power applications
\item Last will message: sent when node is disconnected
\item Allows encrypted data transfer using Transport Layer Security (TLS)/Secure Sockets Layer (SSL)
\item Option for authentication of client
\item Broker/server authorization to restrict the client
\end{enumerate}
\subsection{Hardware Setup}
\begin{figure}[!ht]
\centering
\includegraphics[width=0.9\textwidth, height = 7cm]{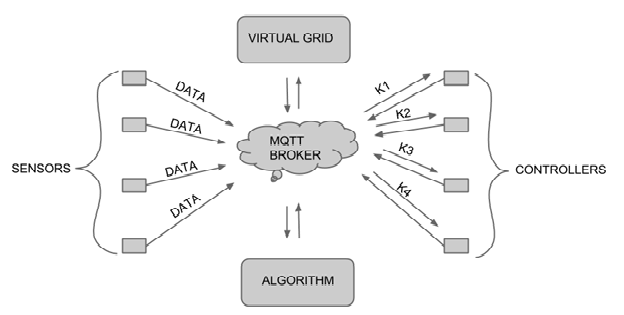}
\caption{Virtual Grid}
\label{Fig:virtual_grid}
\end{figure}
\begin{figure}[!ht]
\centering
\includegraphics[width=0.9\textwidth, height = 9cm]{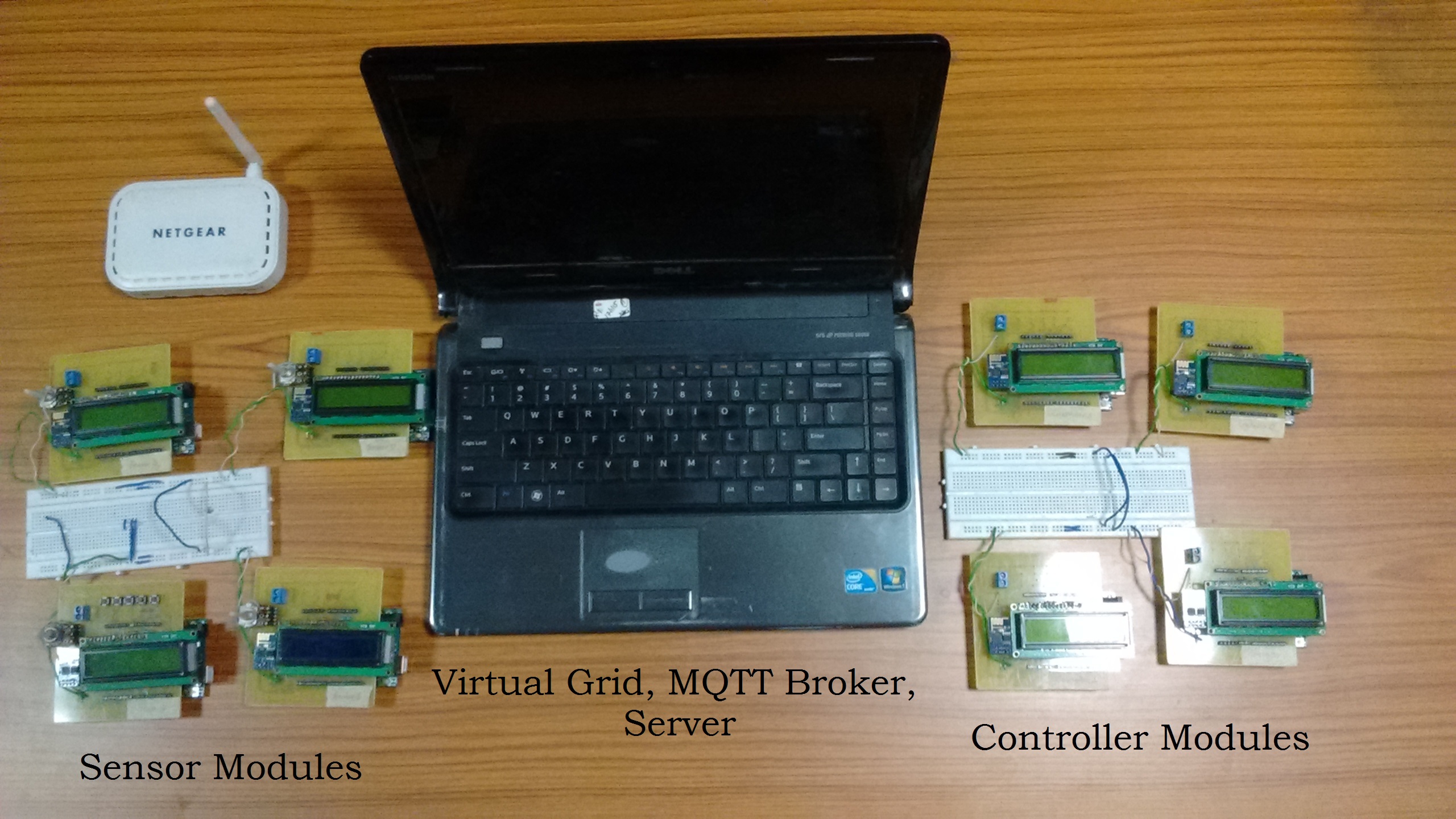}
\caption{Experimental Setup}
\label{Fig:hmw_Copy.jpg}
\end{figure}
Figure \ref{Fig:virtual_grid} depicts the block diagram of components adopted  for realizing the routing algorithms via. virtual grid and the hardware setup can be found in Figure \ref{Fig:hmw_Copy.jpg} . The sensor modules collect the data from the virtual grid running on a server and send them to an MQTT broker which sends the data to the appropriate controllers as per the inputs received from the algorithm to be tested. The controllers deliver the data pertaining to the control action to be taken to the virtual grid which process the control action to manipulate the state variables being sensed.\\

\subsection{Components in the Communication Module}
The overall communication architecture consists of several communication modules which can act as sensors, relays or controllers. Each module consists of following components:\\

~~\textbf{Microcontroller} It performs tasks, processes data and controls the functionality of other components in the individual module (node). Microcontrollers are best choice for embedded systems because of their flexibility to connect to other devices, programmability, less power consumption etc.\\

\textbf{Transceiver} Sensor nodes make use of Wi-Fi 2.4GHz band for communication. The various choices of wireless transmission media are Radio frequency, Optical communication (Laser) and Infrared. Laser requires less energy but needs line-of-sight for communication and also sensitive to atmospheric conditions. Infrared like Laser needs no antenna but it is limited in its broadcasting capacity. Radio frequency based communication is most relevant that fits to most of the Wireless Sensor Network application.\\

\textbf{LCD Display Unit} This components displays the value sensed by the sensor module or the control action taken by the controller module. \\  

\textbf{Sensing Unit} It senses and measures the voltage at PCC and with the help of microcontroller and transceiver the measured data reaches the server for computation and further manipulation. This is attached to the communication module to make it a complete sensor module.\\

\textbf{Controlling Unit} It takes commands from the server and receives it with the help of transceiver and controls the system accordingly. It controls the voltage output of DGs (Distribution Generators) to meet requirements of the Grid. This is attached to the communication module to make it a complete controller module.\\

\begin{figure}[ht!]
\centering
\includegraphics[width=0.9\textwidth, height = 7cm]{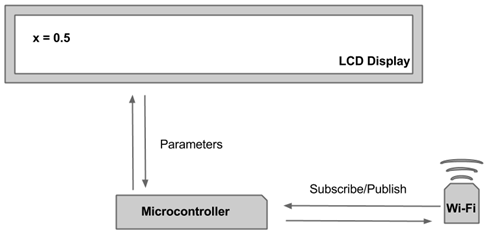}
\caption{Communication module Block Diagram}
\label{Fig:mqtt}
\end{figure}

\begin{figure}[ht!]
\centering
\includegraphics[width=0.9\textwidth, height = 7cm]{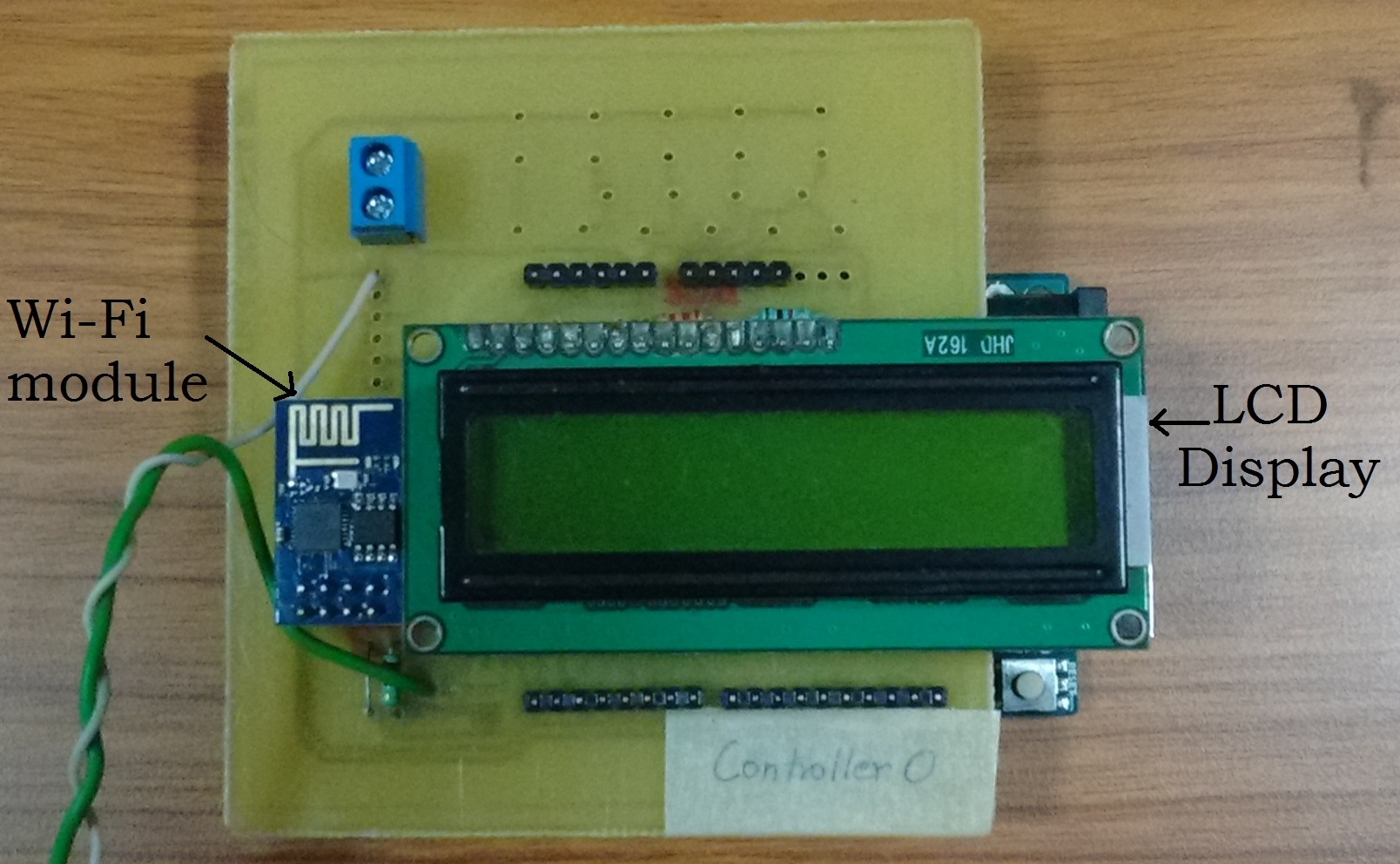}
\caption{Communication module Hardware}
\label{Fig:module}
\end{figure}

Various components in communication module can be found in Figure \ref{Fig:mqtt} \&  \ref{Fig:module}. The microcontroller and Wi-Fi modules used for the realization of communication architecture are Arduino UNO and EPS Wi-Fi module respectively. In Figure \ref{Fig:module} the Arduino board has been placed beneath the LCD display. \\
\\
\section{Summary}
In this chapter we have discussed the development of communication hardware required for networking the sensors and controllers as discussed in the previous sections. A communication test-bed consisting of wireless nodes has been designed and fabricated to function as the cyber system for the cyber-physical smart microgrid. The developed cyber system has been coupled with a virtual grid running on a server to test the developed algorithms for optimal sensor-controller connections. It also can be useful for other cyber physical applications.

 
 \chapter{Robust and Adaptive Formulation with Sensor-Controller Connection Design Algorithm\label{Modeling}}

\section{Introduction}
In this chapter, we have added robust and adaptive Formulation with our proposed sensor-controller connection design algorithm. This formulation is capable to keep our system robust under various critical conditions like fault, node failure, load variation, delay. System model is varied depending on the critical conditions and add adaptivenss to our formulation.
This chapter has been organized as follows. Section \ref{Sec:4MIMO} deals with MIMO system model, where the system model, controller structure and the optimization problem are discussed. The lyapnov formulation based analysis has been introduced in \ref{Sec:LF}. The proposed connection design algorithm has been presented in Section \ref{Sec: GSCD}. Detailed step by step procedure for controller design is presented in Section \ref{Sec:4descon}. Simulation results are presented in \ref{Sec:4results} followed by concluding remarks in Section \ref{Sec:4Conclusion}. 
\section{System Model}\label{Sec:4MIMO} 
 The Cyber-Physical Microgrid(CPMG) structure consists of both physical components like sensors, DGs and controllers which deal with physical parameters like voltage and frequency and communication components like routers which deal with parameters like bandwidth, speed, data loss,etc. Hence, the model adopted for representing the power system must be able to represent parameters both from the physical and communication worlds. Consider an islanded microgrid as given in Figure \ref{Fig:fig1}.The system operates completely in islanded mode. The control hierarchy consists of a basic decentralized control layer among the DGs whose routing is decided through a central server. On a general basis, the decentralized control using the communication network 2 is under operation. It is assumed that the central server using communication network 1 contains an online forecasting tool which predicts various sensitive parameters in the grid in an online manner. The decentralized controller values and their communication design data is relayed to the DGs and sensors whenever the monitoring parameters are expected to go beyond the bounds of existing controller capabilities. This section gives a comprehensive description about the physical and communication aspects of the microgrid along with the controller strcuture being adopted.
 
\subsection{Physical System Model without Delay:}
\begin{figure}[!ht]
\centering
\includegraphics[width=0.9\textwidth, height = 7.5cm, keepaspectratio]{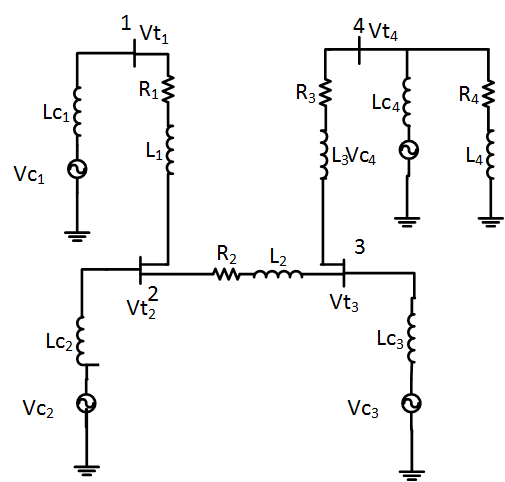}
\caption{Physical Grid}
\label{Fig:physical grid}
\end{figure}

In the CPMG considered, the decentralized controllers present at the DGs sense the bus voltages and come up with a strategy to vary the voltage output of DGs $v_c(t)$ for the bus voltages to reach their reference values $V_{ref}$.
The single phase equivalent diagram of the 4 bus CPMG has been depicted in Figure \ref{Fig:physical grid}. 
\\

The following table shows the parameter values assumed for the microgrid:
\begin{table}[]
\centering
\caption{Microgrid Parameters}
\label{my-label}
\begin{tabular}{|c|c|}
\hline
Paramter & Value  \\ \hline
$R_1$        & 0.175  \\ \hline
$R_2$        & 0.1667 \\ \hline
$R_3$        & 0.2187 \\ \hline
$L_1$        & 0.0005 \\ \hline
$L_2$        & 0.0004 \\ \hline
$L_3$     & 0.0006 \\ \hline
$L_{c1}$     & 0.001  \\ \hline
$L_{c2}$        & 0.001  \\ \hline
$L_{c3}$ & 0.001  \\ \hline
$L_{c4}$        & 0.001  \\ \hline
\end{tabular}
\end{table}
The value of $R_4$ and $L_4$ depend on the amount of load resistance and reactance at a particular point of time.

Applying nodal analysis at bus-1:
\begin{align}
\Big( \frac{L1}{L_{c1}}+1\Big) \dot V_{t1}&-\dot V_{t2}+ \frac{R_1}{L_{c1}}V_{t1}  -\frac{R_1}{L_{c1}}V_{c1} - \frac{L_1}{L_{c1}} \dot V_{c1} =0 \label{Eq:SysDer1} 
\end{align}
\text{At bus-2:}
\begin{align}
\frac{V_{t2}-V_{c2}}{sL_{c2}}+\frac{V_{t2}-V_{t1}}{R_1+sL_1}+\frac{V_{t2}-V_{t3}}{R_2+sL_2}=0 \label{Eq:SysDer2} 
\end{align}
\begin{flalign}
\text{Substituting equ} & \text{ation(\ref{Eq:SysDer1}) in (\ref{Eq:SysDer2}):} \nonumber \\
\frac{V_{t2}-V_{c2}}{sL_{c2}}&+\frac{V_{t1}-V_{c1}}{sL_{c1}}+\frac{V_{t2}-V_{t3}}{R_2+sL_2}=0 \label{Eq:SysDer3} \\
\text{which amounts}&\text{ to} \\
\frac{L_2}{L_{c1}} \dot{V}_{t1} +\hspace{0.1cm}&(\frac{L_2}{L_{c2}}+1)\dot{V}_{t2} -\dot{V_{t3}} + \frac{R_2}{L_{c2}}V_{t2}+  \frac{R_2}{L_{c1}}V_{t1} \nonumber \\ 
-\frac{R_2}{L_{c1}}V_{c1}-&\frac{R_2}{L_{c2}}V_{c2}-\frac{L_2}{L_{c1}}\dot{V}_{c1} -\frac{L_2}{L_{c2}}\dot{V}_{c2}=0 \label{Eq:SysDer8}
\end{flalign}
Upon applying nodal analysis at buses 3 and 4 and performing similar substitutions, the system equations can be written in the following matrix form:
\begin{equation}\label{Eq:sys00}
T \dot{V_t}{(t)}= T_1V_t(t) + T_2V_c(t)+T_3\dot{V_c}{(t)}
\end{equation}
where 
\begin{align}
T&=
\begin{bmatrix} 
    \frac{L_{1}}{L_{c1}}+1 & -1 & 0 & 0  \\
    \frac{L_{2}}{L_{c1}}  & \frac{L_{2}}{L_{c2}}+1 & -1 & 0 \\
    \frac{L_{3}}{L_{c1}}  & \frac{L_{3}}{L_{c2}}  & \frac{L_{3}}{L_{c3}}+1 & -1\\
    \frac{L_{4}}{L_{c1}}  & \frac{L_{4}}{L_{c2}} & \frac{L_{4}}{L_{c3}} & \frac{L_{4}}{L_{c4}}+1
\end{bmatrix} \label{Eq:T}
\\T_1&=
\begin{bmatrix}
    -\frac{R_{1}}{L_{c1}} & 0 & 0 &  0\\
    -\frac{R_{2}}{L_{c1}} & -\frac{R_{2}}{L_{c2}} & 0 & 0 \\
    -\frac{R_{3}}{L_{c1}} & -\frac{R_{3}}{L_{c2}} & -\frac{R_{3}}{L_{c3}} & 0\\
    -\frac{R_{4}}{L_{c1}} & -\frac{R_{4}}{L_{c2}} & -\frac{R_{4}}{L_{c3}} &-\frac{R_{4}}{L_{c4}} 
\end{bmatrix} \label{Eq:T1}
\\T_2&=-T1 \label{Eq:T2}
\\T_3&=
\begin{bmatrix}
    -\frac{R_{1}}{L_{c1}} & 0 & 0 &  0\\
    -\frac{R_{2}}{L_{c1}} & -\frac{R_{2}}{L_{c2}} & 0 & 0 \\
    -\frac{R_{3}}{L_{c1}} & -\frac{R_{3}}{L_{c2}} & -\frac{R_{3}}{L_{c3}} & 0\\
    -\frac{R_{4}}{L_{c1}} & -\frac{R_{4}}{L_{c2}} & -\frac{R_{4}}{L_{c3}} &-\frac{R_{4}}{L_{c4}} 
\end{bmatrix}\label{Eq:T4}
\end{align}\\
Since the system is linear, Equation(\ref{Eq:sys00}) can be written as follows:
\begin{equation}
T \Delta \dot{V_t}(t)= T_1\Delta V_t(t) + T_2 \Delta V_c(t)+ T_3 \Delta\dot{V_c}(t) \label{Eq:sys1}
\end{equation}
Equation (\ref{Eq:sys1}) can further be rewritten as :
\begin{equation}\label{Eq:sys2}
\Delta \dot{V_t}{(t)}= A' \Delta V_t(t) + B' \Delta V_c(t)+M' \Delta \dot{ V_c}(t)
\end{equation}
where  \\
\begin{equation}
 A'=T^{-1}T_{1},\hspace{0.1cm} B'=T^{-1}T_2 \text{ and } M'=T^{-1}T_{3}\\ 
\end{equation}
and
\begin{eqnarray} 
\Delta V_t(t)= V_t(t)-V_{ref}   \\
\end{eqnarray}
A further transformation 
\begin{equation} 
\Delta{V_t(t)}= x(t) + M' \Delta{V_c(t)}
\end{equation}  
has been applied onto Equation (\ref{Eq:sys2}) to get the following state-space representation:
\begin{equation}
\left.
\begin{aligned}
&\dot{x}(t)=Ax(t)+Bu(t)~~\\
&y(t)=Cx(t)\label{Eq:sys111}
\end{aligned}
\right\}
\end{equation}
\text{where}\\ 
$A=A',~B=A'M'+B',~C=I,~x\in R^n, ~A \in R^{n \times n},\\~B\in R^{n \times m},~u\in R^m,~y \in R^p\text{ and }C \in R^{p \times n}$ 

The control vector and the state vector for this formulation are as fikkiws
\begin{eqnarray} 
u(t)&=&\Delta V_c(t) \hspace{0.9cm} \\
x(t)&=& \Delta{V_t(t)}- M' \Delta{V_c(t)} \label{Eq:statevec}
\end{eqnarray}
\subsection{Communication Structure}
Most of the equipment in the CPMG are assumed to be equipped with appropriate communication interfaces- wireless/wired. These interfaces can be clubbed into three types of nodes- sensor nodes, controller nodes and the server nodes.The $n$ sensor nodes placed at individual buses, collect the voltage data coming out of the sensors and send them to the server while $m$ controller nodes placed at DGs receive the reference commands coming from the central server. The central server has receiver nodes to get the data from sensors and a transmitter nodes to send the data to the DGs as well as the sensors. The presence/absence of the connections between these nodes will be decided by the central server.For the sake of simplicity, all sorts of relay nodes are assumed to be absent.\\
\begin{figure}[!ht]
\begin{minipage}[b]{.5\linewidth}
\centering
\includegraphics[width= 1\textwidth, height=6cm]{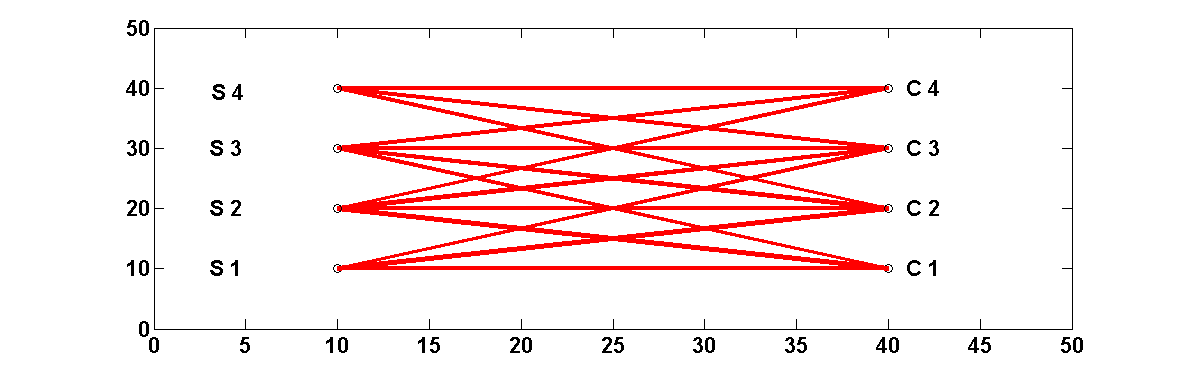}
\caption{cc=2 connections}\label{Fig:cc2}
\end{minipage}%
\begin{minipage}[b]{.5\linewidth}
\centering
\includegraphics[width= 1\textwidth, height=6cm]{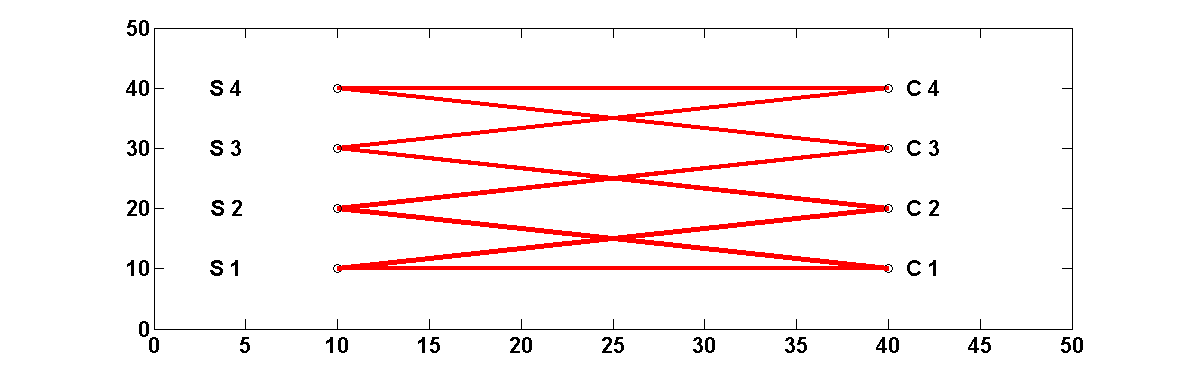}
\caption{bw=2 connections}\label{Fig:bw2}
\end{minipage}
\end{figure}\\
The routing between the nodes would be on a multicast\cite{weigoalie} \cite{kompella1993multicast} basis.
The data flow among these nodes is assumed to be continuous i.e the sampling rate is high, quantization error is low and there is no issue of packet loss. The communication parameters within the scope of this work are as follows:\\
\begin{enumerate}
\item Connection Constraint: This constraint deals with the general topology of the communication network. They specify the availability of a particular node to be connected to other nodes. The value of connection constraint $cc$ signifies the number of neighboring nodes in the previous layer from which a particular node can receive information. For instance, if $cc=1$, then the possible set of connections can be as shown in Figure \ref{Fig:cc2}. 
\item Bandwidth Constraint: This constraint specifies the number of connections that must generate from any node. For example, Figure \ref{Fig:bw2} shows the connections when $bwc=2$ with $bwc$ representing this constraint.
\end{enumerate}

\subsection{Decentralized Control Scheme}
A decentralized state-feedback control scheme 
\begin{equation}\label{Eq: controller} 
u(t)=Kx(t) 
\end{equation}
is used where the matrix $K\in R^{m \times n}$ shows the connection structure between controllers and sensors. If $K_{ij}$  is non-zero, it represents the existence of connection between controller $i$ and sensor $j$. 
Now
\begin{equation} \label{Eq:sys3}
\dot{x}(t)= \bar A x(t)   
\end{equation}
where $\bar{A}= A+BKC$. \\
Without inclusion of the controller, the system stability is defined by the eigenvalues of the matrix $A$ whereas once the controller is added,  the system's stability gets defined through eigenvalues of the matrix  $\bar A$ . If the real parts of all the eigenvalues are negative, then the system matrix  $\bar A $ is stable. Even if one real part of eigenvalue is positive, the system is unstable.

It is seen that substituting Equation(\ref{Eq: controller}) into Equation(\ref{Eq:statevec}) results in the following relation:
\begin{equation}
\Delta{V_t}(t)= (I+MK)x(t) 
\end{equation}
Since, $x$ and $ \Delta V_t $ bear a linear relationship, $ x \to 0$ would mean $\Delta{V_t}\to 0$. Thus, the bus voltage $V_t$ gets stabilized.

\subsection{Physical System Model with Non-Negligible Delay}
In the CPMG, some delay might creep into the system due to distance for which the data needs to be transmitted. Considering this, a system model with non-negligible delay has been adopted from \cite{li2012multicast}
as follows: \\
\begin{equation}
\dot{x}(t)= Ax(t)+ \underset{(n,m)\epsilon R}{\Sigma} b(:,m) k(n,:) Cx(t-d_{nm})
\end{equation}
where $(n,m)\epsilon R $ means that the connection between sensor $n$ and controller $m$ is established, $b(:,m)$ is the $m^{th}$ column of the matrix $B$, $k(n,:)$ is the $n^{th}$ row of matrix $K$, and $d_{nm}$ is the delay between sensor $n$ and controller $m$.  A particular case of delay has been considered in this work where the delay is small but it is non-negligible. This has been taken in light of the speed of the modern communication networks in comparison to the speed of the physical system dynamics. \\
Therefore,
\begin{align}
&x(t-d_{nm}) \nonumber \\
&= x(t) - d_{nm}\dot{x}(t) + o(d_{nm}) \nonumber \\
&\approx x(t) - d_{nm} (Ax(t)+ \underset{(n,m)\epsilon R}{\Sigma} b(:,m) k(n,:) Cx(t)) \nonumber  \\
&= (I- d_{nm}(Ax(t)+ \underset{(n,m)\epsilon R}{\Sigma} b(:,m) k(n,:) C)))x(t)
\end{align}\\
Thus, the dynamics of the system turns out to be 
\begin{equation}\label{Eq: dmodel}
\dot{x}(t)\approx \bar{A} x(t) 
\end{equation}
where  \begin{equation}
\bar{A} = (A+BKC)(I-BDKC)
\end{equation}
where $D$ is the delay matrix formulated as 
\begin{equation}
D_{nm} =\begin{cases}
    &d_{nm} \text{,  if $n$ and $m$ are connected}\\
    &0 \text{,  if $n$ and $m$ are not connected}
  \end{cases}
\end{equation}\\
\section{Lyapunov based Optimization Formulations} \label{Sec:LF}
The bus voltage stability of the microgrid gets affected by many parameters both from the communication  and the physical worlds. These parameters might change either in the natural course of system operation or sometimes  as a matter of choice of the operator. Hence, the set of decentralized controllers as a whole must be capable of stabilizing the system in these scenarios. This section describes various tools adopted and developed for analyzing different aspects of microgrid stability.  

This formulation has been derived with the help of basic Lyapunov stability analysis for assessing the stability of microgrid voltages. Two cases will be dealt with in detail-
\subsection{System without delay}

Considering the linear system model and given a Lyapunov function $V(x)=x^TPx$,it is well known that equilibrium point $x^e$ goes to zero, if the following two inequalities hold simultaneously for all $x \neq 0$
\begin{eqnarray}
V(x)>0 \label{Eq:v}\\
\dot{V}(x)<0 \label{Eq:v.}
\end{eqnarray}
The rate dervitave of the Lyapunov function $V$ for the linear system model is obtained as:
\begin{eqnarray}
\dot{V}(x)=\dot{x}^TPx+x^TP\dot{x}=x^T(\bar{A}^TP +P \bar{A})x \label{Eq:Lyap}
\end{eqnarray}

Given that the matrix $P$ is positive definite, Equation \eqref{Eq:Lyap} implies that $\dot{V}$ will be negative define if  the following condition holds true:
\begin{equation}
\bar{A}^TP+P \bar{A} ~<~0 \label{Eq:Lyap2}
\end{equation}
For the computed $\bar{A}$ to be more stable than A, the matrix $P$ is computed by selecting an arbitrary positive value of $\beta$ such that the following condition holds true:

\begin{equation}
(A-\beta I)P+P(A-\beta I)^T=-I \label{Eq:ABPI}
\end{equation}

It should be noted that for different value of $\beta$, the controller solution will be different. By introducing a design parameter $\gamma$ into the Equation (\ref{Eq:sys3}) , the optimization problem in the form of linear matrix inequality is obtained as:
\begin{equation}
\bar{A}^TP +P\bar{A}+\gamma_1 I<0 \label{Eq:last11}
\end{equation}
One can note that the larger the $\gamma$ value is, larger is the stability margin. As one optimizes the Equation \eqref{Eq:last11}, some constraints arise. First constraint is that $K_{ij}=0$ when the sensor-$i$ is not connected to the controller-$j$. We also would like to see that controller gains should be bounded, i.e. the second constraint is $||K||_2<\rho$ where $\rho$ is a user defined scalar value.\\

\subsection{System with non-negligible delay}
For this, we assume the delay system model delineated in Equation(\ref{Eq: dmodel}). This equation is rewritten after removing the higher order terms as:
\begin{equation}
\centering
\dot{x}= (A+BKC)x + f(x)
\end{equation}
where  
\begin{equation}
\centering
f(x)= -BDK(A+BKC)x
\end{equation}
and 
\begin{equation}
\centering
f^T(x)f(x) \leq \alpha^2 x^Tx
\end{equation}
where 
\begin{equation}
\centering
\alpha = \underset{mn}{max}   c_K  \| B\|_2 (\|A\|_2 + \| B\|_2 c_K)
\end{equation}
Here,  $\alpha$ is representation of the higher delay bound on the system.
For this system, to find the value of $K$ which follows the stability inequality $\bar{A}^TP+P\bar{A}<0$ ,  it should satisfy the following LMI:
\begin{equation}
\centering
\begin{bmatrix}
A^T_KP + P A_K+ \gamma_2 \alpha^2 I     &  P  \\
P                 &  \gamma_2 I
\end{bmatrix}
< 0  \label{Eq: delay_form}
\end{equation}
Further details on derivation of this formulation can be found in \cite{siljak2011decentralized}. The overall optimization formulation for this case would be as follows:
\begin{equation}
\begin{aligned}
\underset {K}{Maximize} \hspace{0.3cm} \gamma_2 \hspace{6cm} \\
{Subject \hspace{0.1cm} to.} \hspace{6.5cm}  \\
\begin{bmatrix}
{A^T_{K}}P + P{A_{K}}+ \gamma_2 {\alpha_2}^2 I     &  P  \\
P                 &  \gamma_2 I
\end{bmatrix} <0 \nonumber \hspace{1cm}\\
\| K \|_2 \leq c_K \hspace{4.7cm} \\  
\gamma_2 <0 \hspace{5.3cm} \\ 
\end{aligned}	
\end{equation}
The algorithm applied for finding the appropriate $K$ value will be described in the upcoming subsection.
\section{Constraint based Sensor- Controller Connection Design Algorithm(CBSCD)} \label{Sec: GSCD}
\begin{figure}[!ht]
\centering
\includegraphics[width=0.8\textwidth, height =7.5cm]{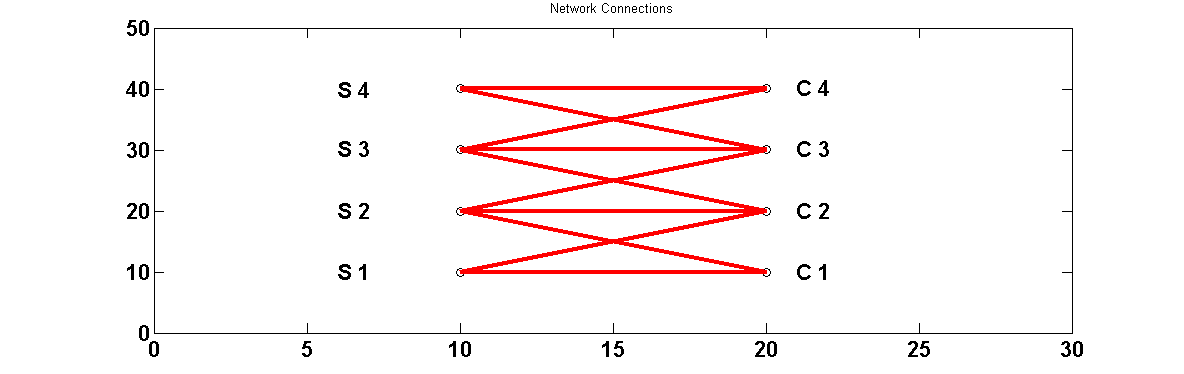}
\caption{Network Connections}
\label{Fig:commconn1}
\end{figure}

Consider the communication structure of the CPMG. It contains sensor nodes placed at every bus and the four controller nodes placed at DG locations that need to be connected in the presence of communication constraints like bandwidth and connection constraints and physical constraints like cost. 
The cost constraint is further divided into peripheral cost constraint $prc$ and central cost constraint $cnc$. These two have been formulated under the assumption that in general, the central generators take maximum load owing to their greater accessibility while the peripheral generators take lesser loads. It is also assumed that central DGs are costly. Here, $prc$ represents the maximum sensors that can be connected to a generator in the periphery i.e., DG1 and DG4 while $cnc$ represents the minimum number of sensors to be connected to a central region DG(DG2 and DG3) for it to be operational. \\  
In this section, a novel algorithm for designing the set of connections between sensors and controllers is presented. \\
\textbf{Step 1:} Enter the structural parameters of the communication network \\ 
Enter the number of sensors $ns$ and number of controllers $nc$. For this case, $ns=nc=4$.\\
\textbf{Step 2:} Enter the connection constraints and bandwidth constraints and generate the possible connections.In this case, $bwc=2$ and $cc=1$. The possible connections for this configuration are shown in Figure \ref{Fig:commconn1} \\
\textbf{Step 3:} Feed in the cost constraints $prc$ and $cnc$ and generate all the possible cost configurations and save them in a set called $cost$. Here $prc=1$ and $cnc=2$.For the given set of constraints, there exists only one cost configuration for the generators i.e, 1331 meaning that first and fourth generators should be connected to one sensor while the second and third generators should be connected to three sensors.\\ 
\textbf{Step 4:} Initialize the set $R$ as an empty set. \\
\textbf{Step 5:} While the set $cost$ is non-empty, go to step 6,or else, go to step 14. \\
\textbf{Step 6:} Select a cost configuration from the set $cost$. Selected 1331.This means that $DG_1$ and $DG_4$ should be connected to one sensor while $DG_2$ and $DG_3$ should be connected to three sensors.\\
\textbf{Step 7:} Sort the non-zero cost generators in ascending order of their cost into a set $cost-sortup$ whose size is given by $count$.  \\
$cost-sortup={1 1 3 3}$ and $count=4$. \\
\textbf{Step 8:} While $count$ $\geq$ $bwc$,go to step 9, else  proceed to step 13.\\ 
\textbf{Step 9:} Pick the first connection available in $cost-sortup$ and \\
\textbf{Step 10:} Find all the possible ways in which the generators can exist with that particular cost and put them in a set $v$.\\
\textbf{Step 11:} Find all the combinations of different elements in set $R$ with that of set $v$ and update set $R$. \\
\textbf{Step 12:} Delete the current connection from $cost-sortup$ and  decrement the count by 1. Go to step 8. \\
\textbf{Step 13:} Pick the first generator from $cost-sortup$, find all the possible ways in which the generator can exist with that particular cost and update set $v$. \\
\textbf{Step 14:} Find the individual elements in set $v$ that can be added to the elements in set $R$ such that empty sensors are served.\\
\textbf{Step 15:} List the co-existable added connections and update set $R$.  \\
\textbf{Step 16:} Create a set $v$ for the final generator and add the elements of this set to set $R$ in such a way that bandwidth constraint gets satisfied. \\
\textbf{Step 17:} End.  	
Figure \ref{Fig:3merge} shows the elements in set $R$ after combining the connections of $DG1$ and $DG4$ at the end of step-12. 
\begin{figure}
\includegraphics[width= 1\textwidth, height=6cm]{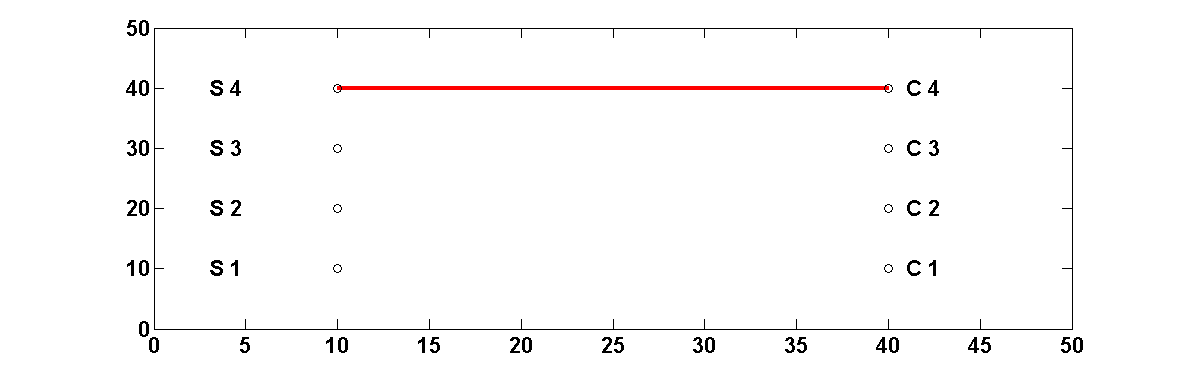}
\caption{DG-1 connections-part 1}
\end{figure}
\begin{figure}
\includegraphics[width= 1\textwidth, height=6cm]{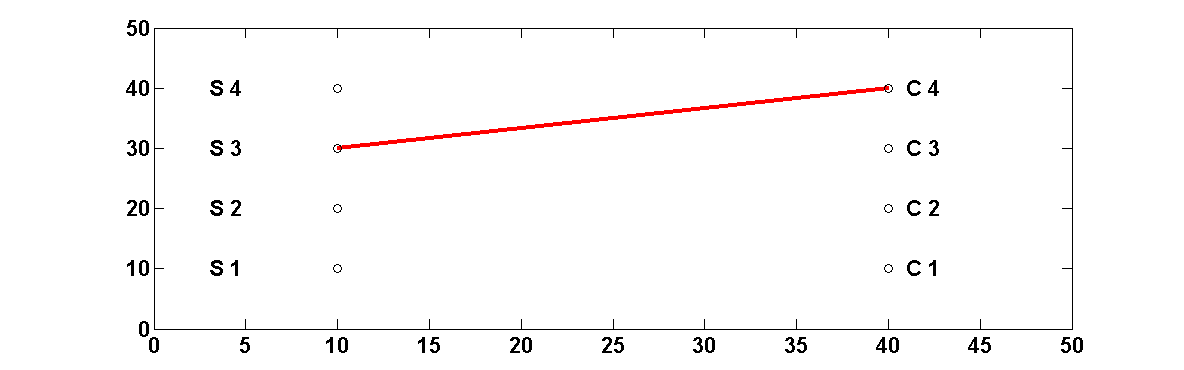}
\caption{DG-1 connections-part 2}\label{fig:DG1con}
\end{figure}
\begin{figure}[!ht]
\begin{minipage}[b]{.5\linewidth}
\centering
\includegraphics[width= 1\textwidth, height=6cm]{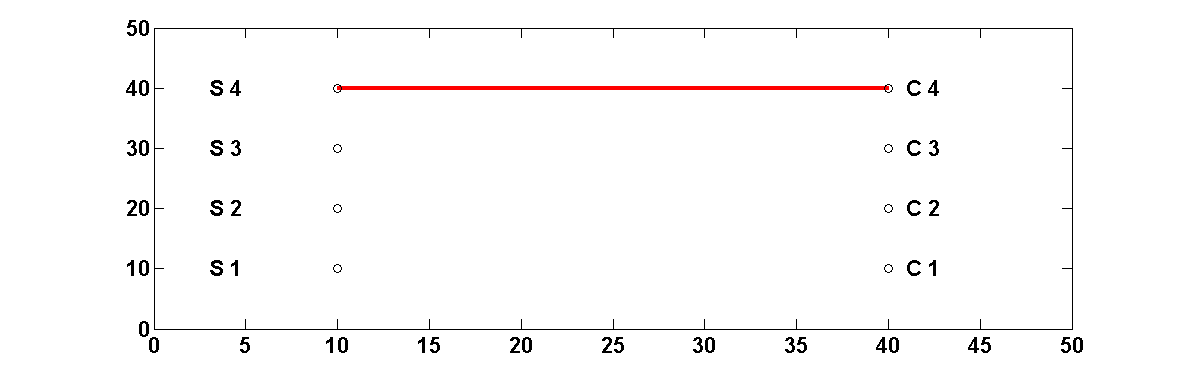}
\caption{DG1 connections-set 1}\label{DG11}
\end{minipage}%
\begin{minipage}[b]{.5\linewidth}
\centering
\includegraphics[width= 1\textwidth, height=6cm]{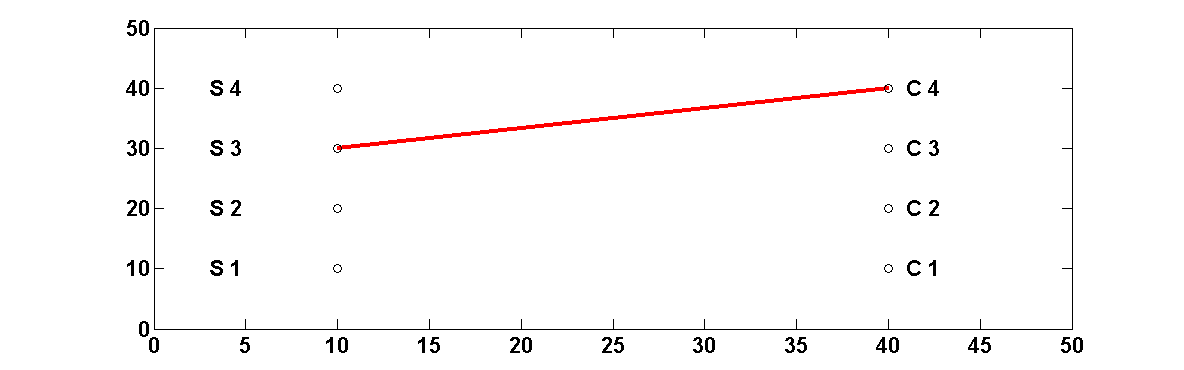}
\caption{DG1 connections-set 2}\label{DG12}
\end{minipage}
\end{figure}\\
\begin{figure}[!ht]
\begin{minipage}[b]{.5\linewidth}
\centering
\includegraphics[width= 1\textwidth, height=6cm]{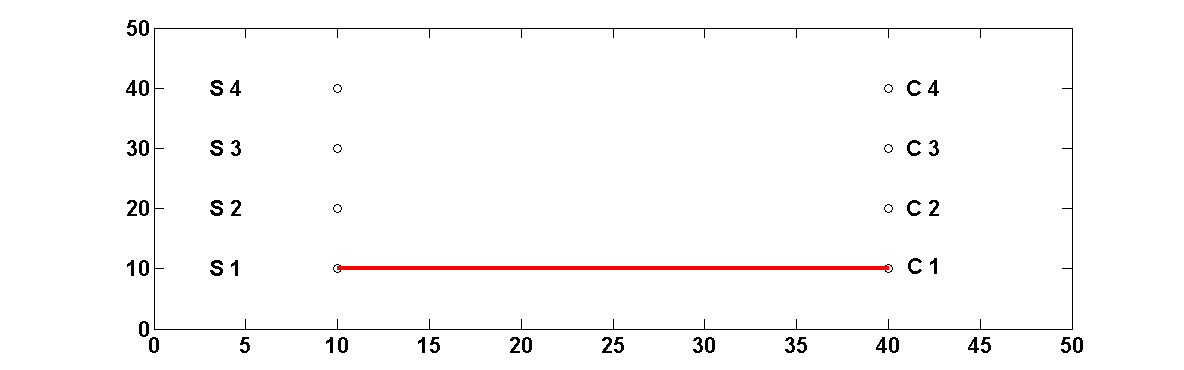}
\caption{DG4 connections-set 1}\label{DG41}
\end{minipage}%
\begin{minipage}[b]{.5\linewidth}
\centering
\includegraphics[width= 1\textwidth, height=6cm]{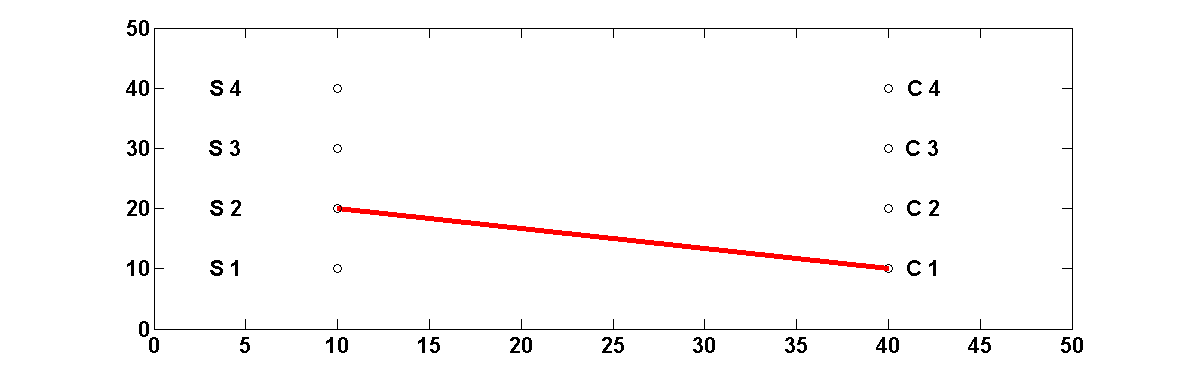}
\caption{DG4 connections-set 2}\label{DG42}
\end{minipage}
\end{figure}\\

\begin{figure}[!ht]
\includegraphics[width=1\textwidth, height =6.0cm]{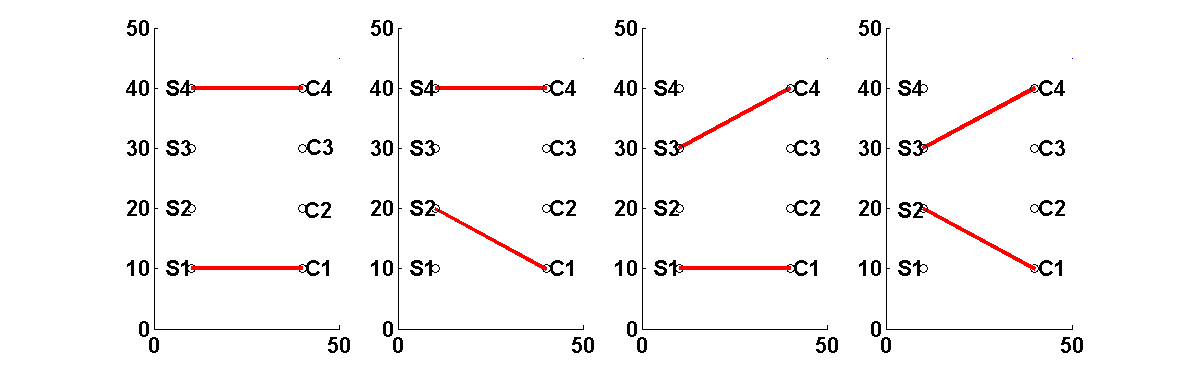}
\caption{Set $R$ after Step-12}
\label{Fig:3merge}
\end{figure}

\begin{figure}[!ht]
\begin{minipage}[b]{.5\linewidth}
\centering
\includegraphics[width= 1\textwidth, height=6cm]{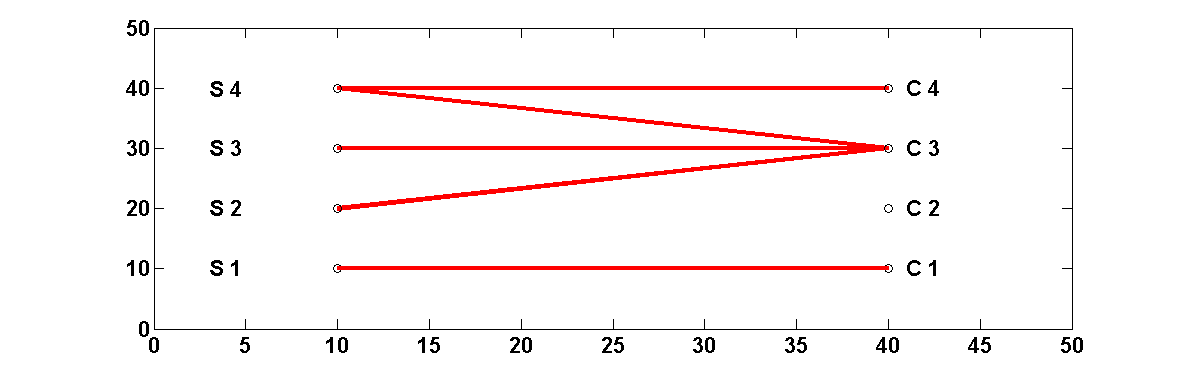}
\caption{Set $R$ after Step-15}\label{step15}
\end{minipage}%
\begin{minipage}[b]{.5\linewidth}
\centering
\includegraphics[width= 1\textwidth, height=6cm]{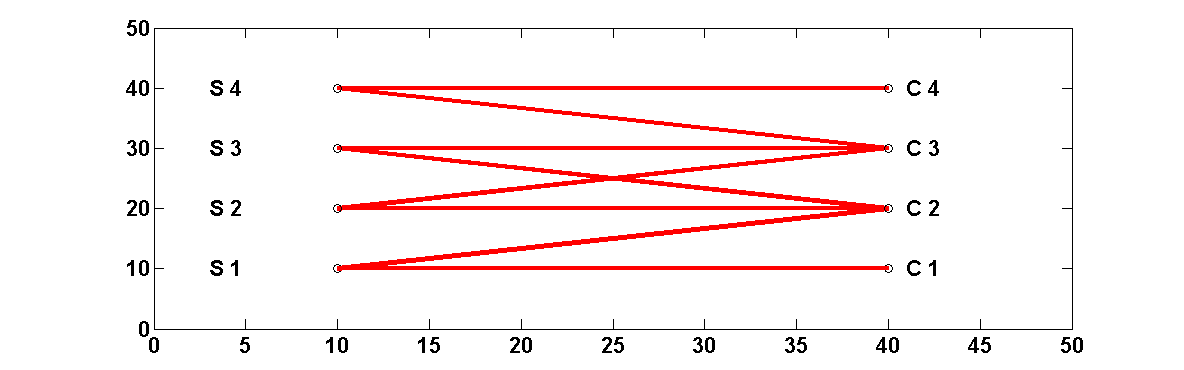}
\caption{Set $R$ after Step-16}\label{step16}
\end{minipage}
\end{figure}
\section{Controller Design with Illustration} \label{Sec:4descon}
With the tools adopted and developed in the previous sections, the central server of the CPMG can be utilized to design basic and adaptive controllers in an online manner in the presence of  various forecast data like load forecast, climate forecast etc. The central server keeps track this forecast data, computes the most appropriate controllers online and updates the K values that govern actions to be taken at individual controllers. Note that this controller design process works with the idea that system matrices and communication parameter ranges are known in advance as per the available forecast data.This section describes how controllers can be designed for increasing the stability of the system to tackle various change in parameters. 

\begin{enumerate}
\item Plot the eigenvalues of the system matrix for the known range of parameters
\item Segregate the range of eigenvalues into individual time zones depending on the time of the day in operation say, morning, noon, night, etc
\item Select a particular timezone depending on the time of the day.
\item Find the worst-case working system matrix for all $n$ individual time zones $A_1, A_2, ...,A_n$.
\item Find the possible sensor-controller combinations for all kinds of constraints as prescribed by the algorithm mentioned in Section \ref{Sec: GSCD}.
\item Find the MIMO state feedback controllers(K matrices) and maximum eigenvalues for all the combinations found in step-2 using the optimization formulation mentioned in  Section \ref{Sec:LF}
\item Define the maximum eigenvalue tolerances $\epsilon_1, \epsilon_2,..., \epsilon_n $for all the individual working zones.
\item Find the least value of constraints that can satisfy the above tolerance value. \\
This would mean that minimum values of constraints would be used while not compromising the stability to a great extent. This could be called as stability enhancement using minimal resources.
\end{enumerate}



\subsection*{Example: Communication constraint manipulation based controller for varying load resistance} 
\begin{enumerate}
\item Plot the eigenvalues of the system matrix for the known range of output resistance variation. 
\begin{figure}[!ht]
\centering
\includegraphics[width=0.7\textwidth, height =6.5cm]{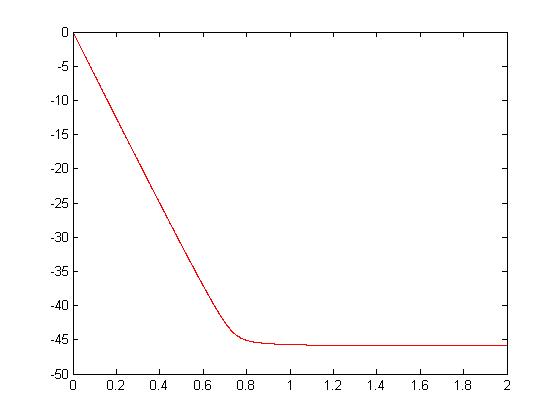}
\caption{Variation of max eigen-value with load resistance}
\label{Fig:maxeigvsr4}
\end{figure}\\
\item Segregate the range of eigenvalues into individual smaller zones.
\begin{figure}[!ht]
\centering
\includegraphics[width=0.7\textwidth, height =6.5cm]{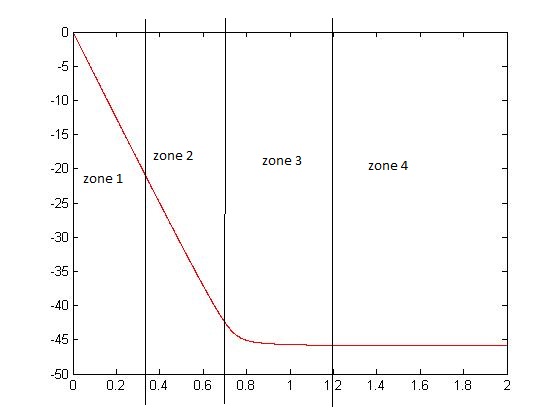}
\caption{Segregated zones}
\label{Fig:maxeigvsr41}
\end{figure}\\
\item Select a particular timezone depending on the time of the day. Let it be $A_1$.
\item Find the worst-case working system matrix for all $n$ individual time zones $A_1, A_2, ...,A_n$. For sample $A_1$ is obtained by putting $R_4=0$ in the mathematical model mentioned in Equation (\ref{Eq:sys1})\\
\begin{eqnarray}
 A_{1} = 
 \begin{bmatrix}
  -150.4375 & -63.0006 & -21.4758  &       0 \\
   -50.6563 & -94.5009 & -32.2137  &       0 \\
    35.6062 &   9.1985 & -53.6896  &       0 \\
   155.0137 & 138.9167 & 100.5830  &       0 \\ 
\end{bmatrix}
\end{eqnarray}
\item Find the possible sensor controller combinations for all kinds of constraints as prescribed by the algorithm in \ref{Sec: GSCD}
The connection parameter of a connection gives an idea of various constraints imposed on the system. In short, $connection\hspace{0.1cm} parameter\hspace{0.1cm} = \hspace{0.1cm} [bwc\hspace{0.1cm} \hspace{0.1cm}  cc \hspace{0.1cm} cnc\hspace{0.1cm}  prc]$.   
\item Find the gamma values and the maximum eigenvalues for all the combinations found in step-5 using the optimization formulation mentioned in Section \ref{Sec:LF} .
\begin{table}[]
\centering
\caption{Step-6 result part-1}
\label{my-label}
\begin{tabular}{|c|c|c|}
\hline
Connection Parameter & $\gamma$ & Max-eigenvalue \\ \hline
3202                 & 2.04     & -2.15          \\ \hline
3212                 & 2.04     & -2.15          \\ \hline
3222                 & 2.04     & -2.15          \\ \hline
3232                 & 2.04     & -2.15          \\ \hline
3242                 & 2.04     & -2.15          \\ \hline
3203                 & 2.7      & -2.838         \\ \hline
3213                 & 2.7      & -2.838         \\ \hline
3223                 & 2.7      & -2.838         \\ \hline
3233                 & 2.7      & -2.838         \\ \hline
3243                 & 2.7      & -2.15          \\ \hline
3204                 & 2.7      & -2.838         \\ \hline
3214                 & 2.7      & -2.838         \\ \hline
3224                 & 2.7      & -2.838         \\ \hline
3234                 & 2.7      & -2.838         \\ \hline
3244                 & 2.04     & -2.15          \\ \hline
3302                 & 3.574    & -3.64          \\ \hline
3312                 & 3.574    & -3.64          \\ \hline
\end{tabular}
\end{table}

\begin{table}[]
\centering
\caption{Step-6 result part-1}
\label{my-label}
\begin{tabular}{|c|c|c|}
\hline
Connection Parameter & $\gamma$ & Max-eigenvalue \\ \hline
3322                 & 3.574    & -3.64          \\ \hline
3332                 & 3.574    & -3.64          \\ \hline
3342                 & 3.574    & -3.64          \\ \hline
3303                 & 4.302    & -4.44          \\ \hline
3313                 & 4.302    & -4.44          \\ \hline
3323                 & 4.302    & -4.44          \\ \hline
3333                 & 4.302    & -4.44          \\ \hline
3343                 & 4.01     & -4.2           \\ \hline
3304                 & 4.302    & -4.44          \\ \hline
3314                 & 4.302    & -4.44          \\ \hline
3324                 & 4.302    & -4.44          \\ \hline
3334                 & 4.302    & -4.44          \\ \hline
3344                 & 4.01     & -4.2           \\ \hline
4301                 & 4.304    & -4.381         \\ \hline
4314                 & 4.304    & -4.381         \\ \hline
4324                 & 4.304    & -4.381         \\ \hline
4334                 & 4.304    & -4.381         \\ \hline
4344                 & 4.304    & -4.381         \\ \hline
\end{tabular}
\end{table}

\item Define the maximum eigenvalue tolerance for $A_1$. Since $A_1$ is near the verge of instability, the $\epsilon_1$ will be chosen to be low like 0.1.
\item Find the least value of constraints that can satisfy the tolerances. \\

This process has been explained in the Figure \ref{Fig:ccc}. 
Find the maximum $\gamma$ in a particular bandwidth. Here only bandwidth-3 and bandwidth-4 exist with respective maximum $\gamma$ values 4.302 and 4.304. The maximum eigenvalue with bandwidth-3 is much better, so further connection constraint-3 is chosen which supports this maximum eigenvalue and within the purview of these two constraints the constraint configuration 3303 is chosen since the sum of its individual digits is least. The constraint configuration is a 4-digit number that gives the idea of the constraints chosen for a particular controller. In this case, 3303 suggests that $bwc=3$, $cc=3$, $cnc=0$ and $prc=3$.
\begin{figure}[!ht]
\centering
\includegraphics[width=0.75\textwidth, height =6.5cm]{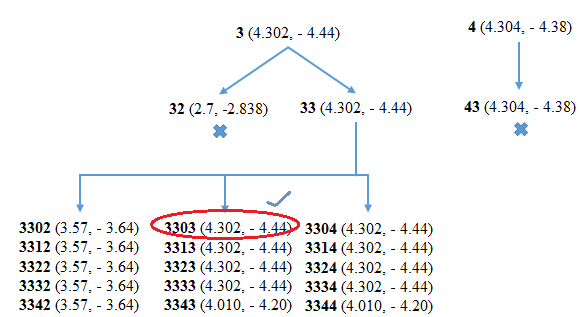}
\caption{Step-8 result}
\label{Fig:ccc}
\end{figure}
\end{enumerate}
\section{Results} \label{Sec:4results}
This section describes various results obtained while designing communication manipulation based controllers for dealing with the following cases:
\begin{enumerate}
\item Change in load resistance
\item Change in communication delay and resistance
\item Failure of communication node
\end{enumerate}
All of these studies have been carried out using the CVX optimization tool\cite{cvx}. These cases can further be combined for describing practical scenarios in the cyber-physical framework. 
The communication manipulation based controllers are found by determining the minimal set of communication constraints that can maximally stabilize the microgrid for a given set of physical and communication parameters.
\subsubsection{Load Variation}
Table \ref{tab:Results-Load Variation} summarizes the results obtained for various scenarios of loading. 
\begin{table}[]
\centering
\caption{Results-Load Variation}
\label{tab:Results-Load Variation}
\begin{tabular}{|c|c|c|c|}
\hline
System Matrix & \begin{tabular}[c]{@{}c@{}}Max-eigenvalue \\ (open)\end{tabular} & \begin{tabular}[c]{@{}c@{}}Max-eigenvalue\\ (closed)\end{tabular} & \begin{tabular}[c]{@{}c@{}}Connection\\  Parameter\end{tabular} \\ \hline
$A_1$         & 0                                                                & -4.44                                                             & 3303                                                            \\ \hline
$A_2$         & -21.9580                                                         & -26.7762                                                          & 3203                                                            \\ \hline
$A_3$         & -42.4541                                                         & -47.7961                                                          & 2303                                                            \\ \hline
\end{tabular}
\end{table}
\subsubsection{Load variation with delay}
Table \ref{tab:Results- Delay Variation} summarizes the results for scenarios when maximum delay of 0.5 millisecond and 1 millisecond is present in communication structure for the same load.
\begin{table}[]
\centering
\caption{Results- Delay Variation}
\label{tab:Results- Delay Variation}
\begin{tabular}{|c|c|c|c|}
\hline
Delay (ms) & \begin{tabular}[c]{@{}c@{}}Max-eigenvalue \\ (open)\end{tabular} & \begin{tabular}[c]{@{}c@{}}Max-eigenvalue\\ (closed)\end{tabular} & \begin{tabular}[c]{@{}c@{}}Connection\\  Parameter\end{tabular} \\ \hline
0.5        & 0                                                                & -28.9729                                                          & 2303                                                            \\ \hline
1          & 0                                                                & -31.0441                                                          & 2302                                                            \\ \hline
\end{tabular}
\end{table}
\subsubsection{Load Variation with communication nodefailure}
Table \ref{tab:Results- Load Variation with communication node failure} summarizes the results obtained for various scenarios of l. The case where there has been a problem with two connections -one between controller-1 and sensor-1 and the other between controller-4 and sensor-3 has been studied.
\begin{table}[]
\centering
\caption{Results- Load Variation with communication node failure}
\label{tab:Results- Load Variation with communication node failure}
\begin{tabular}{|c|c|c|c|}
\hline
System Matrix & \begin{tabular}[c]{@{}c@{}}Max-eigenvalue \\ (open)\end{tabular} & \begin{tabular}[c]{@{}c@{}}Max-eigenvalue\\ (closed)\end{tabular} & \begin{tabular}[c]{@{}c@{}}Connection\\  Parameter\end{tabular} \\ \hline
$A_1$         & 0                                                                & -1.0633                                                           & 3300                                                            \\ \hline
$A_2$         & -21.9580                                                         & -23.4642                                                          & 3300                                                            \\ \hline
$A_3$         & -42.4541                                                         & -46.8994                                                          & 3300                                                            \\ \hline
\end{tabular}
\end{table}
\\ An important finding in this method is that a even if the system works with lower values of bandwidth and other constraints, that is, with lower number of connections, comparable stabilities are achieved to that of highly connected cases. This is in contrary to the general notion that system stability is proportional to number of communication connections present. 
\section{Summary}\label{Sec:4Conclusion}
The issue of adding stability to a microgrid system using communication has been dealt with. Systematic optimization framework using the Lyapnov Function has been developed which can enhance the voltage stability of communication intensive decentralized control architecture. For sparse architecture based control, a heuristic constraint based sensor-controller communication design algorithm has been developed which finds the cheapest set of constraints within the maximum constraint range to operate with. The efficacy of the technique has been demonstrated in constrained situations of load change, delay change and communication link failure.

 \chapter{Peak Load Shaving through Real Time Modeling for Cyber Physical Smart Grid Stability\label{NFTSMC}}
\section{Introduction}
In this chapter, real time peak load shaving concept has been added as another objective along with our primary objective of smart grid stability. This is meant to make our architecture user-centric by fulfiling user-requirement of economical and reliable smart grid. This formulation handles various tasks with different deadlines under many types of constraints like delay, scability requirements, communication limitations.
This chapter has been organized as follows. Section \ref{Sec:MIMO} deals with MIMO system model, where the system model, controller structure and the optimization problem are discussed. The cyber physical smart grid system has been introduced in \ref{Sec:application}. The proposed connection design algorithm has been presented in Section \ref{Sec:MIMO_proposed_algo}.Simulation results are presented in \ref{Sec:results}. Results of real time scheduling algorithms have been presented in next Section. Our novel framework to achieve peak load shaving as additional objective along with stability as the primary objective have been presented in next Section. Final results with this novel framwrok are presented in the next section followed by concluding remarks in Section \ref{Sec:5co}.
\section{System Model} \label{Sec:MIMO}
\begin{figure}[!ht]
\centering
\includegraphics[width=0.9\textwidth, height=7cm, keepaspectratio]{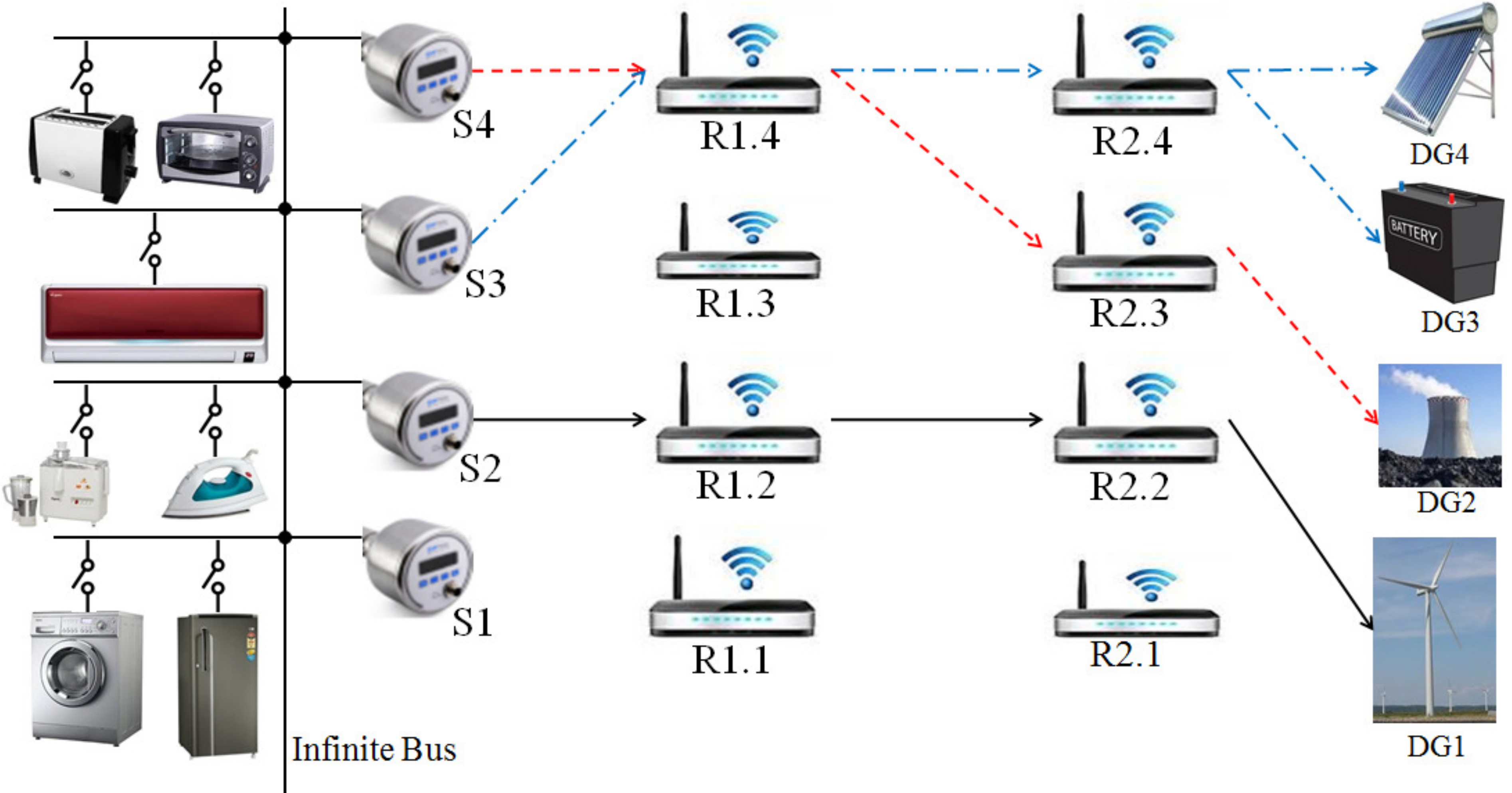}
\caption{An illustration of communications in CPES}
\label{fig:figure1}
\end{figure}

Let's refer Figure \ref{fig:figure1} which schematically depicts a four-bus distribution system operating in islanded mode. So, the infinite bus is disconnected as shown in this figure. This figure shows that the system consists of four sensors measuring the voltage at the points of common coupling (PCCs) (the point where the DGs are integrated into network) and load demand in that region, different types of loads present in the system, four DGs of varying capacities and two array of relay nodes for communication between sensors and controllers. The objective is to design a proper communication link between the controllers and sensors such that installed DGs can stabilize the voltage at a pre-defined reference value and shave peak load using real time modelling algorithms. 
This system has four inputs and four outputs. Such systems can be modeled as multi-input multi-output (MIMO) system. In this paper we are concerned only with linear models.
\subsection{MIMO System Model} \label{Sec:MIMO_model}
The generic model of a linear MIMO system is given as:
\begin{equation}
\left.
\begin{aligned}
&\dot{x}(t)=Ax(t)+Bu(t)~~\\
&y(t)=Cx(t)\label{Eq:sys}
\end{aligned}
\right\}
\end{equation}
where $x \in R^n, A \in R^{n \times n}, B \in R^{n \times m}, u \in R^m, y \in R^p$ and $C \in R^{p \times n}$\\
This model takes into account that there are $p$ sensors and $m$ control inputs (DGs for smart grid system).
\subsection{Controller Structure} \label{Sec:MIMO_Control}
The control law is given as:
\begin{align}
u(t) = Ky(t)
\label{Eq:ut}
\end{align}
where the matrix $K$ gives the structure of connections between the control input vector $u(t)$, i.e. DGs and sensor vector $y(t)$. This matrix $K$ has the following structure:
\begin{equation}
\nonumber
  k_{ij}\begin{cases}
    &=0 \text{, if no connection between sensor $j$ - controller $i$}\\
    &\neq 0 \text{, otherwise}.
  \end{cases}
\end{equation}	
\subsection{Optimization Problem} \label{Sec:MIMO_optimization}
Considering a Lyapunov function $V(x)=x^TPx$, following ineqauility has been used as the optimization problem to maintain stability.
\begin{equation}
A^TP +\left(BKC\right)^TP+PA+PBKC+\gamma I<0 \label{Eq:unknown}
\end{equation}
One can note that larger the $\gamma$ value, larger is the stability margin. As one optimizes equation \eqref{Eq:unknown}, various constraints arise because of practical limitations of the Cyber-Physical System. We also would like to see that controller gains should be bounded, i.e. the second constraint is $||K||_2<\rho$ where $\rho$ is a user defined scalar value.
\section{Cyber Physical System Structure} \label{Sec:application}
This work adopts an IEEE 4 bus distribution feeder \cite{rp:ieee_4bus} as test system which has been illustrated in Fig. \ref{fig:figure1}. 
Entire smart grid system has been modeled as linear model as proposed in section 2-a and 2-b. \cite{rp:multicast_paper}. $m=n=p=4$ in the considered model of the four-bus system. The Cyber-physical system structure has been divided into two parts: physical smart grid system and the communication network. 
\subsection{Physical Smart Grid System}
The model of a power network operating in island mode has source voltage $V_s~=~0$. Here the feeder is disconnected from the infinite bus. Power Electronic (PE) interfaces consisting of an inverter and a DC link capacitor are used to connect DGs to the network at the PCCs. A coupling inductor separates inverter from the rest of the system.\\
Laplace transform for the four bus system produces nodal voltage equations as:
\begin{equation} 
Y(s)v_t(s)=\frac{1}{s}Lv_c(s)
\label{Eq:yvsl}
\end{equation}
where $Y(s)$ is the admittance matrix, $v_t \in R^p$ represents the vector of output voltages at PCC which is also known as sensor output vector, $v_c \in R^m$ represents the vector of DG output voltages which is also known as input vector in the current formulation, and $L \in R^m$ represent the coupling inductors between DGs and the feeder network. 
The Laplace transform model of equation \eqref{Eq:yvsl} has been converted to linear state-space representation in time-domain \eqref{Eq:sys} of section 2.
Voltages at PCCs $x~=~v_t-V_{ref}$ have been taken as the state vector $x$. As these voltages also represent output $y$, the output matrix $C=I$ where $I \in R^{n\times n}$ is the identity matrix. It is to be noted that the perturbations as noise are not considered in this model. The objective of the voltage control in the feeder network is to maintain all PCC voltages to a pre-defined reference value $V_{ref}$ and at the same time shave peak load.
\subsection{Communication Network} \label{Sec:communication network structure}
The communication network consists of sensors, relays and controllers. It has been assumed that communication interfaces are attached with sensors and controllers. 
The following two assumptions have been made for easier analysis considering practical limitations in communication. \cite{rp:multicast_paper}:
\begin{enumerate}
\item The communication speed is very fast as compared to the physical system dynamics. So, data that flows from sensors to controllers are continuous i.e. there is no delay, no packet loss and no network uncertainties.
\item For a link between any two nodes say $n_1$ and $n_2$, the transmission of data from one sensor requires one unit of bandwidth. Here it is assumed that the bandwidth constraint equals 1. So any node or link would accept and transmit data from only one sensor. But controllers are allowed to receive data from multiple sensors.
\item The data of a particular node can be transmitted only to its next hop neighbors.
\end{enumerate}
Thus communication from sensors to controllers is established via two arrays of relay nodes, thus forming a $4\times4$ array of nodes. The structure of $K$ matrix decides the communication topology as well as the control command required for maintaining bus voltage in the smart grid. Further it is used to distribute loads to achieve peak load shaving.
The generalized connection design process has been discussed in the following section.
\section{Proposed Communication Design Algorithm} \label{Sec:MIMO_proposed_algo} 
\begin{figure}[!ht]
\begin{minipage}[b]{.5\linewidth}
\centering
\includegraphics[width=0.9\textwidth, height = 4cm]{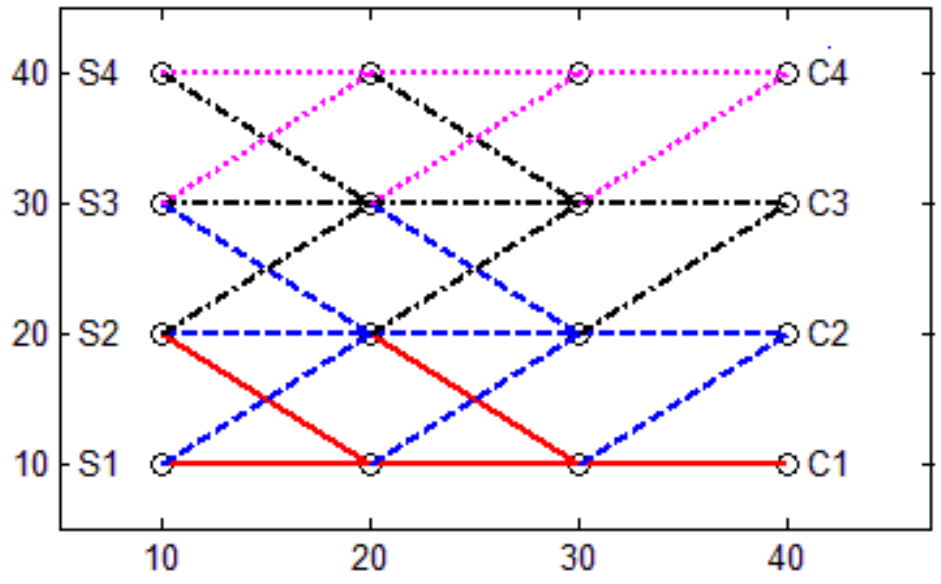}
\caption{Possible connections}\label{Fig:Network_Connections}
\end{minipage}%
\begin{minipage}[b]{.5\linewidth}
\centering
\includegraphics[width=0.9\textwidth, height =4.2cm]{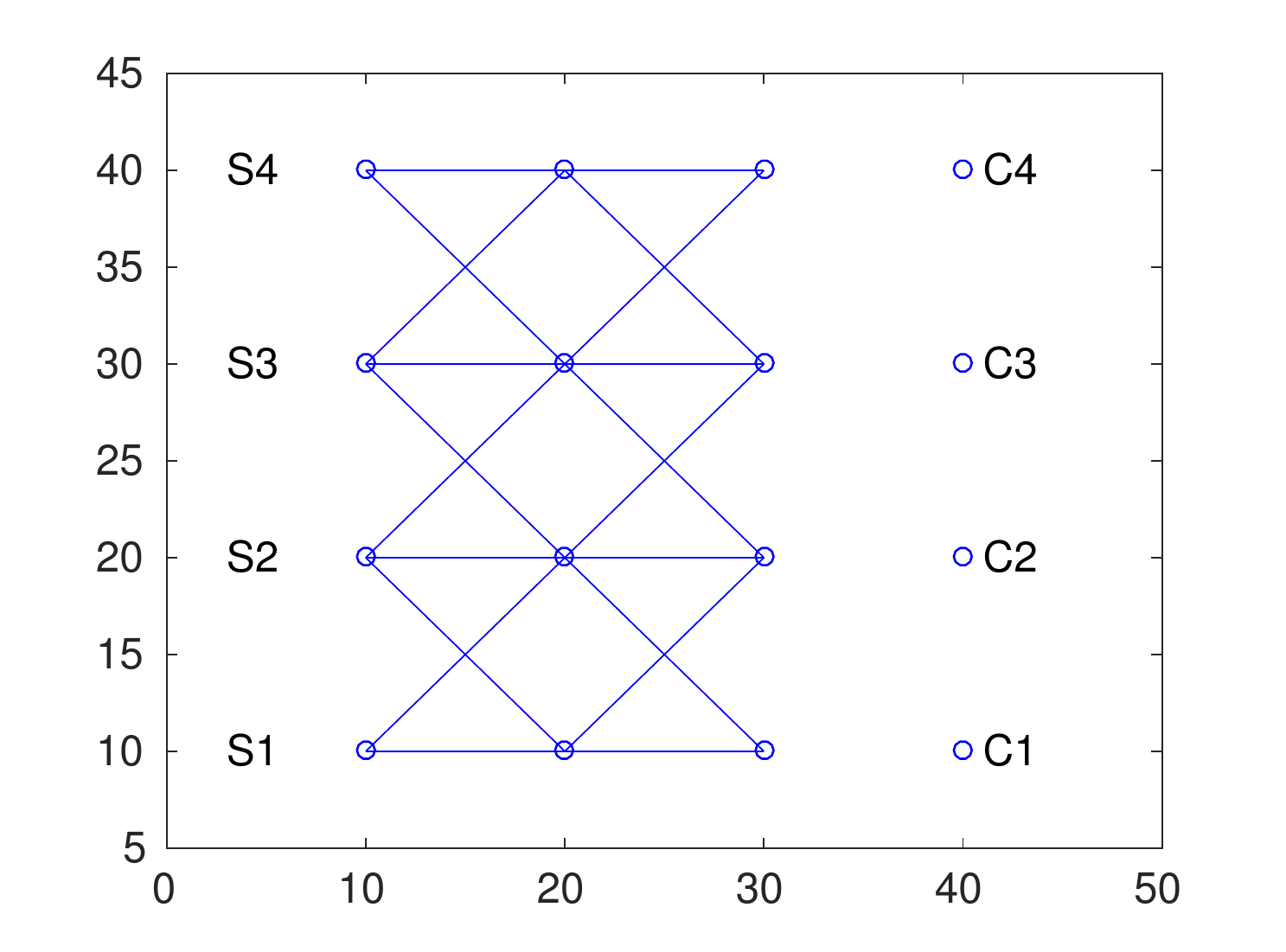}
\caption{Step 6 output}\label{Fig:step6out}
\end{minipage}
\end{figure}

\begin{figure}[!ht]
\centering
\includegraphics[width=0.9\textwidth, height = 5cm]{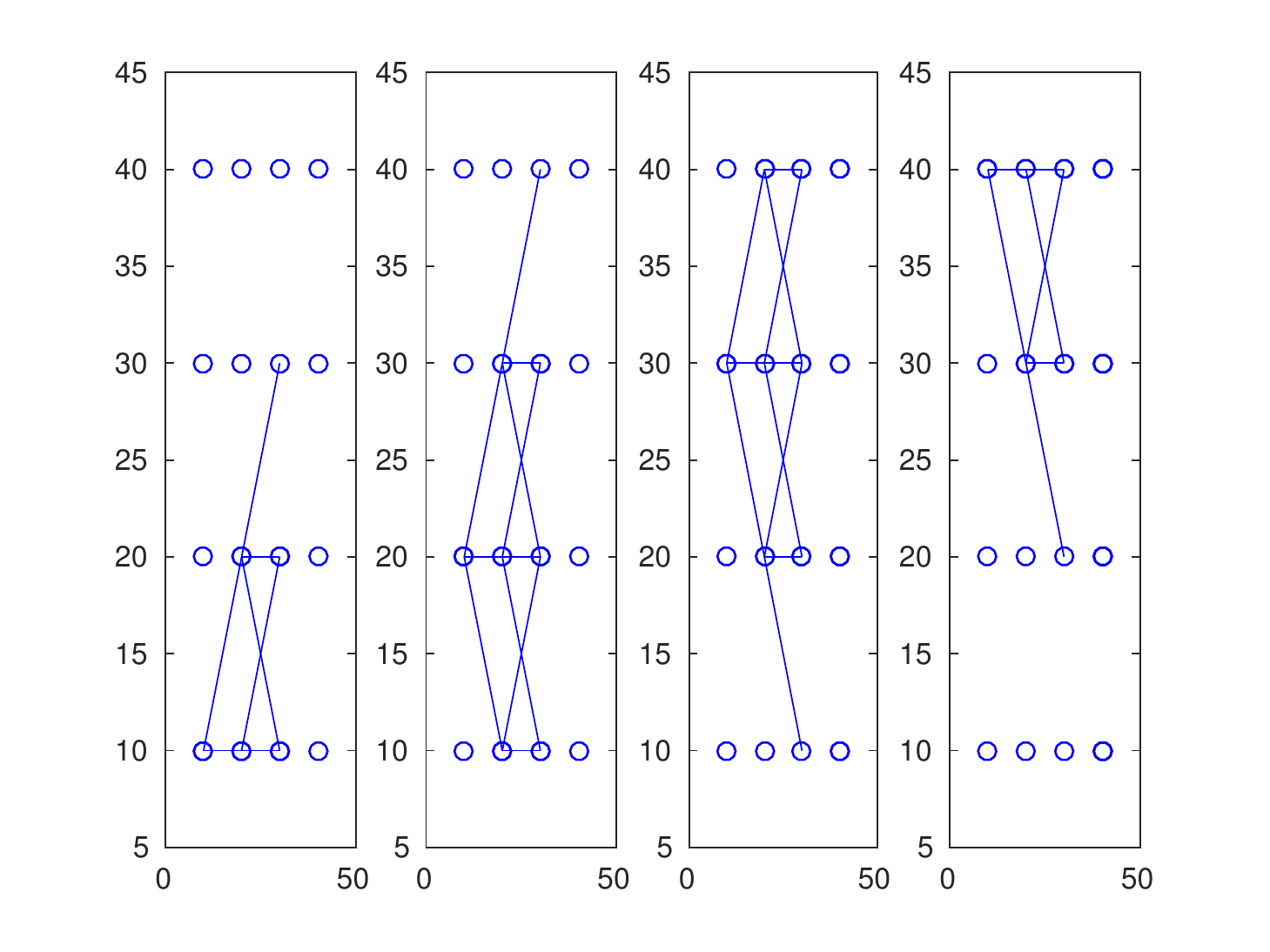}
\caption{Grouping of connections}
\label{Fig:step7}
\end{figure}

\begin{figure}[!ht]
\begin{minipage}[b]{.65\linewidth}
\centering
\includegraphics[width=0.9\textwidth, height = 5cm]{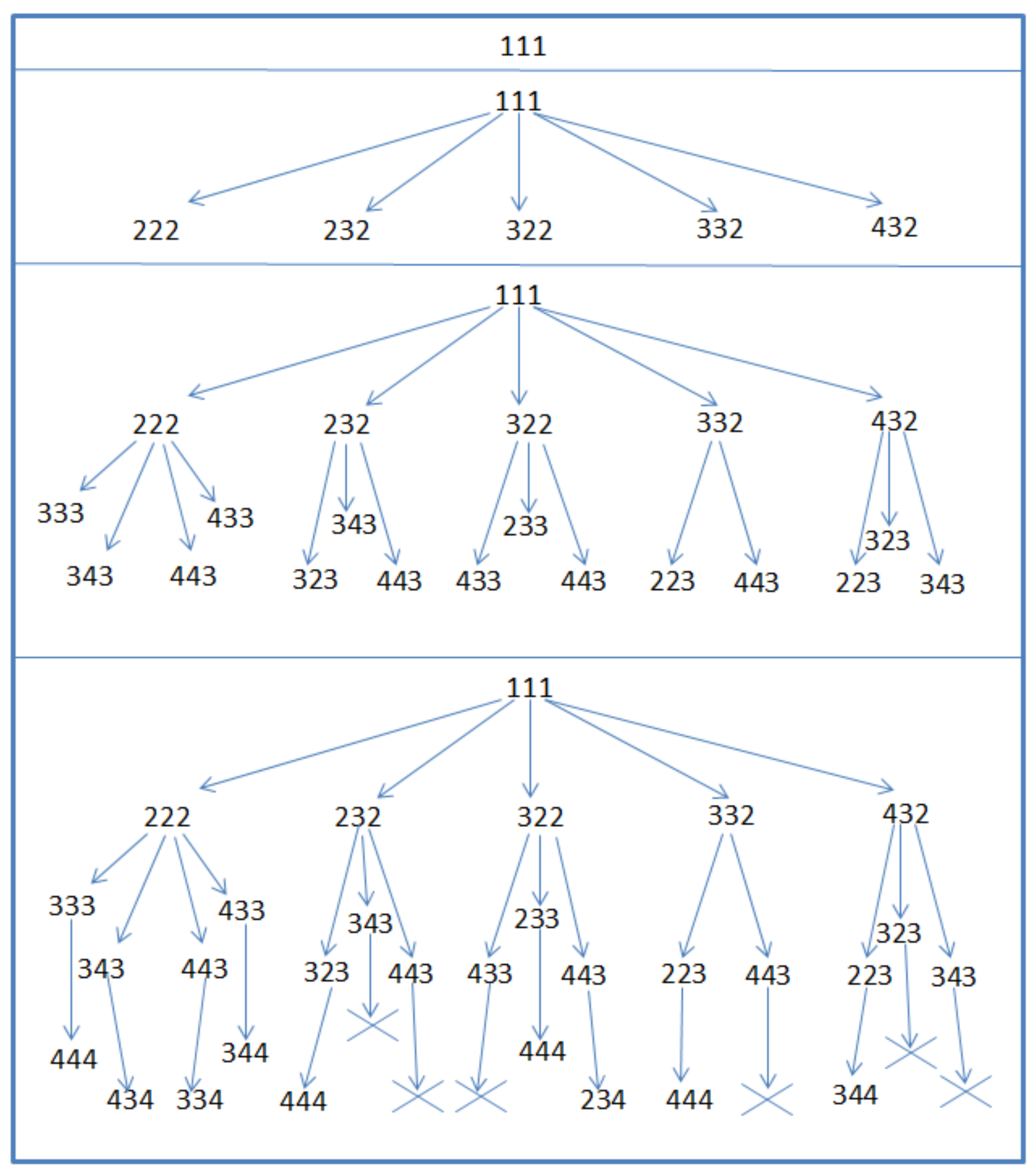}
\caption{Group of sub groups}\label{Fig:Group of sub groups}
\end{minipage}%
\begin{minipage}[b]{.35\linewidth}
\centering
\includegraphics[width=0.9\textwidth, height = 5cm]{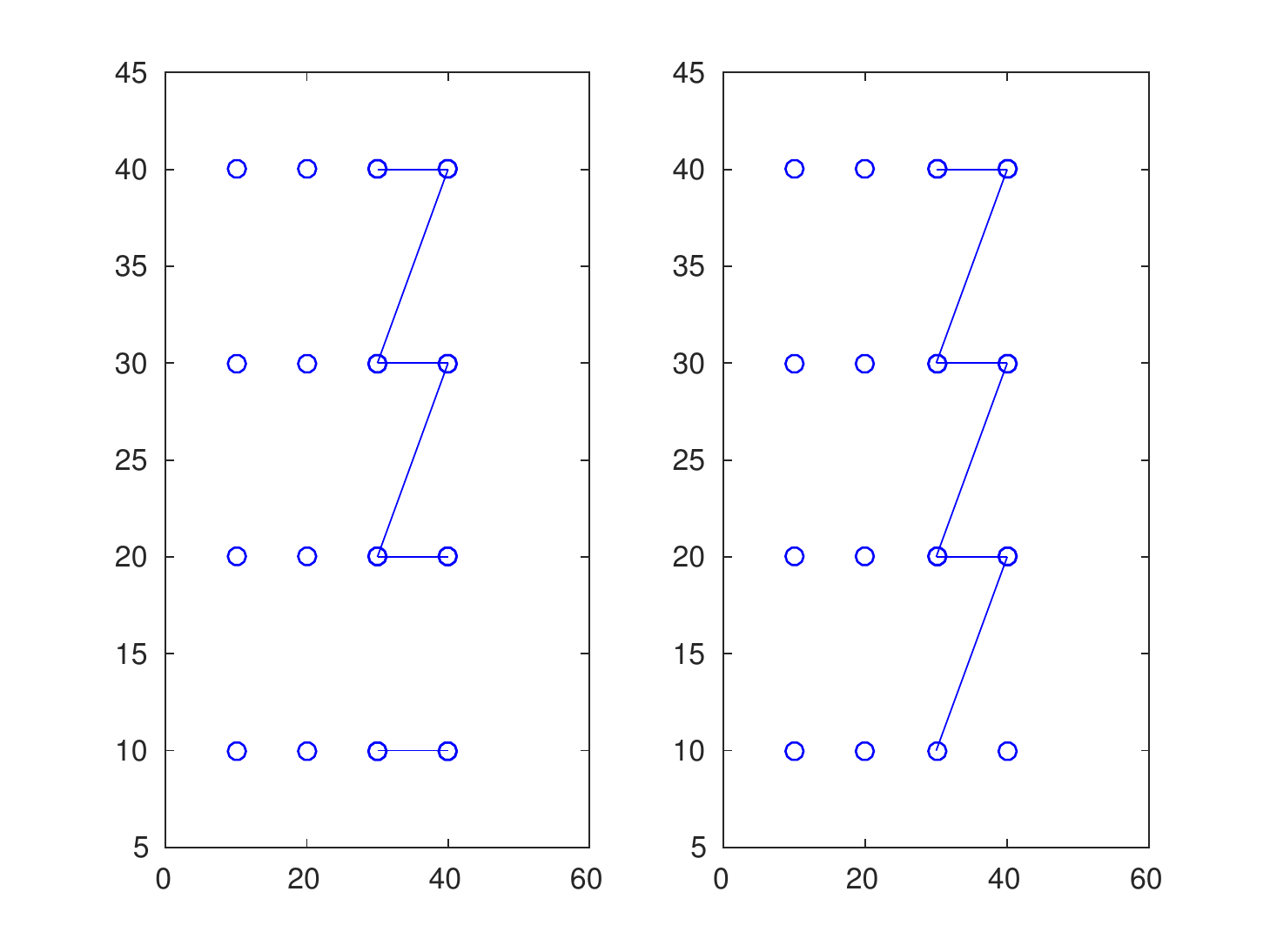}
\caption{Set w}\label{Fig:Step9}
\end{minipage}
\end{figure}

In this section, a novel generalized framework for most stable sensor-controller connection design is presented.\\
\textbf{Step 1:} Enter Specifications \\ 
Enter the number of sensors, controllers and intermediate relays present in the system.\\
\textbf{Step 2:} Define connection constraints imposed on the cyber-physical system.\\
For the application considered the connection constraints are defined based on the assumptions made in section \ref{Sec:communication network structure}.\\ 
\textbf{Step 3:} Find all possible connections with connection constraints.\\
Starting from the senser layer, successively map possible connections to previous layers till controller layer following the connection constraints.\\ 
Executing this results in a set of $47$ connections as shown in the Figure \ref{Fig:Network_Connections} below
\\
\textbf{Step 4:} Define the system's physical resource constraints.\\
For the application considered, physical resource constraints are defined as bandwidth constraint.\\
\textbf{Step 5:} Define user-requirement.\\
Here, user-requirement is optimized scheduling of loads to achieve peak load shaving. (Assumption is relay 24 and relay 21 cover lesser load regions as compared to relay 23, relay 22. Lesser load regions can be managed with 1 controller connected where as higher load regions should have 2 controllers connected. Depending on the secondary objective, assumptions can be changed.)\\
\textbf{Step 6:} Find the set of connections which can exist together without violating user-requirement constraint.\\
\emph{for} $i=g\footnote{represents the index of final layer, in this case $g = 4$}:-1:1$\\
$\left\{\right.$\\
$~~~$\emph{if}$~(i\neq h\footnote{index of layer where user-requirement constraints are present. For the application considered, user-requirement constraints are at controllerss, hence h=4})$\\
$~~~$Add connections to its previous layer considering connection constraints.\\
$\left.\right\}$\\
From this, we get $26$ connections as shown in Figure \ref{Fig:step6out}.\\
\textbf{Step 7:} Form groups from the results of step 6 in such a way that every group covers till the layer nearest to $h$ as shown in Figure \ref{Fig:step7} (in this case the nearest layer is that of relay 2).\\
\\
$~$\textbf{Step 8:} Find all sets of co-existing sub-groups without violating any constraints and place them in set $v$.\\
$~$\textbf{Step 8.1:} Define different groups in such a way that every group converges to the layer nearest to $h$ and divide the group in to $d$ divisions based on the sensor (layer which has physical resources constraint). The divisions are tabulated in Table \ref{Tab:Divisons}.\\
\begin{table}[!h]
\centering
\caption{Divisons}
\label{Tab:Divisons}
\begin{tabular}{|c|c|c|}
\hline
\multirow{3}{*}{d1} & d1.1 & 111,121     \\ \cline{2-3} 
                    & d1.2 & 211,221     \\ \cline{2-3} 
                    & d1.3 & 321         \\ \hline
\multirow{4}{*}{d2} & d2.1 & 112,122     \\ \cline{2-3} 
                    & d2.2 & 212,222,232 \\ \cline{2-3} 
                    & d2.3 & 322,332     \\ \cline{2-3} 
                    & d2.4 & 432         \\ \hline
\multirow{4}{*}{d3} & d3.1 & 123         \\ \cline{2-3} 
                    & d3.2 & 223,233     \\ \cline{2-3} 
                    & d3.3 & 323,333,343 \\ \cline{2-3} 
                    & d3.4 & 433,443     \\ \hline
\multirow{3}{*}{d4} & d4.2 & 234         \\ \cline{2-3} 
                    & d4.3 & 334,344     \\ \cline{2-3} 
                    & d4.4 & 434,444     \\ \hline
\end{tabular}
\end{table}
\begin{table}[!h]
\centering
\centering
\caption{Co-existing subgroups set $v$} \label{Tab: subgroups set}
\begin{tabular}{|c|c|} \hline
v1 & 111, 222, 333, 444\\ \hline
v2 & 111, 222, 343, 434\\ \hline
v3 & 111, 222, 433, 344\\ \hline
v4 & 111, 222, 443, 334\\ \hline
v5 & 111, 232, 323, 444\\ \hline
v6 & 111, 322, 233, 444\\ \hline
v7 & 111, 322, 443, 234\\ \hline
v8 & 111, 332, 223, 444\\ \hline
v9 & 111, 432, 223, 344\\ \hline
v10 & 121, 212, 333, 444\\ \hline
v11 & 121, 212, 343, 434\\ \hline
v12 & 121, 212, 433, 344\\ \hline
v13 & 121, 212, 443, 334\\ \hline
v14 & 211, 122, 333, 444\\ \hline
v15 & 211, 122, 343, 434\\ \hline
v16 & 211, 122, 433, 344\\ \hline
v17 & 211, 122, 443, 334\\ \hline
v18 & 211, 332, 123, 444\\ \hline
v19 & 211, 432, 123, 344\\ \hline
v20 & 221, 112, 333, 444\\ \hline
v21 & 221, 112, 343, 434\\ \hline
v22 & 221, 112, 433, 344\\ \hline
v23 & 221, 112, 443, 334\\ \hline
v24 & 321, 112, 233, 444\\ \hline
v25 & 321, 112, 443, 234\\ \hline
\end{tabular}
\end{table}
\textbf{Step 8.2:}\emph{for }$i = 1:5$\\
$\left\{\right.$ $~~~$\textbf{Step 8.2.1:} Select $i^{\text{\tiny th}}$ element of d1 \\ \emph{for }$j = 2:d$\\
$\left\{\right.$\\
$~~~$\textbf{Step 8.2.2:} Select feasible conections from d$j$ without violating any constraint (sensor and relay-1 i.e. first and second number of the connection should be different from the existing connnections in the set as illustrated iteration-wise in Figure \ref{Fig:Group of sub groups}).\\
$\left.\right\}$\\
$\left.\right\}$.\\
\\
\begin{table}[]
\centering
\caption{w set elements}
\label{W set}
\begin{tabular}{|c|c|}
\hline
w1 & 11,22,23,33,34,44 \\ \hline
w2 & 12,22,23,33,34,44 \\ \hline
\end{tabular}
\end{table}
%
\textbf{Step 8.3:} Write each set of subgroups into set $v$ found at the end of step 8.2 if they are complete i.e. if they have $d$ number of subgroups in their set as shown in Table \ref{Tab: subgroups set}.\\
\textbf{Step 9:} Find the available complete coexisting elements set $w$ in the layer where user-requirement constraints are present to connect to each of the element of set $v$.\\ 
\emph{for} $i = 1:d$ \\
$\left\{\right.$\\
$~~$\textbf{Step 9.1:} Find the number of elements available in the layer where user-requirement constraints are present (controller layer) following the connection constraint to connect to each of the complete co-existing subgroups set.\\
$~~$\textbf{Step 9.2:} Connect them to the each of the elements of set $w$ following the user-requirement constraint of step-6.\\
$\left.\right\}$\\
The elements of set $w$ have been shown in Figure \ref{Fig:Step9} and tabulated in Table \ref{W set}.\\
\textbf{Step 10:} Find the complete set of connections by connecting each member of set $v$ to that of set $w$.\\
From Tables \ref{Tab: subgroups set} \& \ref{W set} we obtain $50$ sets of connections (for case-1) that can simultaneously exist. Each connection set consists of $6$ paths between sensors and controllers.\\ 
\textbf{Step 11:} Solve the matrix equation in \eqref{Eq:ut} for all complete set found in Step 10.\\
\textbf{Step 12:} Solve optimization problem in equation \eqref{Eq:unknown} using LMI and obtain $\gamma$.\\
\textbf{Step 13:} List the connection sets with maximum $\gamma$.\\
\textbf{Step 14:} From step 13 choose the connection with shortest path and obtain the corresponding $K$.\\
\section{Real time scheduling algorithms}\label{realtime algo}
The main idea is to use real time scheduling algorithms to schedule and shift the peak electric loads in smart grid. In this framework we are representing our physical systems in a more detailed way by adding loads in the form of tasks present in different houses. Similar to our previous work, these informations are carried from sensors to controllers and controllers take action using our proposed framework to save peak load. Cyber physical smart grid system and real time scheduling algorithms are related by adding user-requirement section in our formulation to achieve smart grid stability. By modeling loads as tasks with priority and time constraints, real time scheduling algorithms are used to handle execution of various tasks. \cite{rp:facchinetti2010reducing}. The condition is obviously that sum of powers should be less than the maximum power capacity of the grid. Our framework is meant to reduce peak loads which is undesirably present in many homes\cite{rp:lee2011power}. 
\subsection{Algorithm Introduction}\label{ssec:algos}
\subsubsection{Earliest Deadline First}
Earliest deadline first (EDF) is a dynamic scheduling algorithm used in real-time operating systems to place processes in a priority queue\cite{wiki:edf}. At the starting of every time slot, the process with the closest deadline is selected for scheduling. Note that in this process total power should not exceed the capacity. It is also called as least time to go algorithm. This is because EDF scheduling algorithm can complete all the tasks within deadlines obviously if tasks are not enabling CPU to be utilized more than 100\%,

\subsubsection{Least Laxity First}
Least laxity first assigns priority based on the slack time of a process, i.e., the amount of time left after a job if the job was started now \cite{wiki:llf}. At the start of each time slot, it selects the tasks with lowest slack time for scheduling. Obviously in this process, total power should not exceed the capacity. 
Mathematically, slack time is\cite{wiki:llf} 
\begin{equation}
(d_{i} - t) - {c_{i}}'
\end{equation} 
where $d_{i}$ is the deadline of task $i$, $t$ is the real time since the starting of the task $i$, and ${c_{i}}'$ is the remaining execution time of the task $i$ at time $t$.
\subsubsection{Dynamic Rate Priority}
Dynamic Rate Priority Scheduling uses a dynamic rate function to change priority of various scheduled tasks. This function considers the difference between deadline and time required by the task to complete the job at time t. So, basically it provides the rate with time t \cite{wiki:drp}. Mathematically,
\[P_{i} = P_{i} \ast  \left (  1 - \frac{d_{i} - {c_{i}}'}{t}\right )\]
where $P_{i}$ is the  priority of task $i$, $d_{i}$ is the deadline of task $i$ and ${c_{i}}'$ is the remaining execution time of task $i$ at time $t$.
\subsection{System Architecture}
\subsubsection{Model Developement}
 A task model comprises of task arrival time, scheduling time or execution time and task deadline. So, each task $T_i$ is modelled as a set of \[< r_i , d_i , s_i >\] where $r_i$ is the arrival time of task $T_i$, $d_i$ is the deadline of task $T_i$, and $s_i$ denotes the scheduling unit.

\subsubsection{State of Tasks}
Each task can obtain one of four states which is \textit{completed}, \textit{hold}, \textit{off} and \textit{running}. The significance of these terms are very much straightforward. When the task is not arrived i.e. not yet started by the system, the task is in the \textit{off} state. Currently if the application is running, task attains the status as \textit{running}. If the task has arrived but not running, then the task is in \textit{hold} state. This may be due to many reasons like absence of enough power to execute the task or low priority. The \textit{completed} state indicates the task has completed.	
\subsubsection{Power Constraints}
The power constraints imposed by the system on the scheduling problem has been already briefed before in previous sections. Sum of powers of all the tasks running at a time slot must be less than total power capacity, otherwise the task will obtain \textit{hold} status. The only way to deal with this is to satisfy the power constraints or increase the priority of our desire task manually or in an automated way.
\begin{equation}
P^{\ total} \geq \sum_{i\  \in \ running} P_{i}^{\ t}
\label{Eq:constraint}
\end{equation}
where $P^{\ total}$ is the total available power and $P_{i}^{\ t}$ is the power consumed by $i^{th}$ running application at time $t$.
\subsubsection{Peak Load Scheduling Algorithms}
Basically real time peak load scheduling algorithm varies based on the way priority changes. It uses the K matrix of the generalized sensor-controller connection design algorithm. K matrix is used to distribute loads among sensors to enable the aplication of decentralized control. At each controller, the sum of electric powers of tasks already arrived (which are not completed yet) are calculated. If the sum is less than the power constraint in \ref{Eq:constraint} present the system, the algorithm will execute all the tasks for given time slot as given in equation \ref{Eq:running}. 
\begin{equation}
pt = pt + P_{i},
\quad
C_{i} = C_{i} - 1, 
\quad
Status_{i} = running, 
\quad
\label{Eq:running}
\end{equation}
where $C_{i}$ is remaining execution time of task $i$.

If power requirement is above threshold, these tasks are executed based on priority. They are sorted according to priority, highest to lowest, and lower priority tasks will enter to \textit{hold} state according to equation \ref{Eq:hold} till it satisfies the power constraint in equation. \ref{Eq:constraint}
\begin{equation}
pt = pt - P_{i},
\quad
C_{i} = C_{i} + 1,
\quad
Status_{i} = hold,
\quad
\label{Eq:hold}
\end{equation}
When the task arrives, natually it is in \textit{off} state, it is directly copied into the running queue and its state changes to \textit{running}. When remaining computation time of a task becomes zero, it is removed from the running queue and attain \textit{completed} status.
\begin{equation}
Status_{i} = completed\  \vert \ C_{i} = 0\ \forall\ i \in\ running
\end{equation}
Among the tasks in \textit{hold} state, those which satisfy the power constraint in equation \ref{Eq:constraint} are sorted based on the priority. Higher priority tasks will enter into the running state till the total power stays below the given threshold.

Priorities changee as per the standard algorithms which have been already discussed in section \ref{ssec:algos}.
\section{Simulation results} \label{Sec:results}
Simulations have been performed to find the $K$ matrix corresponding to the most stable connection with achievement of peak load shaving in the form of user requirement. In our previous work \cite{mishra2015generalized} we have developed an algorithm to find the most stable connection. But the focus was not on peak load shaving. The algorithm of section \ref{Sec:MIMO_proposed_algo} has been developed to incorporate specific user-requirement apart from the primary user-requirement of stability. So, to validate efficiacy of our algorithm we have compared with previous algorithms. There are 4 following scenarios which we have compared through simulation.
\begin{enumerate}
\item Greedy algorithm based connection design to find the most stable connection with reliabillity as the user requirement. 
\item Optimal connection design algorithm to find most stable and reliable connection.
\item Optimal connection design algorithm to find most stable and economic connection..
\item Proposed connection design algorithm to maximize stability and achieve peak load shaving as user-requirement.
\end{enumerate} 
All simulations have been carried out using MATLAB and YALMIP toolbox \cite{rp:yalmip}. In all scenarios, $\rho=5$ and the value of $\beta$ is chosen as 5000.
\subsection{Scenario 1} \label{ssec:s1}
\begin{figure}[!ht]
\centering
\includegraphics[width= 0.9\textwidth, height=7cm, keepaspectratio]{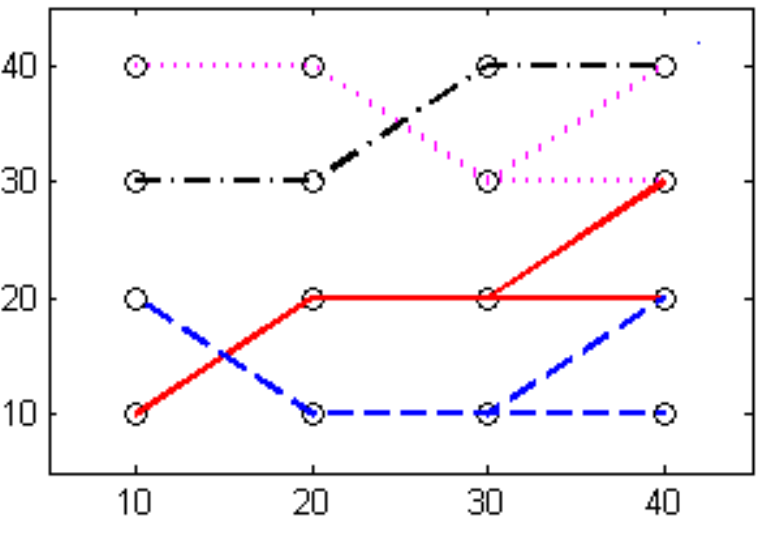}
\caption{Final topology for scenario 1}\label{Fig:Scenario1_plot}
\end{figure}

\begin{figure}[!ht]
\centering
\includegraphics[width= 0.9\textwidth, height=7cm, keepaspectratio]{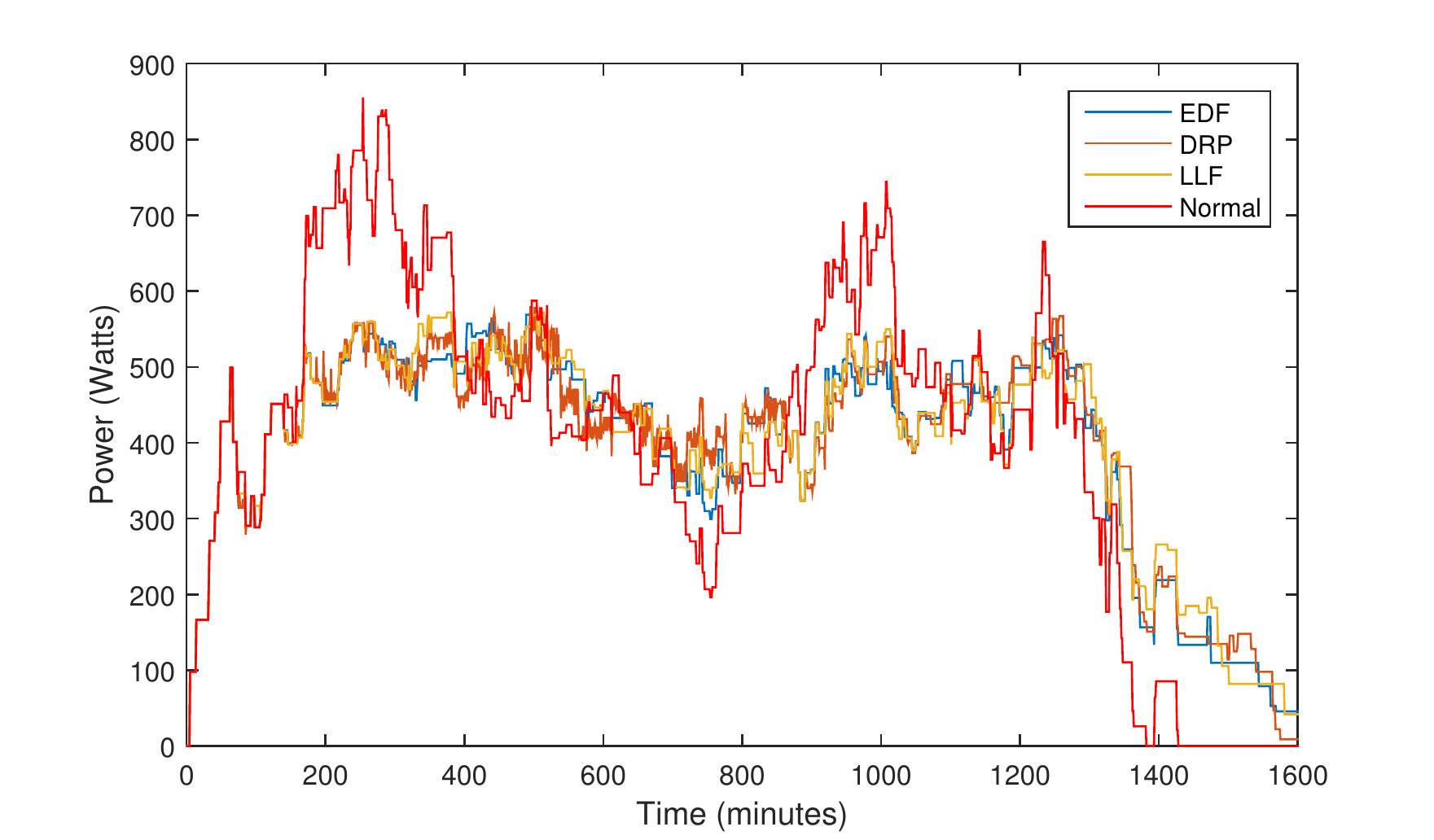}
\caption{Power plot for Scenario 1}\label{Fig:Scenario1_power_plot}
\end{figure}

\begin{figure}[!ht]
\centering
\includegraphics[width= 0.9\textwidth, height=7cm, keepaspectratio]{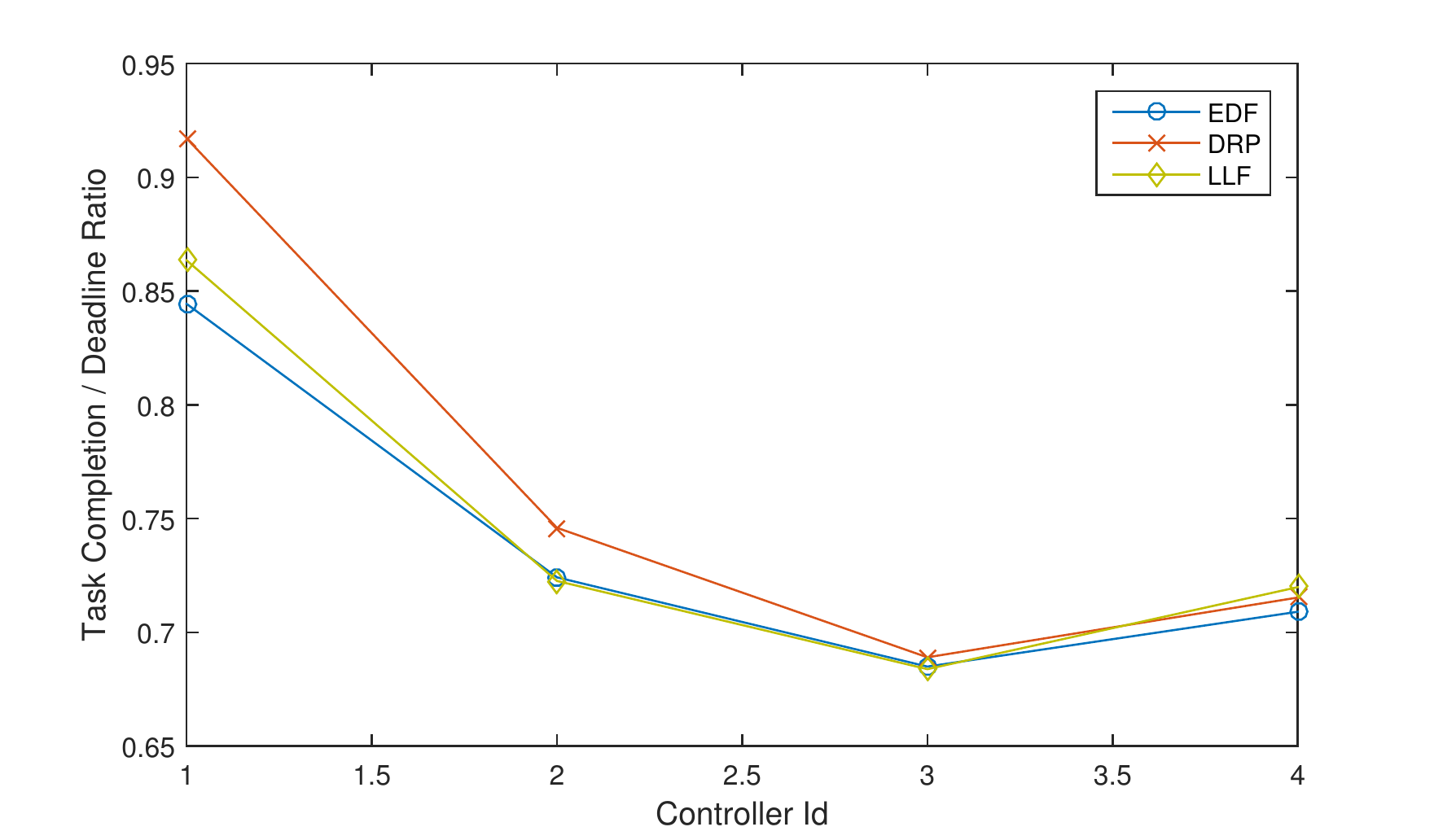}
\caption{Algorithm comparison for Scenario 1}\label{Fig:Scenario1_comp_plot}
\end{figure}

Results obtained for this case\cite{mishra2015generalized} are maximum eigenvalue value: -213.2269, $\gamma$ = 30.7836 with $K$ matrix in \eqref{Eq:K_scenario1} and the final topology is presented in Figure \ref{Fig:Scenario1_plot}. The power plot for scheduling algorithms are shown in Figure \ref{Fig:Scenario1_power_plot} and algorithms are compared in Figure\ref{Fig:Scenario1_comp_plot} on Task Completion / Deadline Ratio.\\ 
\begin{equation}
K=\left(
\begin{tabular}{cccc}
0 & 1.3568 & 0 & 0 \\
-0.9309 & -0.7023 & 0 & 0 \\
-0.4902 & 0 & 0 & -2.1480 \\
0 & 0 & 0.2385 & 0.2955\\
\end{tabular}
\right)\label{Eq:K_scenario1}
\end{equation}

%
%
\subsection{Scenario 2} \label{ssec:s2}
\begin{figure}[H]
\centering
\includegraphics[width= 0.9\textwidth, height=7cm, keepaspectratio]{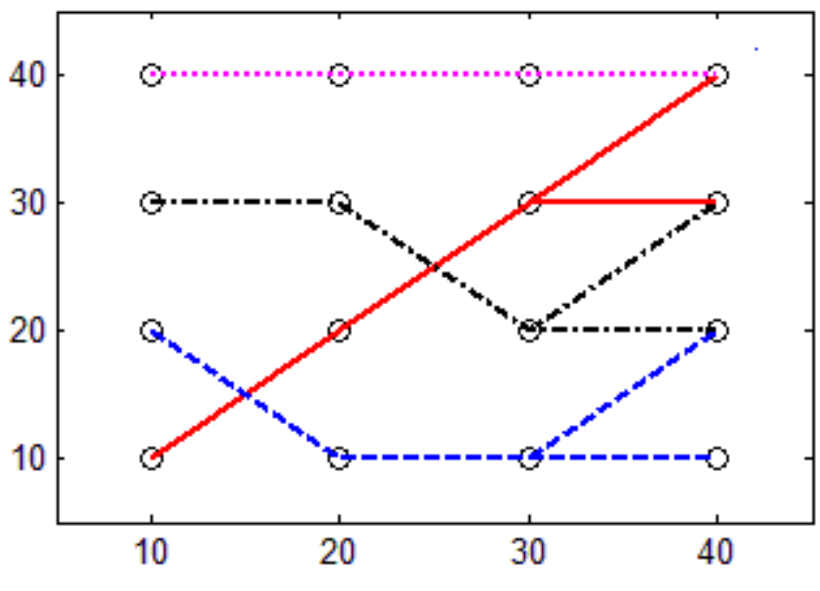}
\caption{Final topology for scenario 2}\label{Fig:Scenario2_plots_reduced}
\end{figure}

\begin{figure}[H]
\centering
\includegraphics[width= 0.9\textwidth, height=7cm, keepaspectratio]{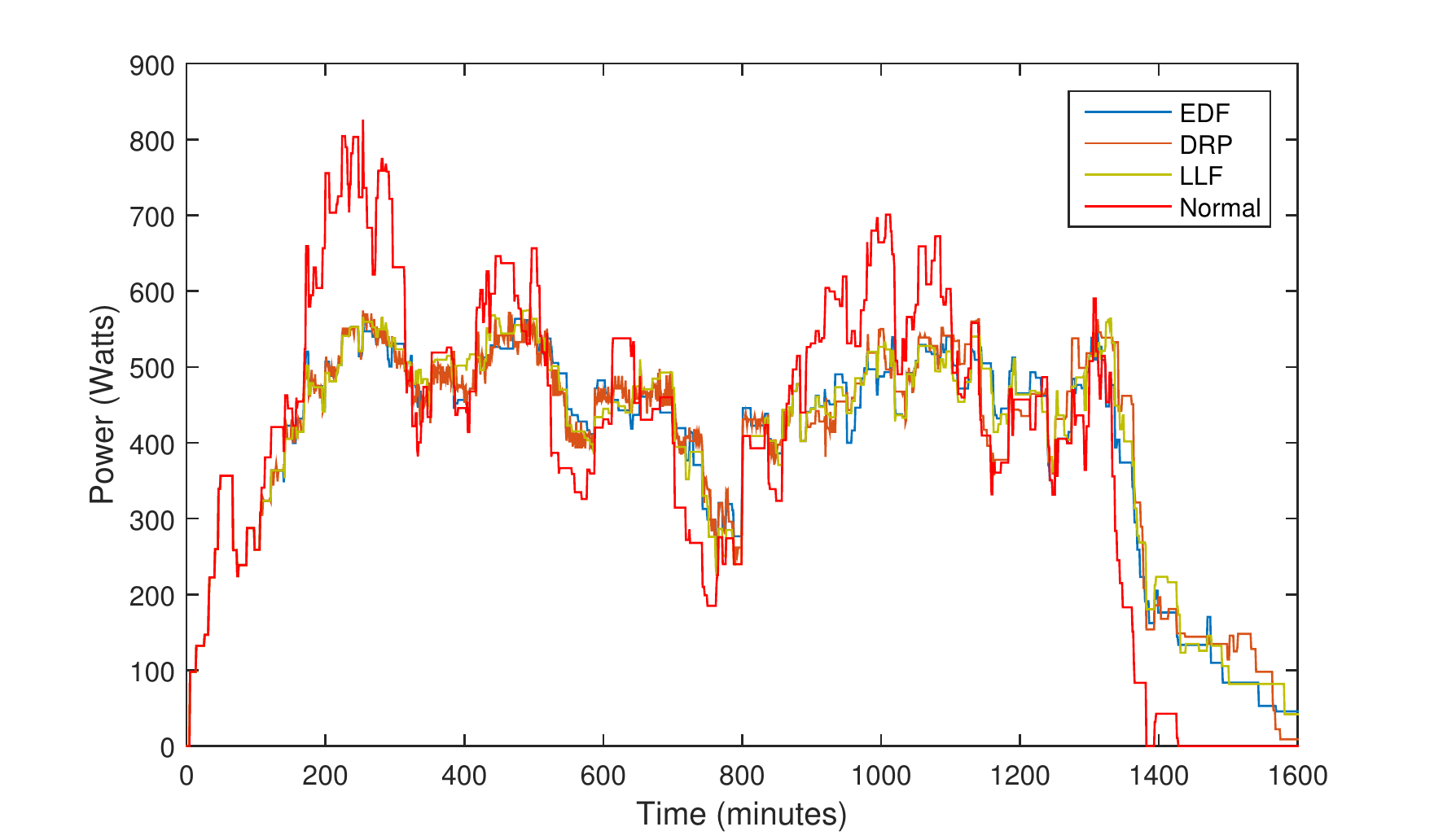}
\caption{Power Plot for Scenario 2}\label{Fig:Scenario2_power_plot}
\end{figure}

\begin{figure}[H]
\centering
\includegraphics[width= 0.9\textwidth, height=7cm, keepaspectratio]{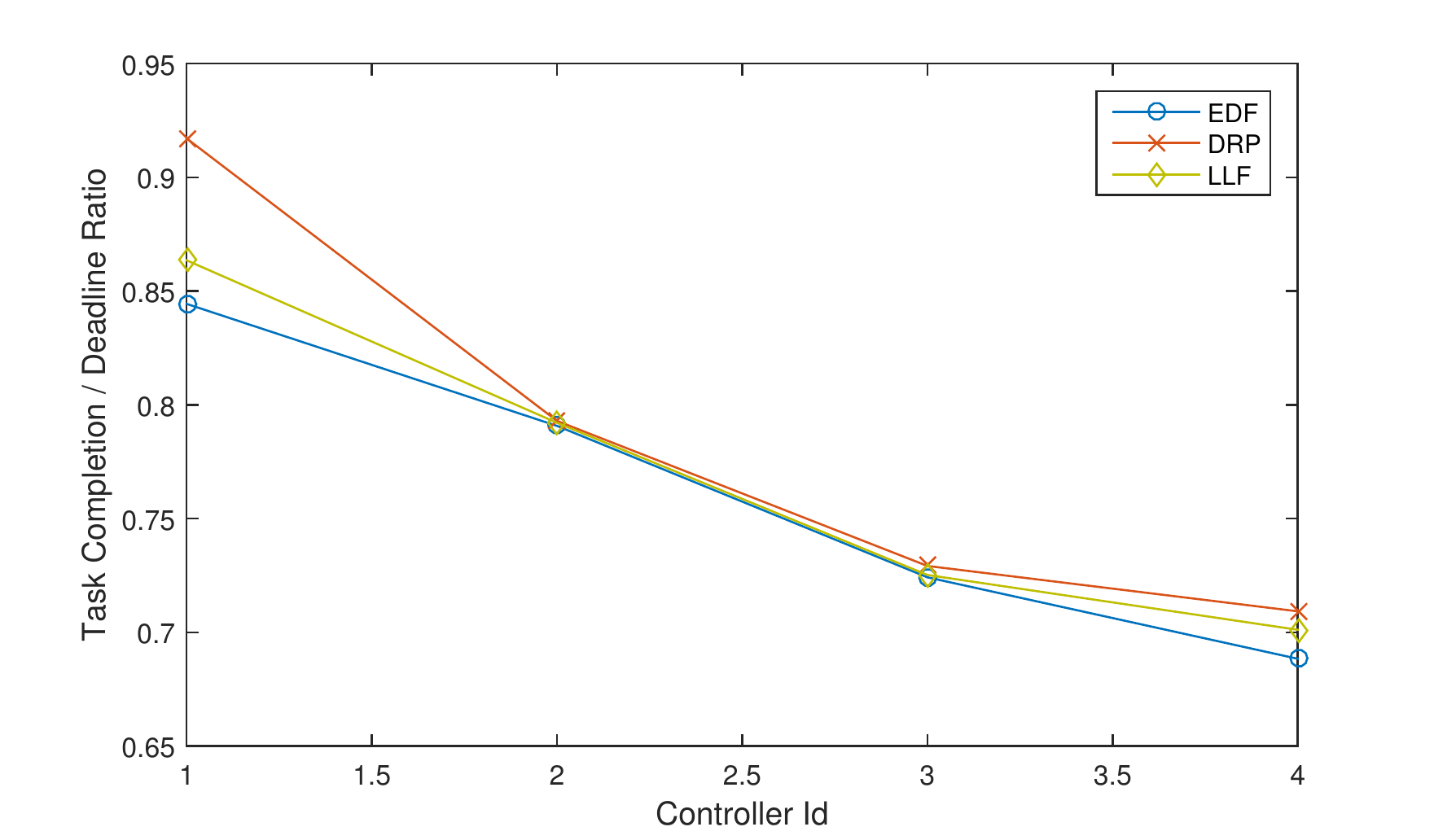}
\caption{Algorithm comparison for Scenario 2}\label{Fig:Scenario2_comp_plot}
\end{figure}

In this scenario $\gamma$ has increased to 74.3584 and maximum eigenvalue reduced to -431.7517. So, this finds a more stable solution when compared to the greedy approach\cite{mishra2015generalized}. The final topology obtained is shown in Figure \ref{Fig:Scenario2_plots_reduced} and $K$ matrix as in \eqref{Eq:K_scenario2}.The power plot for scheduling algorithms are shown in Figure \ref{Fig:Scenario2_power_plot} and algorithms are compared in Figure\ref{Fig:Scenario2_comp_plot} on Task Completion / Deadline Ratio.
\begin{equation}
K=\left(
\begin{tabular}{cccc}
0 & 1.3877 & 0 & 0 \\
0 & -0.7842 & -1.3757 & 0 \\
0.2672 & 0 & 1.6004 & -0\\
1.4442 & 0 & 0 & 0.8173\\
\end{tabular}
\right)\label{Eq:K_scenario2}
\end{equation}
\subsection{Scenario 3} \label{ssec:s3}
\begin{figure}[H]
\centering
\includegraphics[width= 0.9\textwidth, height=7cm, keepaspectratio]{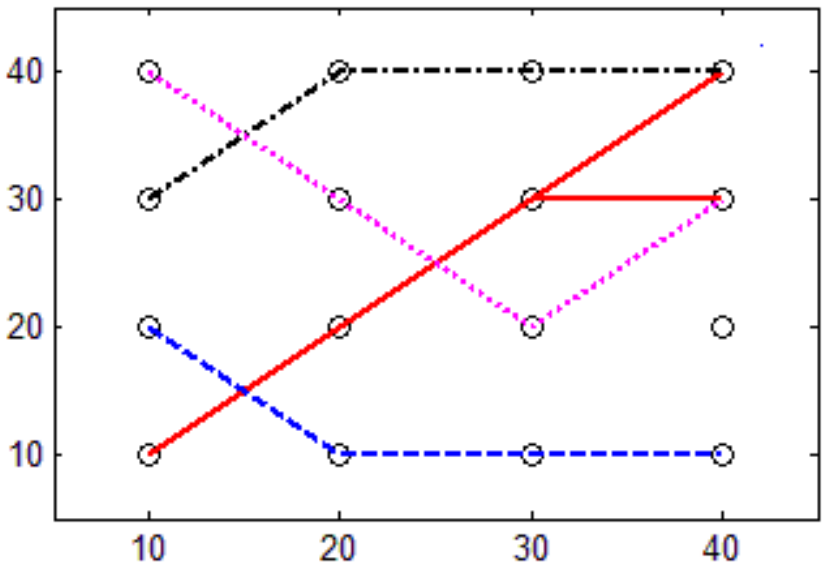}
\caption{Final topology for scenario 3}\label{Fig:Scenario32_plots}
\end{figure}

\begin{figure}[H]
\centering
\includegraphics[width= 0.9\textwidth, height=7cm, keepaspectratio]{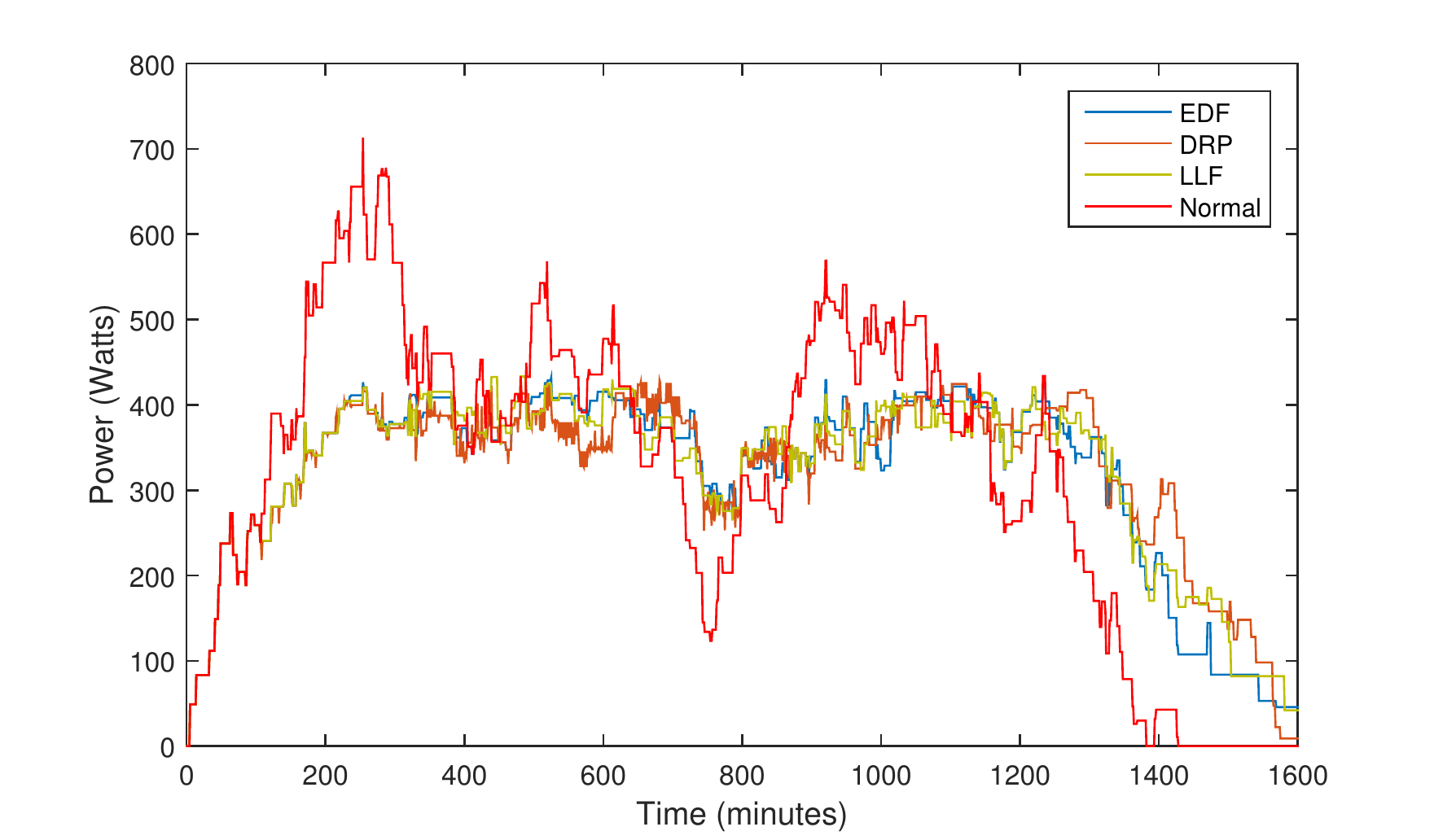}
\caption{Power Plot for Scenario 3}\label{Fig:Scenario3_power_plot}
\end{figure}

\begin{figure}[H]
\centering
\includegraphics[width= 0.9\textwidth, height=7cm, keepaspectratio]{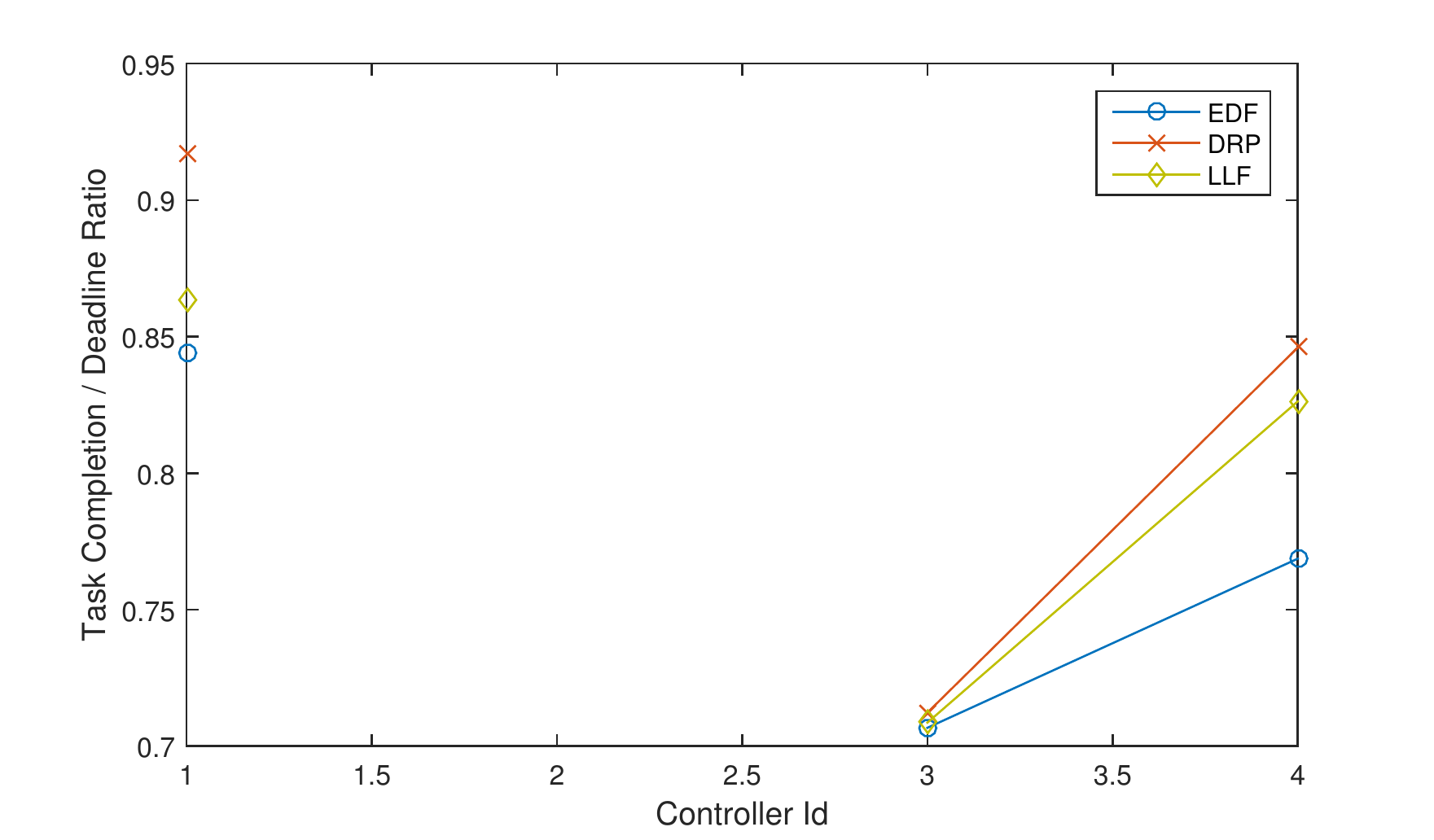}
\caption{Algorithm comparison for Scenario3}\label{Fig:Scenario3_comp_plot}
\end{figure}

Here $\gamma$ has been 19.9634 and reducing eigenvalue to -127.297 thus, making system stable. This is acceptable because final number of connections are lesser here. The final topology is shown in Figure \ref{Fig:Scenario32_plots} and $K$ matrix is mentioned in \eqref{Eq:K_scenario3}. The power plot for scheduling algorithms are shown in Figure\ref{Fig:Scenario3_power_plot} and algorithms are compared in Figure\ref{Fig:Scenario3_comp_plot} on Task Completion / Deadline Ratio.
\begin{equation}
K=\left(
\begin{tabular}{cccc}
0 & 1.0332 & 0 & 0 \\
0 & 0 & 0 & 0 \\
-0.2090 & 0 & 0 & -2.0210\\
0.2409 & 0 & -0.0137 & 0\\
\end{tabular}
\right)\label{Eq:K_scenario3}
\end{equation}





\subsection{Scenario 4} \label{ssec:s4}
\begin{figure}[H]
\centering
\includegraphics[width= 0.9\textwidth, height=7cm, keepaspectratio]{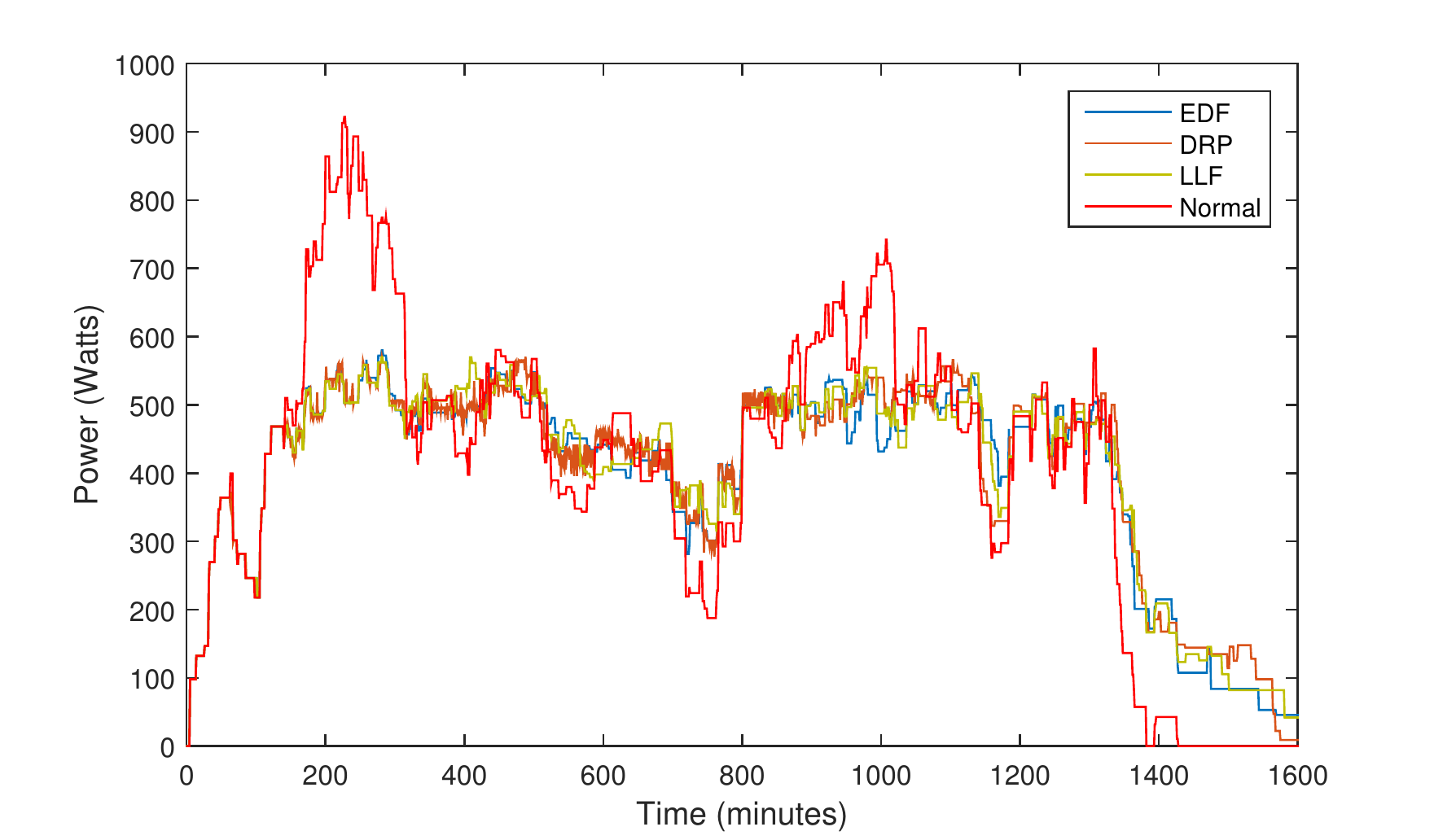}
\caption{Power Plot for Scenario4}\label{Fig:Scenario4_power_plot}
\end{figure}

\begin{figure}[H]
\centering
\includegraphics[width= 0.9\textwidth, height=7cm, keepaspectratio]{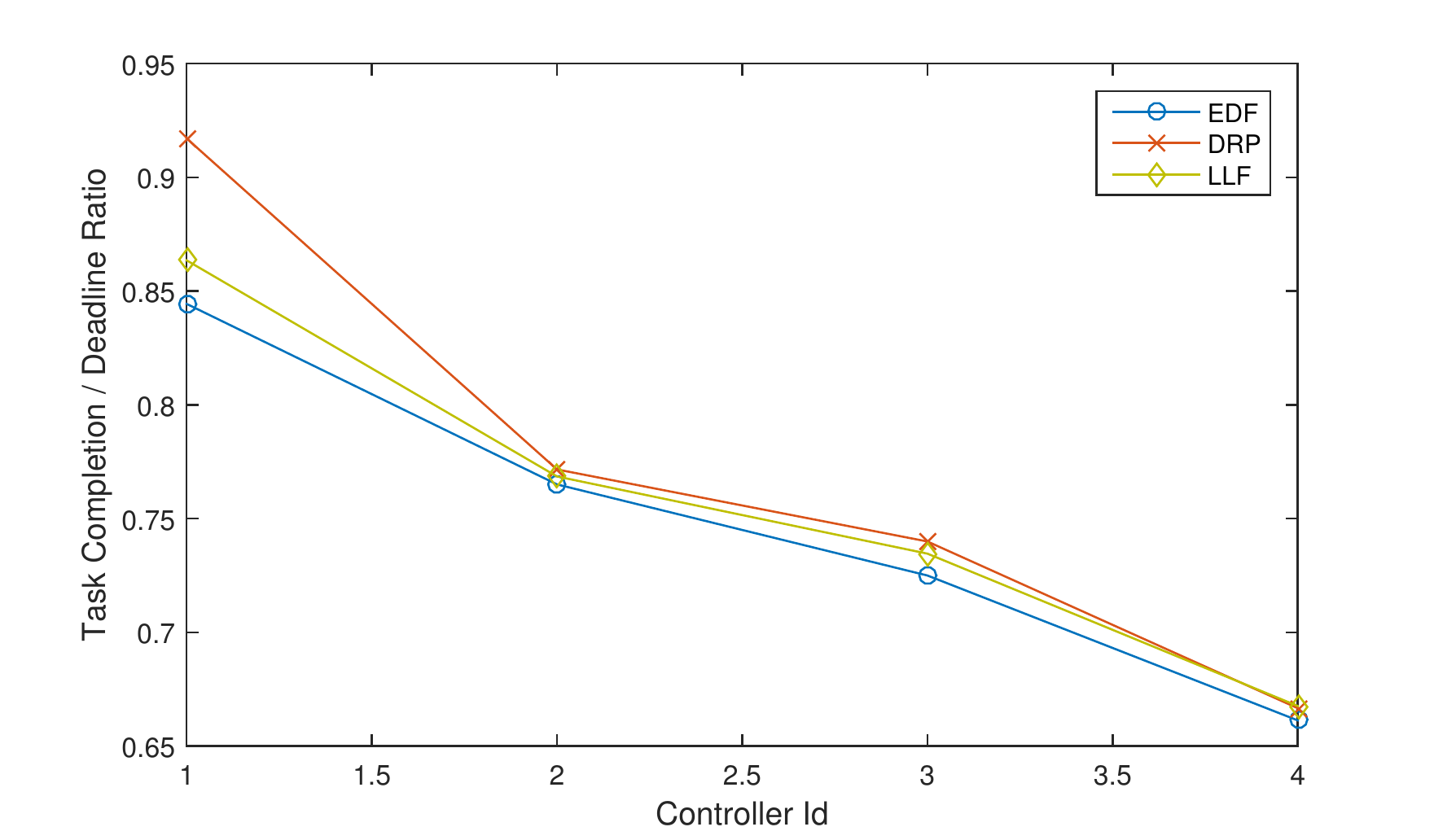}
\caption{Algorithm comparison for Scenario4}\label{Fig:Scenario4_comp_plot}
\end{figure}

In this case $\gamma$ has been  62.3844 and eigenvalue is -469.9 thus, which makes system very much stable. $K$ matrix is mentioned in \eqref{Eq:K_scenario4}. The power plot for scheduling algorithms are shown in Figure\ref{Fig:Scenario4_power_plot} and algorithms are compared in Figure\ref{Fig:Scenario4_comp_plot} on Task Completion / Deadline Ratio.

\begin{equation}
K=\left(
\begin{tabular}{cccc}
0 & 1.6767  & 0 & 0 \\
0 & 0 & -1.8287  & 0 \\
 0.4631 & 0 & 1.2485& 0\\
 1.0132  & 0 & 0 & 1.0854\\
\end{tabular}
\right)\label{Eq:K_scenario4}
\end{equation}



\subsection{Scenario comparison}
Though \nameref{ssec:s1} is best in terms of peak load shaving, its stability is very low because of lesser gamma. \nameref{ssec:s4} is best in terms of both peak load shaving as well stability. \nameref{ssec:s2} is best in terms of stability and \nameref{ssec:s3} is best with minimum number of connections. So proper scenario needs to be selected depending on our point of interest.
\subsubsection{Earliest Deadline First Algorithm}
\begin{figure}[H]
\centering
\includegraphics[width= 0.9\textwidth, height=7cm, keepaspectratio]{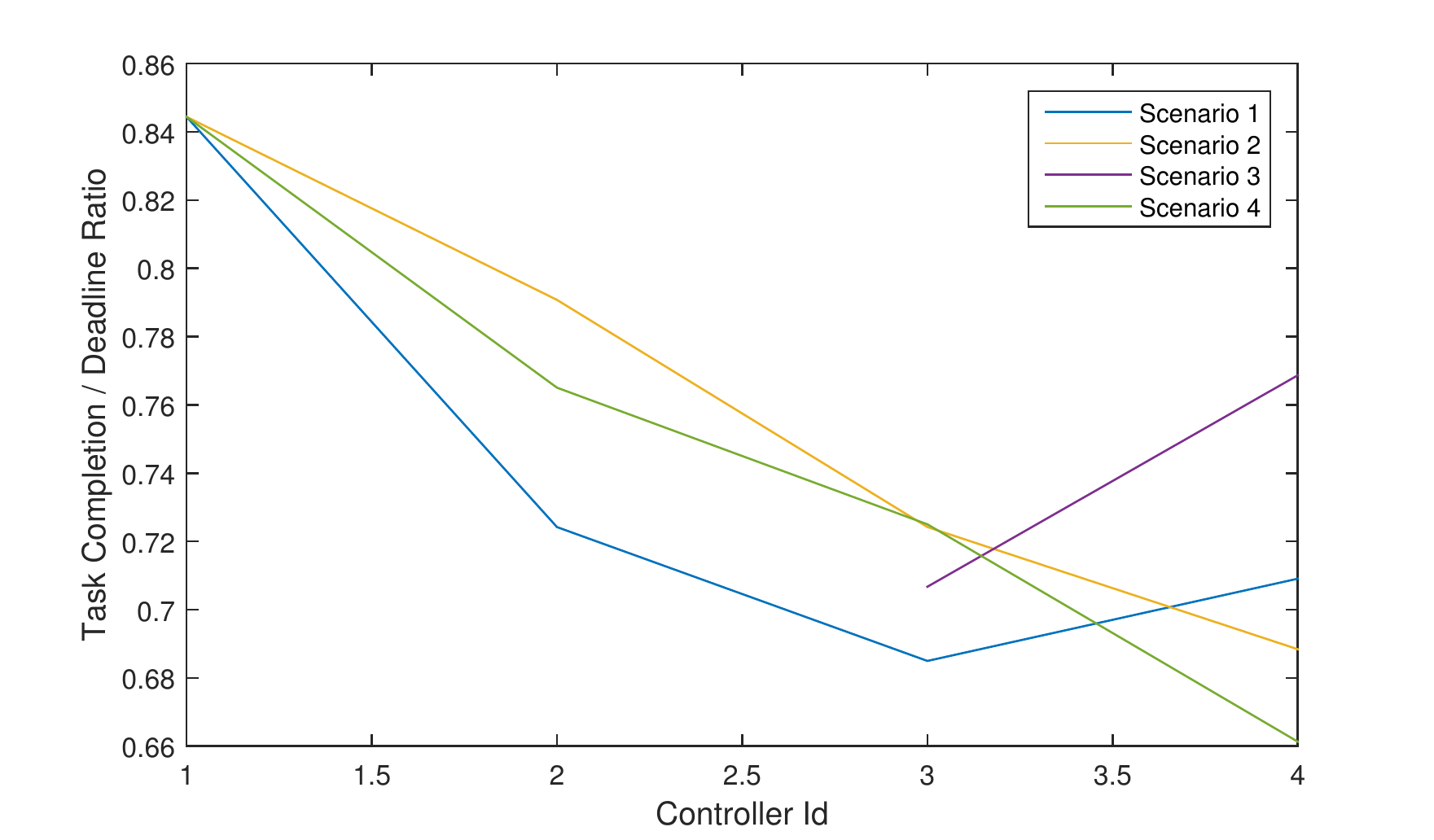}
\caption{Scenario comparison for EDF}\label{Fig:Scenario_edf_plot}
\end{figure}

The comparision between \nameref{ssec:s1}, \nameref{ssec:s2}, \nameref{ssec:s3} and \nameref{ssec:s4} for Earliest Deadline First Algorithm is shown in Figure \ref{Fig:Scenario_edf_plot}. 
\subsubsection{Least Laxity First Algorithm}
The comparision between \nameref{ssec:s1}, \nameref{ssec:s2}, \nameref{ssec:s3} and \nameref{ssec:s4} for Least Laxity First Algorithm is shown in Figure \ref{Fig:Scenario_edf_plot}. 
\begin{figure}[H]
\centering
\includegraphics[width= 0.9\textwidth, height=7cm, keepaspectratio]{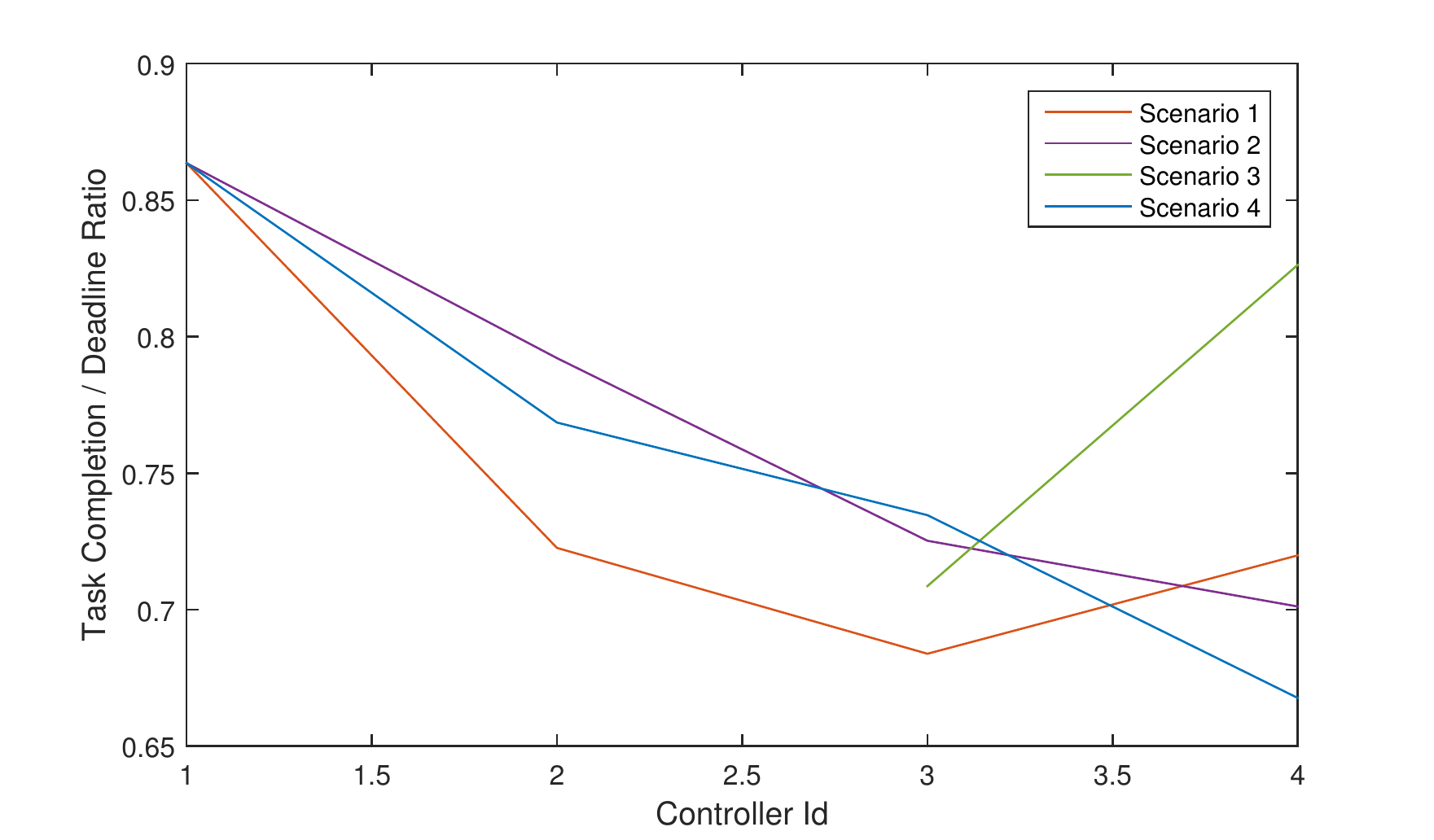}
\caption{Scenario comparison for LLF}\label{Fig:Scenario_llf_plot}
\end{figure}
\subsubsection{Dynamic Rate Priority Algorithm}
The comparision between \nameref{ssec:s1}, \nameref{ssec:s2}, \nameref{ssec:s3} and \nameref{ssec:s4} for Dynamic Rate Priority Algorithm is shown in Figure \ref{Fig:Scenario_edf_plot}. 
\begin{figure}[H]
\includegraphics[width= 0.9\textwidth, height=7cm, keepaspectratio]{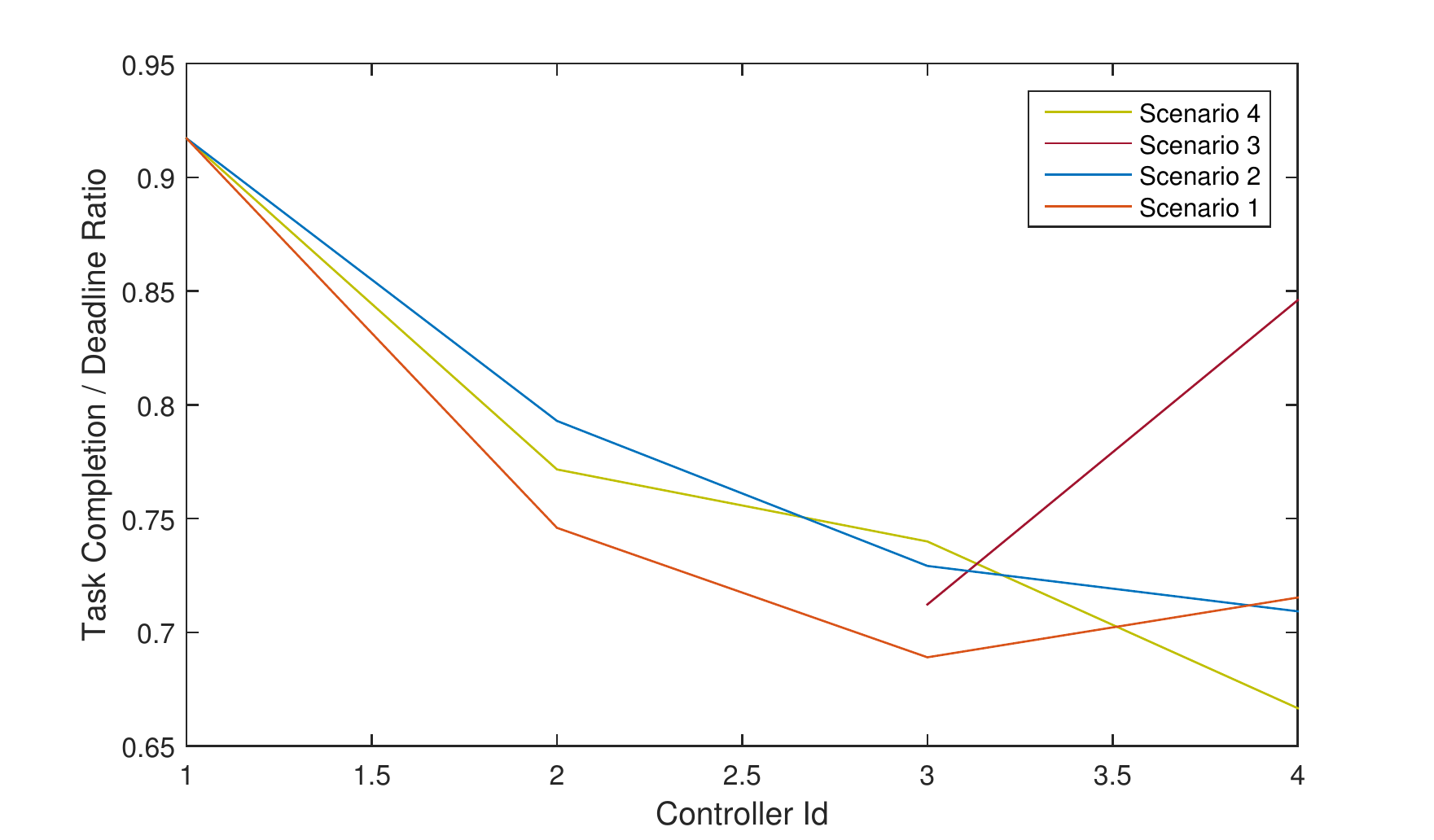}
\caption{Scenario comparison for DRP}\label{Fig:Scenario_drp_plot}
\end{figure}
As already mentioned at the beginning of this section, proper scenario needs to be selected depending on our point of interest.
\section{Scalability to 13 bus system}\label{Sec:scale}
\begin{figure}[H]
\includegraphics[width= 0.9\textwidth, height=7cm, keepaspectratio]{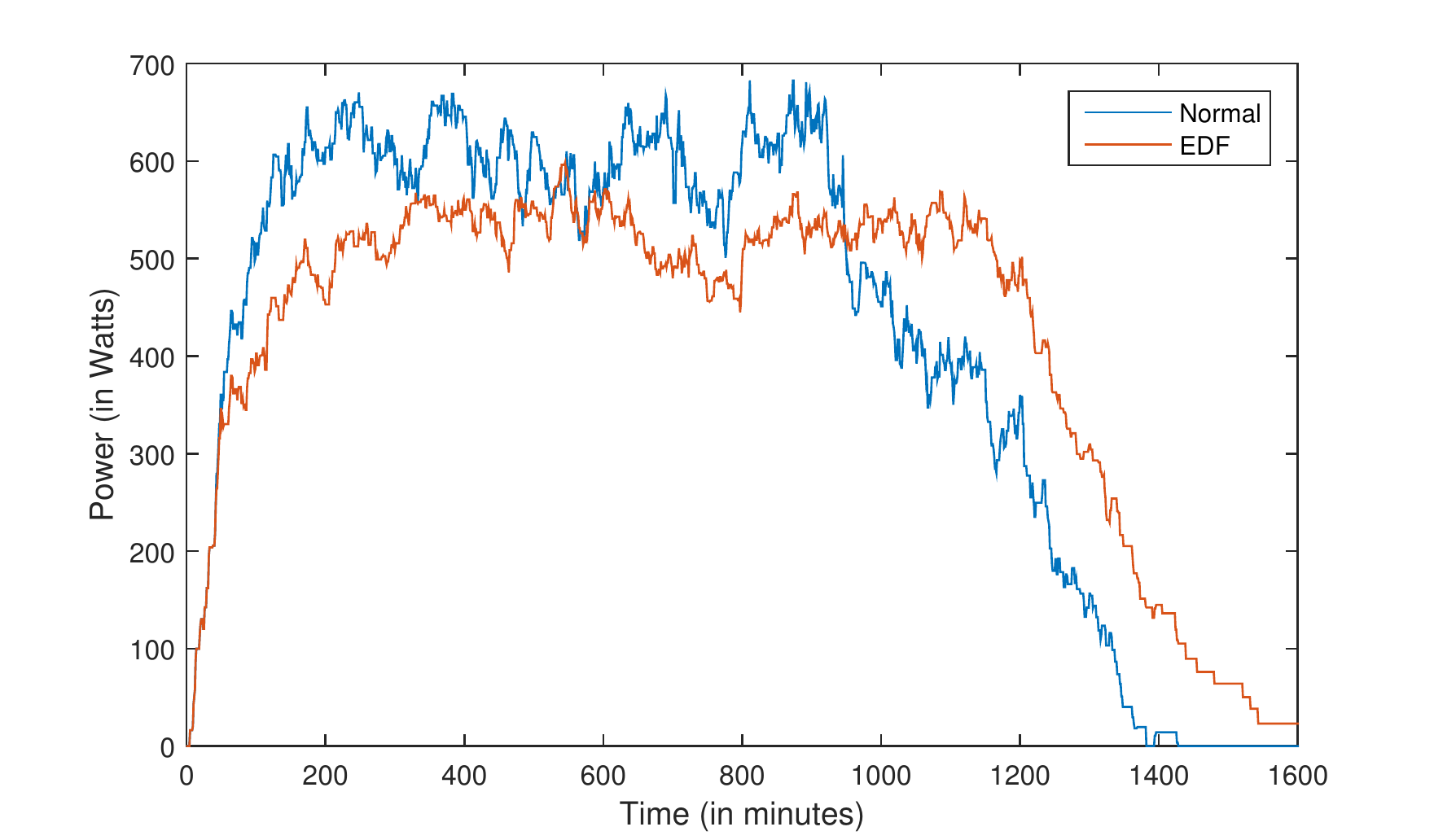}
\caption{EDF for 13 bus system}\label{Fig:Scenario5_edf_plot}
\end{figure}

\begin{figure}[h]
\includegraphics[width= 0.9\textwidth, height=7cm, keepaspectratio]{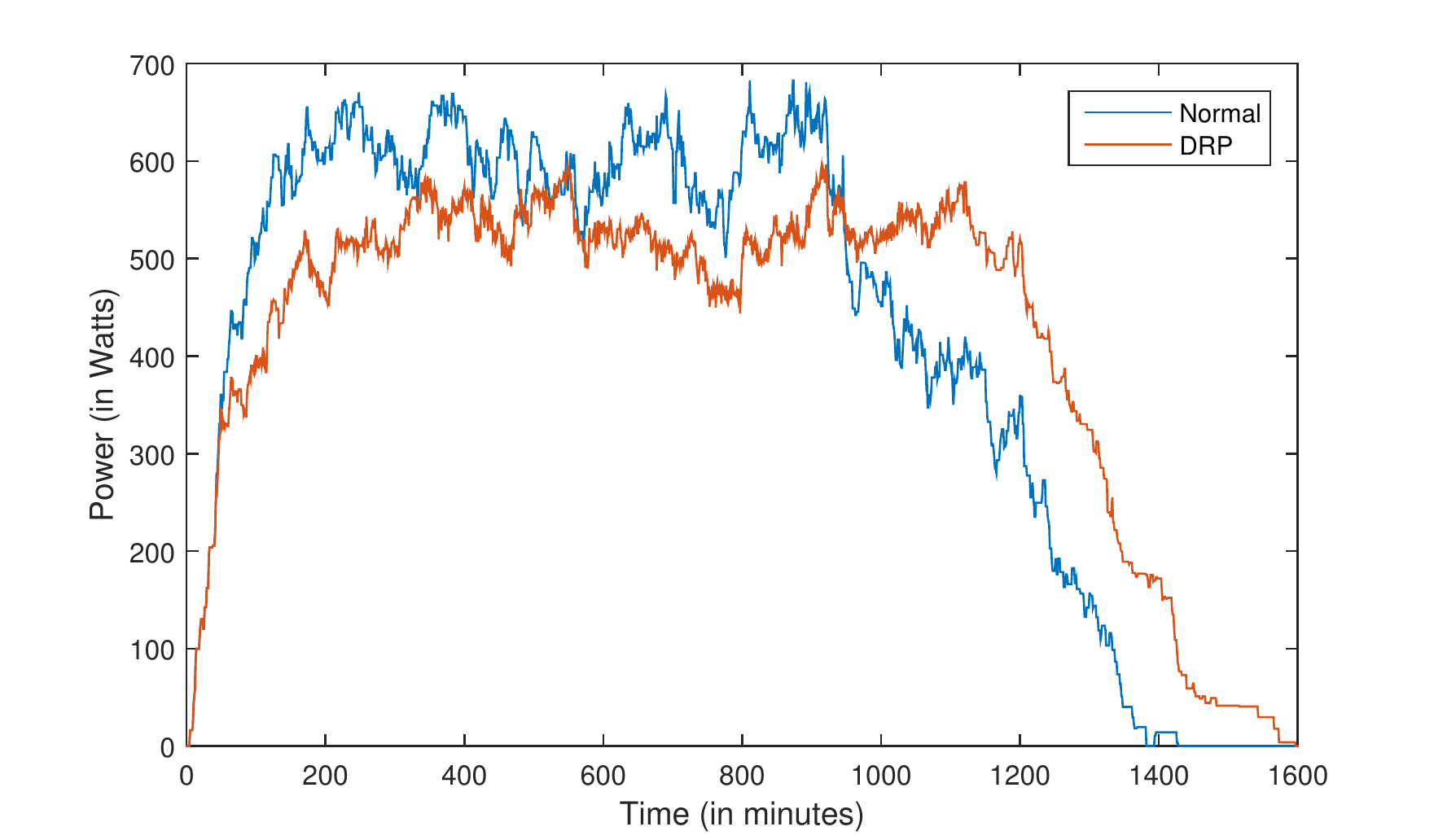}
\caption{DRP for 13 bus system}\label{Fig:Scenario5_drp_plot}
\end{figure}

\begin{figure}[h]
\includegraphics[width= 0.9\textwidth, height=7cm, keepaspectratio]{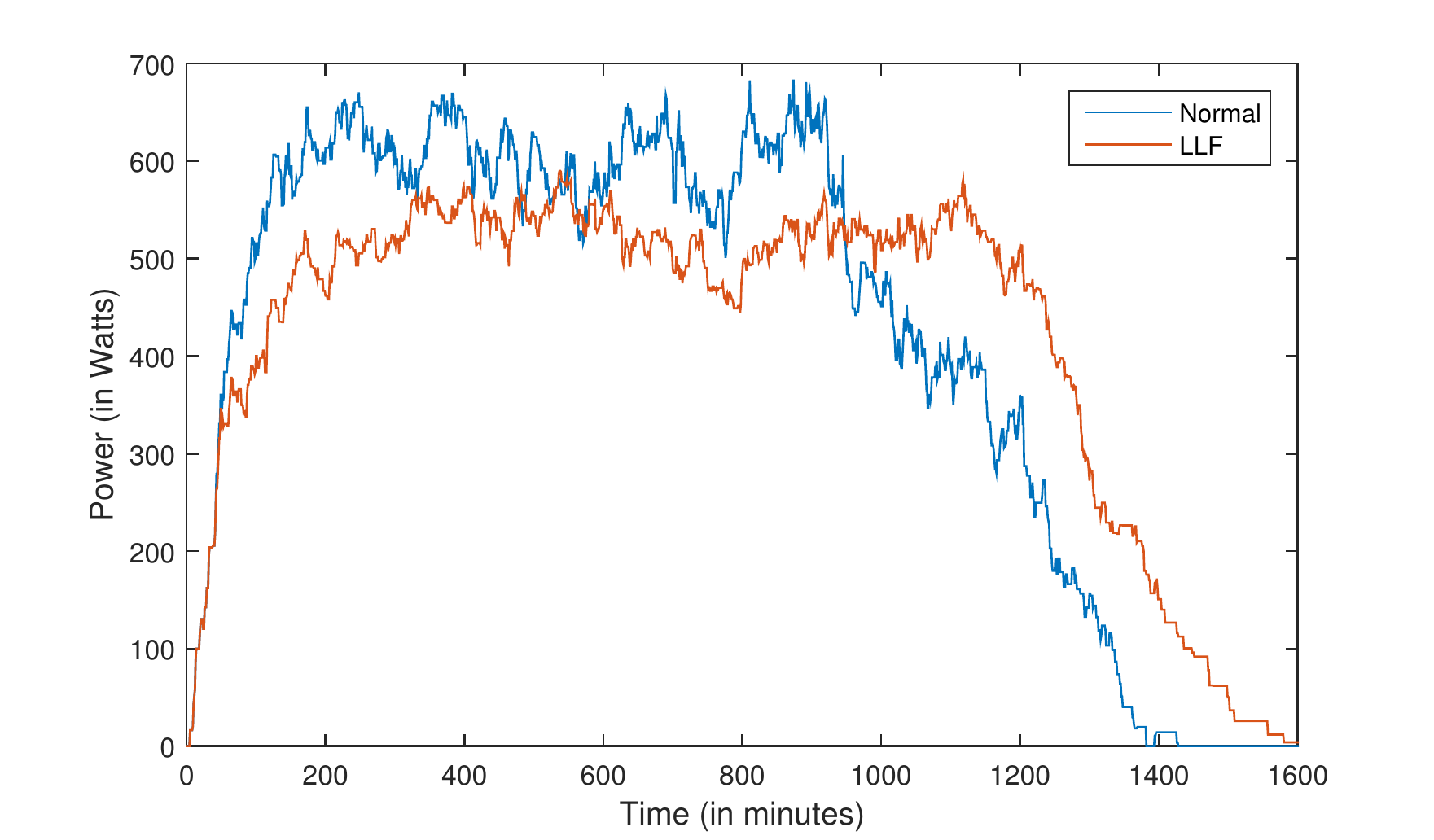}
\caption{LLF for 13 bus system}\label{Fig:Scenario5_llf_plot}
\end{figure}

\begin{figure}[h]
\includegraphics[width= 0.9\textwidth, height=7cm, keepaspectratio]{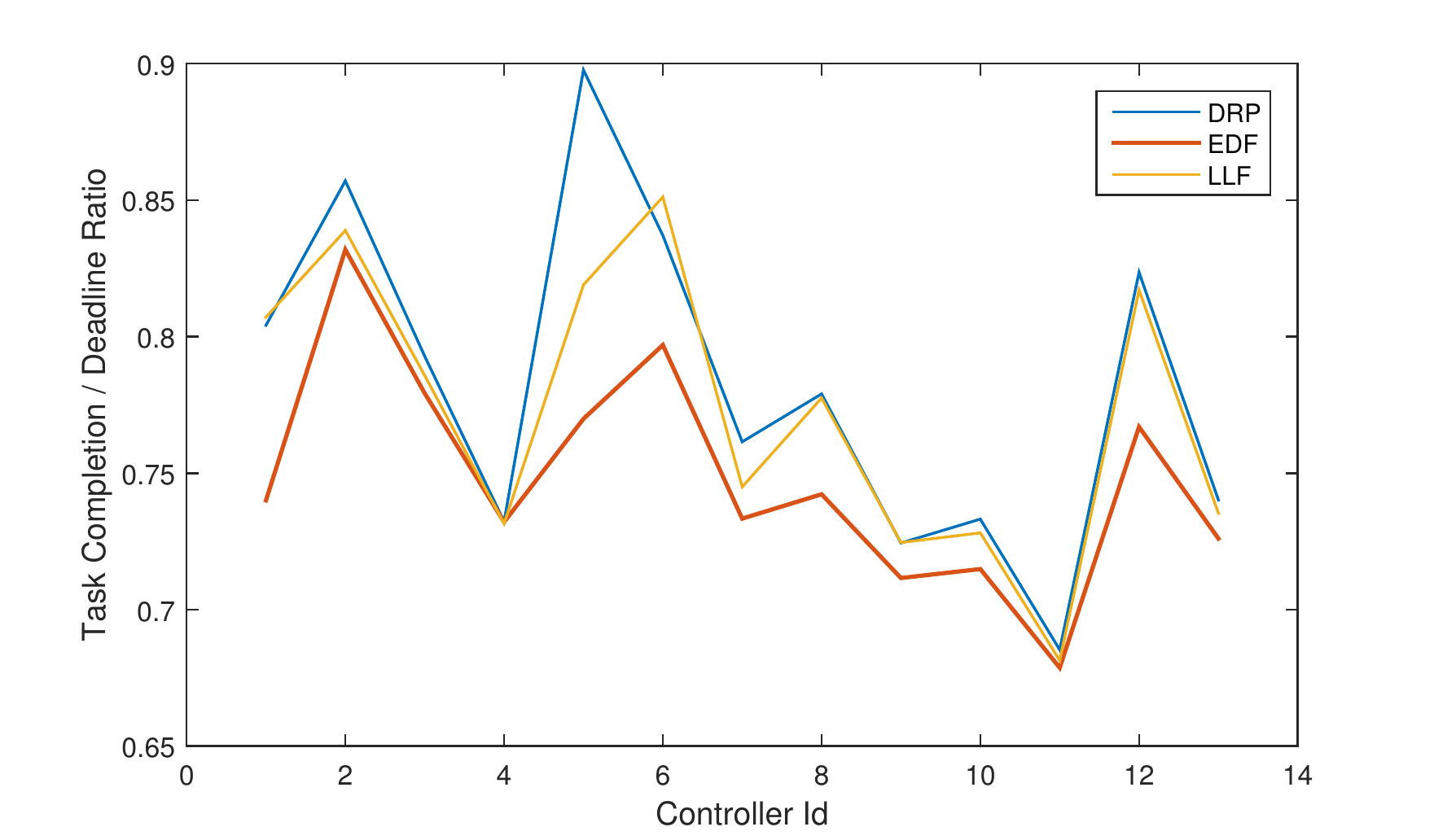}
\caption{Algorithm comparison for 13 bus system}\label{Fig:Scenario_comp_plot}
\end{figure}

To justify scalability of our proposed framework, we simulated IEEE standard 13 bus system with modified parameters $\rho=5e+10$ and the value of $\beta$ as 5000.
Here $\gamma$ has been  4246035.3125 and eigenvalue is -2.2296e+05 thus, system is largely stable here. $K$ matrix is mentioned in Table \eqref{Eq:K_3}. The power plot of peak load shaving for Earliest Deadline First Algorithm is shown in Figure \ref{Fig:Scenario5_edf_plot}, Dynamic Rate Priority Algorithm is shown in Figure  \ref{Fig:Scenario5_drp_plot} and Least Laxity First Algorithm is shown in Figure \ref{Fig:Scenario5_llf_plot}. Also, the above algorithms are compared in Figure \ref{Fig:Scenario_comp_plot} for 13 bus system. Similar to the previous sections, it can be observed here that output of various algorithm varies depending on the point of interest. So it need to be selected based on the user-requirement.	
\begin{table*}[h]
\centering
\caption{13 Bus K-matrix}
\resizebox{\linewidth}{!}{
\begin{tabular}{cccccccccccccc}
\multirow{13}{*}{K = 1e+5 *} & -2.2358 & 0       & 0       & 0       & 0       & 0       & 0       & 0       & 0       & 0       & 0       & 0       & 0       \\
                             & 0       & -2.2209 & 0       & 0       & 0       & 0       & 0       & 0       & 0       & 0       & 0       & 0       & 0       \\
                             & 0       & -0.0316 & -2.1924 & 0       & 0       & 0       & 0       & 0       & 0       & 0       & 0       & 0       & 0       \\
                             & 0       & 0       & -0.0568 & -2.1691 & 0       & 0       & 0       & 0       & 0       & 0       & 0       & 0       & 0       \\
                             & 0       & 0       & 0       & -0.0814 & -2.1459 & 0       & 0       & 0       & 0       & 0       & 0       & 0       & 0       \\
                             & 0       & 0       & 0       & 0       & -0.0983 & -2.1353 & 0       & 0       & 0       & 0       & 0       & 0       & 0       \\
                             & 0       & 0       & 0       & 0       & 0       & -0.1075 & -2.1298 & 0       & 0       & 0       & 0       & 0       & 0       \\
                             & 0       & 0       & 0       & 0       & 0       & 0       & -0.1109 & -2.1303 & 0       & 0       & 0       & 0       & 0       \\
                             & 0       & 0       & 0       & 0       & 0       & 0       & 0       & -0.0918 & -2.1634 & 0       & 0       & 0       & 0       \\
                             & 0       & 0       & 0       & 0       & 0       & 0       & 0       & 0       & -0.0433 & -2.2073 & 0       & 0       & 0       \\
                             & 0       & 0       & 0       & 0       & 0       & 0       & 0       & 0       & 0       & -0.0071 & -2.2194 & 0       & 0       \\
                             & 0       & 0       & 0       & 0       & 0       & 0       & 0       & 0       & 0       & 0       & 0.0523  & -1.6937 & 0       \\
                             & 0       & 0       & 0       & 0       & 0       & 0       & 0       & 0       & 0       & 0       & 0       & 0.7724  & -0.2368
\end{tabular}
}
\end{table*}\label{Eq:K_3}

\section{Results with delay model}
\begin{figure}[!h]
\centering
\includegraphics[width= 0.9\textwidth, height=7cm, keepaspectratio]{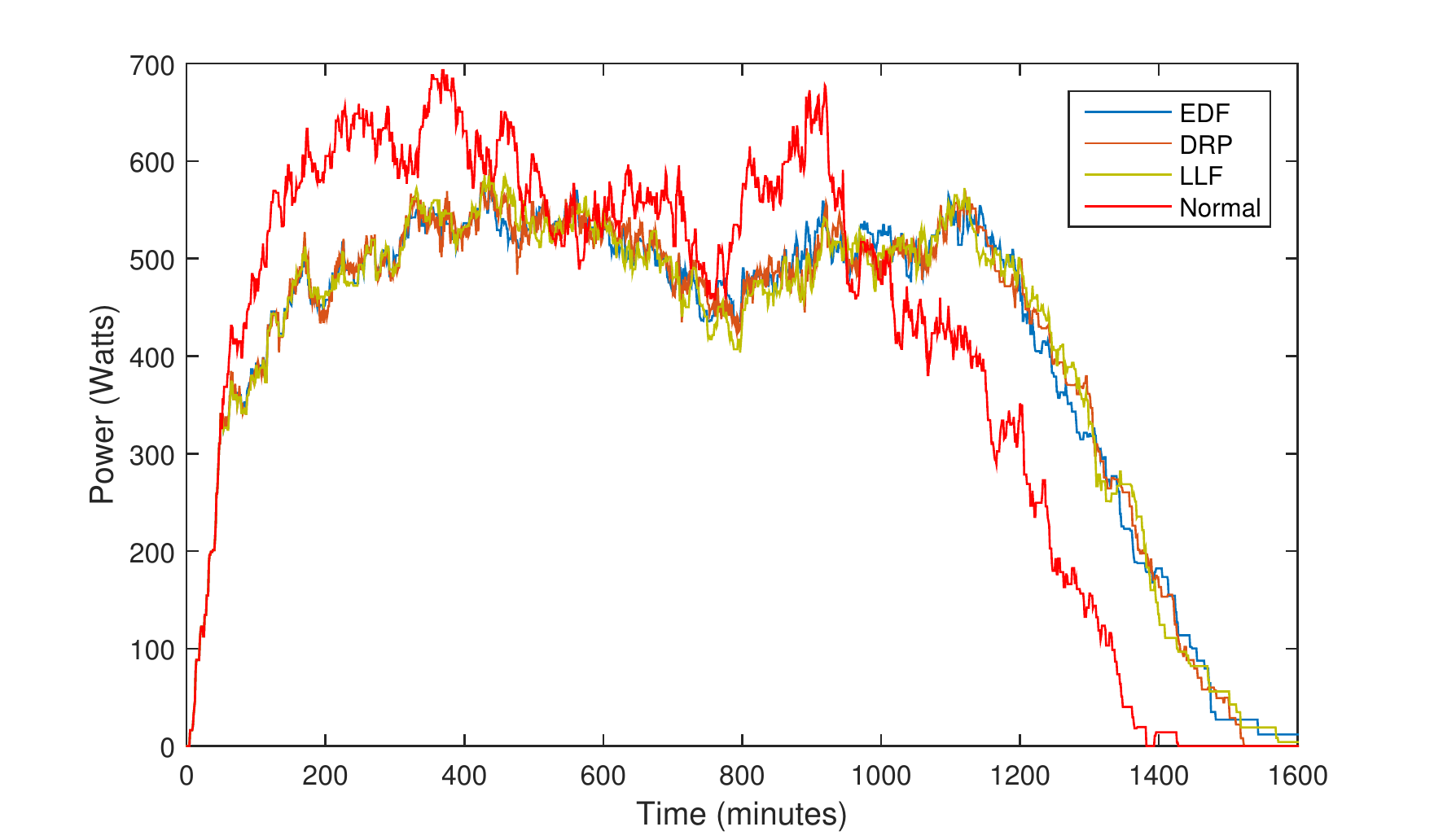}
\caption{Power plot for dealy case}\label{Fig:Scenario6_power_plot}
\end{figure}

\begin{figure}[!h]
\centering
\includegraphics[width= 0.9\textwidth, height=7cm, keepaspectratio]{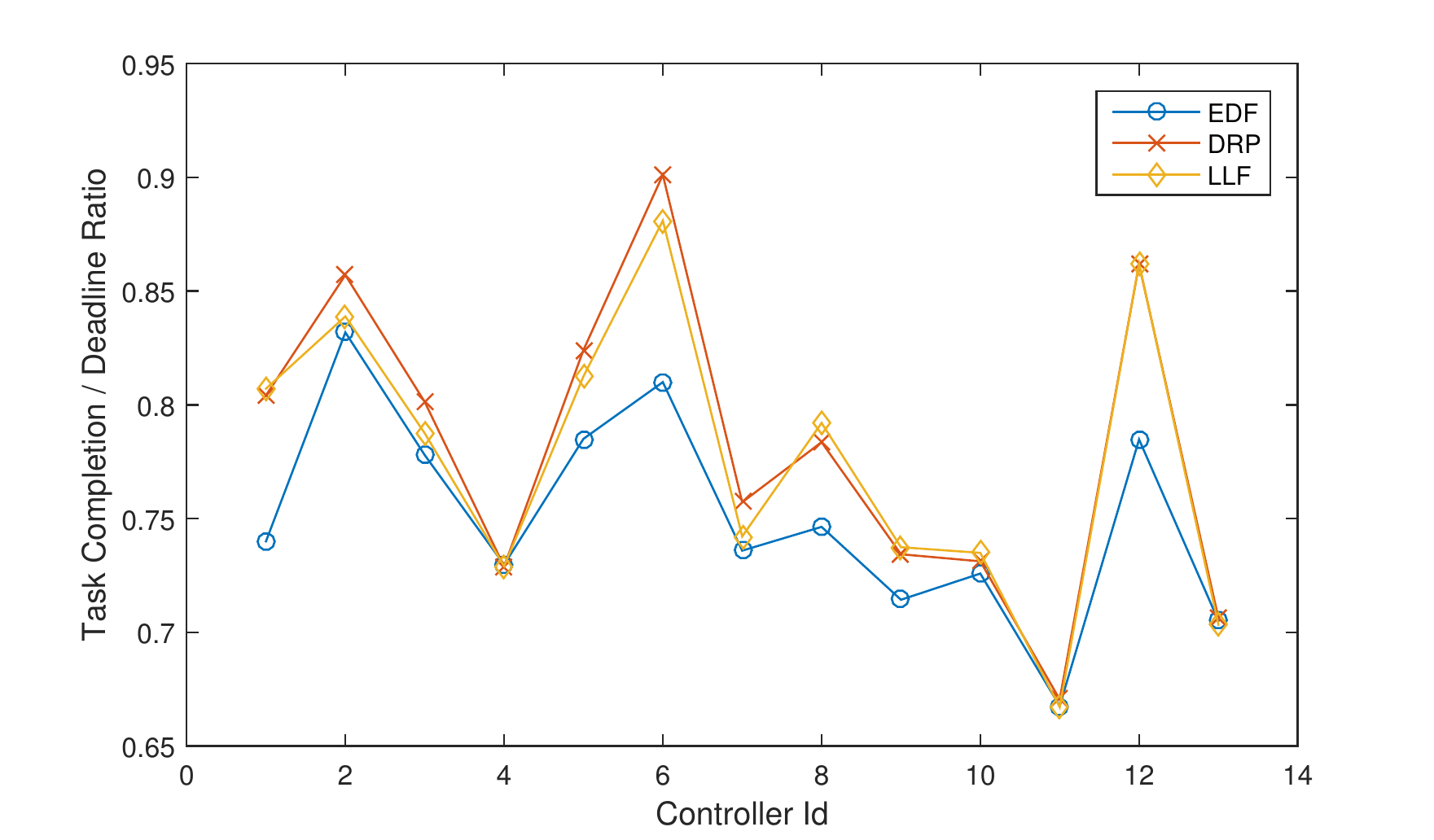}
\caption{Algorithm comparison for delay case}\label{Fig:Scenario6_comp_plot}
\end{figure}

To test adaptiveness of our framework under critical condition like delay, we formulated delay as in \cite{li2012multicast} and simulated in IEEE standard 13 bus system with modified parameters $\rho=5$ and the value of $\beta$ as 5000.
Here $\gamma$ has been  -4.8459e-12and eigenvalue is -31.6300 thus and it making system stable. 
As the delay model has been already explained in a previous section, we have only presented results here. The power plot of peak load shaving for three scheduling algorithms is shown in Figure \ref{Fig:Scenario6_power_plot} and the algorithms are compared in Figure \ref{Fig:Scenario6_comp_plot} for 13 bus delay model. It can be easily seen that, here also algorithm need to be selected based on user-objective like other cases.

\section{Summary}\label{Sec:5co}
We have added real time peak load shaving concept as additional objective along with our primary objective of smart grid stability in this chapter. This is meant to make our architecture user-centric by fulfiling user-requirement of economical and reliable smart grid. This formulation handles various tasks with different deadlines under many types of constraints like delay, scability requirements, communication limitations. Simulation results for four types of scenarios, IEEE 4 bus systems, IEEE 13 bus systems with delay have been performed.  Efficiacy of our proposed framework has been proved with these simulation results 
 
 \chapter{User-Friendly Cognitive Network Design using an Emotion-based Algrotihm\label{Chap4}}
\section{Introduction}
After realizing the importance of intelligent algorithms in every sector (including daily life activities), the main vision behind this paper is to develop a user-friendly intelligent platform which a common public can use (without having any knowledge about mathematical tools), can easily implement his/her ideas to solve real-life problems and improve the way of living. This network has been designed based on our analysis about learning and its relation to emotion in humans. To help users self-analyze and grow personally, reverse concept can be used to track user attributes based on the tuning process selected by the user from the set of tuning parameters. To demonstrate the user-friendly feature of this network to add cognition, moral finding application (as a part concept teaching) have been demonstrated. For any type of story and characters, the network can find the moral(s) from a set of morals. The beauty of this network is the user-friendly design. Unlike existing networks which require some skills to use and implement own ideas, this network simply requires the skill of a nice human (with the ability to understand emotion). Some interesting applications of this network have also been mentioned. This can be extended to develop an emotional language which can help us get rid of this world of divisive language and stop end up machines working with other machines.  
\section{Network Description} \label{Sec:Network Description}
\begin{figure}[h!]
\centering
\includegraphics[width=0.9\textwidth, height=3.5in]{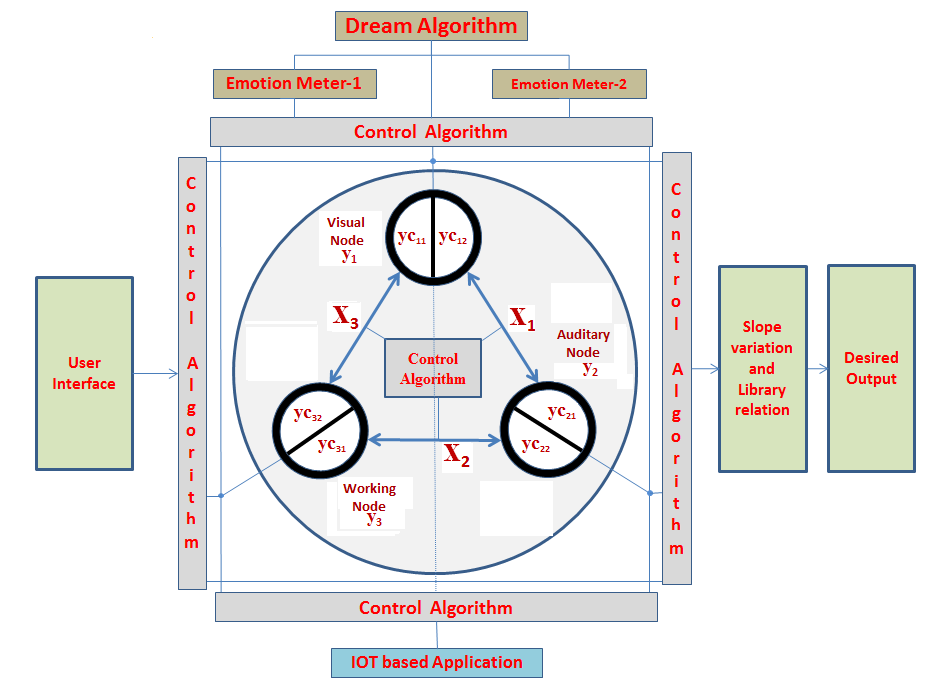}
\caption{The Cognitive Network}
\label{figure1}
\end{figure}
The proposed network as shown in Figure \ref{figure1} has following features.
\subsection{Network Structure} \label{Sec:Network structure}
The sphere houses the network. The growth of the network is bounded by the volume of the sphere.This comprises of 3 nodes which represent visual node, auditary node and working node respectively. More nodes like tactile neuron etc. can be added depending on the application. To make a unified comprehension of all inputs, all these nodes are inter-connected. Links represent "confidence values" between connecting neurons.
 \begin{figure}[h!]
\centering
\includegraphics[width=0.9\textwidth, height=1.8in]{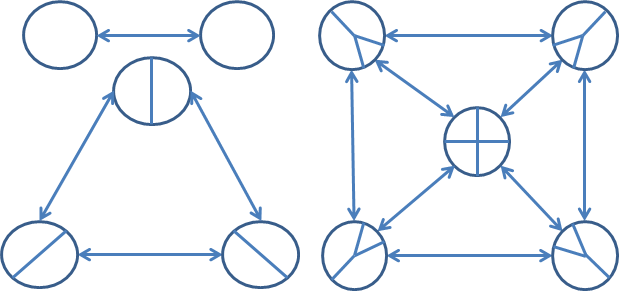}
\caption{Contribution factor based network for 2,3, 5 nodes}
\label{figure2}
\end{figure}
\subsection{Network Dynamics} \label{Sec:Network Dynamics}
Each node is divided into sub-nodes whose values named “contribution values” are directly proportional to normalized link values. Contribution factor based network for 2, 3 and 5 nodes have been shown in Figure \ref{figure2}. Representing "confidence values" of the links connecting the node by ${ x }_{ i }$ where i=1,2..n and node value by y, "contribution values" of sub-nodes ${ yc }_{ i }$ become 
\begin{equation}
\left.
\begin{aligned}
&{ yc }_{ i }=\frac {nxy }{ \sum { { x }_{ i } }}~~\\
&\sum _{ i=1 }^{ n }{ { yc }_{ i } } =y \label{Eq:cv}
\end{aligned}
\right\}
\end{equation}
For multi-tasking application, "contribution values" are set proportional to normalized priority of the task. Denoting priority indexes(on a scale of (0-5)) by ${ z }_{ i }$ where i=1,2..n, corresponding "contribution values" ${ ycm}_{ i }$ becomes 
\begin{equation}
\left.
\begin{aligned}
&{ ycm}_{ i }=\frac {nz.yc }{ \sum { { z }_{ i } } }~~\\
&\sum _{ i=1 }^{ n }{ { ycm }_{ i } } =yc ~~\\
&{ycme}_{ i }=1.1.{ycm}_{ i } \label{Eq:cv}
\end{aligned}
\right\}
\end{equation}
This way multi-tasking processes are treated as superposition of individual tasks.ycme represrents effective multitasking based contribution factor. 10\%\ of the allocated multitasking index has been added to take in to account additional parameters as already dissussed in the literature survey section.\\
Contribution values update as: 
\begin{equation}
\left.
\begin{aligned}
& { yc }_{ new }={ yc }_{ old }.(x/5)~~\\ 
&if\ { y }_{ avg }\ (i.e.(\sum _{ i=1 }^{ n }{ { y }_{ i } } )/n)~~\\ 
&for\ present\ scene ( > 1\ \&\ increasing)~~\\
& or\ for\ present\ scene (< 1 \ \&\ decreasing)~~\\\\
&{ yc }_{ new }={ yc }_{ old }.(5/x)~~\\
&if\ { y }_{ avg }\ (i.e.(\sum _{ i=1 }^{ n }{ { y }_{ i } } )/n)~~\\ 
&for\ present\ scene (> 1 \ \&\ decreasing)~~\\
& or\ for\ present\ scene (< 1 \ \&\ increasing)\label{Eq:cv}
\end{aligned}
\right\}
\end{equation}
 Emotion meters are attached to relevant nodes and dream meters to relevant emotion meters. Various relevant emotional states like sad, frustrated etc. are defined corresponding to the error between updated and  old node values depending on the desired application. Same way dream state is defined based on the change in relevant emotion levels.
Update rule of confidence values of connecting links with respect to different emotional states, dream state are set based on the desired application. 
\begin{equation}
\left.
\begin{aligned}
&\Delta x=5.\left[ 2.{ E }_{ iavg } \right] \% \quad~~\\ 
&where\quad { E }_{ iavg }={ (\sum _{ i=1 }^{ n }{ { (y }_{ inew }- }  }{ y }_{ iold }))/n~~\\ 
&if\ { y }_{ avg }\ for\ present\ scene (> 1 \ \&\ increasing)~~\\
& or\ for\ present\ scene\ (< 1 \ \&\ decreasing)~~\\\\
&\Delta x=(-5).\left[ 2.{ E }_{ iavg } \right] \% \quad~~\\
&where\quad { E }_{ iavg }={ (\sum _{ i=1 }^{ n }{ { (y }_{ inew }- }  }{ y }_{ iold }))/n~~\\ 
&if\ { y }_{ avg }\ for\ present\ scene( < 1  \ \&\ increasing)~~\\
& or\ for\ present\ scene (< 1  \ \&\ increasing)\label{Eq:cv}
\end{aligned}
\right\}
\end{equation}
So,for nodes errors as[0.5,1),[1,1.5),[1.5,2),[2,2.5),emotional states become ``This is not fair”,``sad",``crying",``frustrated” and confidence values of connecting links increase by 5\%,10\%,15\%,20\% respectively. When error crosses 2.5, then ``dream" states get activated and confidence values of connecting links increase by 25\% for 1st case and same way just opposite happens for the 2nd case. 
\section{Network Operation} \label{Sec:Network Operation}
\begin{table}[]
\centering
\caption{User Interface}
\label{UI}
\resizebox{\linewidth}{!}{
\begin{tabular}{|l|c|}
\hline
1 & \begin{tabular}[c]{@{}c@{}}Answer from a scale of 1:5.\\     for i=1: n(last scene)\\ \{\\ 1. What is the ratio of impact of characters now? (Virtual Eye)\\ 2. What is the ratio of power of voice of characters now? (Virtual Ear)\\ 3. What is the ratio of energy of characters now? (Virtual Work-ability)\\ \}\end{tabular} \\ \hline
2 & \begin{tabular}[c]{@{}c@{}}Select your question\\ 1. What is the moral of this story?\\ 2. Write a new story based on the message given by scenes?\\ 3. Write a story based on multi-tasking applications?\\ 4. Form running schedule of IOT based appliances to better describe a story?\end{tabular}                            \\ \hline
3 & \begin{tabular}[c]{@{}c@{}}Answer from a scale of 1:5. depending on your liking in these situations\\ 1. Looking at pictures of your favorite places. (Attachment to Eye)\\ 2. Listening your favorite music. (Attachment to Ear)\\ 3. Doing the work which you love most.(Attachment to work)\end{tabular}                       \\ \hline
4 & Form library of elements (keywords, instructions etc.) in terms of which, answers will be expressed.                                                                                                                                                                                                                                                            \\ \hline
5 & \begin{tabular}[c]{@{}c@{}}Select your area of liking from the table\\ 1. Science, Nature, Animal planet, Action, News, Movies, Dance, Music, Yoga.\end{tabular}                                                                                                                                                                  \\ \hline
6 & \begin{tabular}[c]{@{}c@{}}Add photo,video, story, taglines and as many details as possible in the section of \\ the following table representing the reason for which your remember the moment.\\ 1. Looks good\\ 2. Voice is good.\\ 3. Work done as desired\end{tabular}                                                       \\ \hline
7 & \begin{tabular}[c]{@{}c@{}}Add any other thing you want to relate with this Network in the above sections.\\  (E.g. Data of IOT based appliances)\end{tabular}                                                                                                                                                                    \\ \hline
\end{tabular}
}
\end{table}
\begin{figure}[h!]
\centering
\includegraphics[width=0.9\textwidth, height=1.8in]{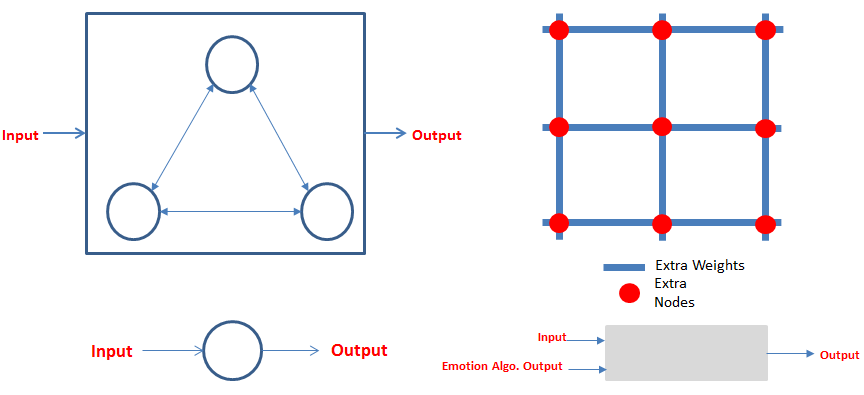}
\caption{Network, Control Algorithm, Neuron and Weight model}
\label{figure3}
\end{figure}
\subsection{Network Interaction with External Inputs}
 Links represent internal structure of humans, so this is a one-time input for a particular person. Node values are completely dependent on external inputs and these are set based on the answer to question no. 1 of the user interface mentioned in Table  \ref{UI}.
Confidence values are set proportional to normalized natural affinity for the emotion variables of connecting nodes. This is based on the answers to question no. 3 of the user interface. If the answers to question no. 3 for two emotion variables(which are represented by the nodes connected by this link) are x and y, then confidence values become $\frac { x+y }{ 2 } $.\\
Sub structures of the main network is available in Figure \ref{figure3}.
\begin{table}[]
\centering
\caption{Forming a library and relating it to slope variations}
\label{Algo}
\resizebox{\linewidth}{!}{
\begin{tabular}{|c|c|c|c|c|}
\hline
\multicolumn{5}{|c|}{Sample library for concept-teaching case}                                                                                                                                                                                                                                                                                                                                                                                                                                                                                                                                                                                                                                                                                                                                                              \\ \hline
Power                                                                                                                                                        & Brain                                                                                                                                                          & Work                                                                                                                                                            & Thing                                                                                                                                                       & Other                                                                                                                                                       \\ \hline
Is                                                                                                                                                           & Has                                                                                                                                                            & Not                                                                                                                                                             & No                                                                                                                                                          &                                                                                                                                                             \\ \hline
Very                                                                                                                                                         & Much                                                                                                                                                           & Than                                                                                                                                                            & With                                                                                                                                                        &                                                                                                                                                             \\ \hline
Good                                                                                                                                                         & Natural                                                                                                                                                        & Actually                                                                                                                                                        & Do                                                                                                                                                          & Be                                                                                                                                                          \\ \hline
\multicolumn{5}{|c|}{\begin{tabular}[c]{@{}c@{}}Algorithm to relate library with slope-variation\\ (For finding conclusion of a story)\\ If slope is increasing first and then decreasing\\ \{\\ If ($ | \text{slope at any point}|$\textgreater3.5), add E44, E23, E45, E34, E23, E43\\ ("Do not be with not actually")\\ Else add E12, E11, E31, E31, E41, E33, E13, E11\\ ("Brain power very very good than work power")\\ \}\\ Else if slope is decreasing first and then increasing\\ \{\\ If ($ | \text{slope at any point}|$\textgreater3.5), add E23, E42, E13, E21, E23, E41\\ ("Not natural work is not good")\\ Else add E31, E41, E42, E13, E34, E24, E15, E14, E22, E31, E32, E11\\ ("Very good natural work with no other thing has very much power")\\ \}\\ NOTE: EAB means element of Ath row \& Bth coloumn of library\end{tabular}} \\ \hline
\end{tabular}
}
\end{table}
\subsection{Decision Taking Process}
Node errors multiplied by weights(selected based on personal guess for the desired application)are connected to points(e.g.relevant nodes, combination of nodes after proper transformation)whose slope variation with respect to different scenes are marked and this is further related to the user-defined library to produce our desired outputs. This library is designed based on our desired output of interest.E.g. Library of set of words(user favorite words) are formed for concept teaching application. Relation of slope variation to library is established after analyzing slope variations of different extreme cases of the application. For concept teaching case, this step is shown in Table \ref{Algo}  as an example.
\subsection{Algorithm similarity with human experience}
Algorithm is based on a concept analogous to equilibrium state. In mechanics it happens when net force becomes 0. In a general context, When there is only one type of elements present, the equilibrium state will be achieved when there is no direction involved for further change in either way. Else system will be unstable.Basically each update equation is defined for a single sensory organ based syste. If present state is higher (i.e >1 as ratio is involved) and next state makes it further higher, then a direction is coming, so system is becoming unstable. The same can be verified through expression of an auidance (e.g. eye opening width)while watching a scene where hero is being beaten. So part of the input from our senses will be passed further and that's why its value decrease. Same way,just opposite happens when next state reduces the effect of present state. Again the expression of smile or clap can be verified from an audiance Confidence factor changes based on the same concept. It increases if stability increases else decreases. These are consistent as per the common definition of these terms (like confidence) and can be verified by personal experiences as well.\\\\
Algorithm takes all the processed values (i.e. updated values) as input and uses the same concept to reach at a particular conclusion. As we discussed in the previous section, if the POV (value at point of observation) is increasing, then it is going towards instability. Similarly if POV decreases, it moves towards stability. So when both increasing and decreasing process happens smoothly, intelligence has to be involved to bring back stability from the unstable condition irrespective of other things of the story. If in any part of this process of stability establishment, high slope (slope beyond the 50 \% threshold) is observed, then a doubt arises naturally (as the process seems forcible and drastic due to the high slope value). So some-part of pretending behavior has to be present and that's why the conclusion "Don't pretend" (after decoding it from the language of the algorithm). Similarly if POV decreases first, then it is also becoming unstable. But this time the instability has happened in the just opposite direction. So, if it increases later to neutralize the instability, it becomes favorable (thus stable) using the same concept. Increasing valuer (thus positive slope) at the beginning was unstable in a negative way, but decreasing at the beginning is comparatively less unstable  as instability though has happened, still happened in a positive way. And with the positive slope later, finally it recoveres to a stable stage. This is analogus to ups and downs of life. To cross hurdles (down part) and come again to the balance state happens only when the person follows his/ her heart. So the conclusion is "follow your heart or stay natural." Here also if the slope limit is crossed (and the slope variation curve becomes rough and not smooth), it implies something unnatural the person has peformed for sure. As the overall process (decreasing first and then increasing) undergoes through postive unstale system, so positive version of pretend is used. The conclusion thus becomes "Anything unnatural can't last long." Similarly other morals can be added.
\subsection{Additional Information}
Minimization of number of terms in the library as well as number of possible cases need to be ensured to reduce space and time usage of the algorithm. Extension of these few cases to huge number of real life situations can be done by changing weights, points of observation. E.g. for the Pancha-tantra stories, point of observation is kept fixed at all nodes(slope of node error average is related with the library) and variation is done in arranging scenes depending on our point of interest in the story. For bigger story having multiple maxima, minima in the slope variation curve, point of interest need to selected intelligently. (E.g. All maximas as point of observation).Minimization of terms in the library produces outputs with simple words which is similar to "speech of a villager in English". Additional conversion table can be formed by relating words of this library to sophisticated words to get outputs in our desired form.\\\\
Playing with this network with different connections through personal experience along with addition of data from technologies like automation, cyber-physical system etc. in the IOT based application section of this NN can further produce amazing applications. Application of various scientific techniques(e.g. control algorithms, optimization techniques)can be done further using the control algorithm section of this network to get interesting results and additionally this will be helpful for students to learn these scientific concepts in a better way. One simple example is use of links as additional weights, junctions as some parameters of NN (e.g. node, weight etc.) and control algorithm to establish relation between them.
 \begin{table}[]
\centering
\caption{User Input for Lion-Rabbit Story}
\label{Lr}
\resizebox{\linewidth}{!}{
\begin{tabular}{|c|c|l|c|c|}
\hline
1                      & Scenes                                                                                                                                                                                                                                                                                                   & \multicolumn{1}{c|}{Q1}  & Q2                       & Q3                       \\ \hline
\multicolumn{1}{|l|}{} & \begin{tabular}[c]{@{}c@{}}S1 (Introduction of characters: Lion \& Rabbit.Once upon a time there lived a ferocious lion in \\the forest. It was a greedy lion and started killing animals in the forest \\indiscriminately. Seeing this, the animals gathered and decided to approach the lion\\ with the offer of one animal of each species volunteering to be eaten by the lion \\every day. So every day it was the turn of one of the animals and in the end came the rabbits’\\ turn. The rabbits chose an old rabbit amongst them. The rabbit was wise and old. It \\took its own sweet time to go to the lion. The lion was getting impatient on not seeing any\\ animal come by and swore to kill all the animals the next day.)\end{tabular}                                                                                                                                                                                                               & \multicolumn{1}{c|}{3.5} & 4                        & 4                      \\ \hline
\multicolumn{1}{|l|}{} & \begin{tabular}[c]{@{}c@{}}S2 (Lion becomes angry\\ after seeing rabbit which is very small for his meal.)\end{tabular}                                                                                                                                                                                  & \multicolumn{1}{c|}{3.8} & 4.3                      & 4.2                      \\ \hline
\multicolumn{1}{|l|}{} & \begin{tabular}[c]{@{}c@{}}S3 (Rabbit says actually\\ 6 rabbits were sent for his meal, but on the way 1 lion ate 5 of them, I only\\ escaped from him to reach here, that lion was saying “I am the real king”.)\end{tabular}                                                                           & \multicolumn{1}{c|}{3.6} & 4.2                      & 4.1                        \\ \hline
\multicolumn{1}{|l|}{} & \begin{tabular}[c]{@{}c@{}}S4 (Lion becomes violent\\ and its roar shook the forest. He desperately wants to kill the other lion and\\ it orders the rabbit to take him to the other lion.)\end{tabular}                                                                                                 & 4.5                     & \multicolumn{1}{l|}{4.6} & \multicolumn{1}{l|}{ 4.7} \\ \hline
\multicolumn{1}{|l|}{} & \begin{tabular}[c]{@{}c@{}}S5 (The rabbit took him to the well and the lion sees into it and finds another lion inside(which\\ is simply his reflection), it roars and finds response from the other lion in the same intensity as \\ his(which is again his echo only), the lion goes mad)\end{tabular} & \multicolumn{1}{c|}{4.8} & 4.9 & 4.8 \\ \hline
\multicolumn{1}{|l|}{} & \begin{tabular}[c]{@{}c@{}}S6 (To attack the other lion he found, this lion jumps into the well and died, rabbit becomes\\ happy.)\end{tabular}                                                                                                                                                          & \multicolumn{1}{c|}{0}   & 0                        & 0                        \\ \hline
2                      & 1. What is the moral of the story?                                                                                                                                                                                                                                                                       & \multicolumn{3}{l|}{}                                                          \\ \hline
3                      & \multicolumn{1}{l|}{}                                                                                                                                                                                                                                                                                    & 3.5                        & \multicolumn{1}{l|}{4} & \multicolumn{1}{l|}{4.5} \\ \hline
4                      & Brain Power, Work Power, Good                                                                                                                                                                                                                                                                            & \multicolumn{3}{l|}{}                                                          \\ \hline
5                      & Not required                                                                                                                                                                                                                                                                                             & \multicolumn{3}{l|}{}                                                          \\ \hline
6                      & Not required                                                                                                                                                                                                                                                                                             & \multicolumn{3}{l|}{}                                                          \\ \hline
7                      & No & \multicolumn{3}{l|}{}                                                          \\ \hline
\end{tabular}
}
\end{table}
\begin{table}[]
\centering
\caption{Set and updated values for Lion-Rabbit Story}
\begin{minipage}{.38\linewidth}
      \centering
\label{Sv}
\resizebox{\linewidth}{!}{
\begin{tabular}{|c|c|c|c|c|c|c|}
\hline
\multicolumn{7}{|c|}{Initially ${ x }_{ 1 }$=3.75, ${ x }_{ 2}$= 4.25, ${ x }_{ 3 }$= 4} \\ \hline
Scene no.  & ${ yc }_{ 11 }$  & ${ yc }_{ 12 }$  & ${ yc }_{ 21 }$  & ${ yc }_{ 22 }$  & ${ yc }_{ 31}$ & ${ yc }_{ 32}$ \\ \hline
S1         & 3.61  & 3.39  & 3.75  & 4.25  & 4.12 & 3.88 \\ \hline
S2         & 4.03  & 3.57  & 3.91  & 4.70  & 4.32 & 4.1  \\ \hline
S3         & 3.82  & 3.39  & 3.82  & 4.58  & 4.22 & 3.98 \\ \hline
S4         & 4.77  & 4.23  & 4.18  & 5.01  & 4.84 & 4.56 \\ \hline
S5         & 5.09  & 4.51  & 4.45  & 5.34  & 4.95 & 4.65 \\ \hline
S6         & 0     & 0     & 0     & 0     & 0    & 0    \\ \hline
\end{tabular}
}
\end{minipage}%
    \begin{minipage}{.62\linewidth}
      \centering
        \label{Uv}
\resizebox{\linewidth}{!}{
\begin{tabular}{|c|c|c|c|c|c|c|c|c|c|}
\hline
Scene no. & ${ yc }_{ 11n }$ & ${ yc }_{ 12n }$ & ${ yc }_{ 21n }$ & ${ yc }_{ 22n }$ & ${ yc }_{ 31n}$ & ${ yc }_{ 32n}$ & ${ x }_{ 1n }$  & ${ x }_{ 2n}$  & ${ x }_{ 3n }$  \\ \hline
S1        & 2.89  & 2.54  & 2.81  & 3.61  & 3.50  & 3.10  & 2.71 & 3.25 & 3.1  \\ \hline
S2        & 2.47  & 1.93  & 2.11  & 3.05  & 2.81  & 2.50  & 1.52 & 1.83 & 1.72 \\ \hline
S3        & 1.31  & 1.03  & 1.16  & 1.67  & 1.54  & 1.37  & 0.86 & 1.03 & 0.97 \\ \hline
S4        & 0.92  & 0.72  & 0.72  & 1.03  & 1.0   & 0.89  & 0.48 & 0.58 & 0.54 \\ \hline
S5        & 0.55  & 0.43  & 0.43  & 0.62  & 0.57  & 0.51  & 0.27 & 0.32 & 0.31 \\ \hline
S6        & 0     & 0     & 0     & 0     & 0     & 0     & 0    & 0    & 0    \\ \hline
\end{tabular}
}
\end{minipage}
\end{table}
\section{Results} \label{Sec:Results}
\subsection{Concept teaching}
Lack of memory is a big problem for which machines especially computers have been used a lot to assist humans. Humans like to think and deal things separately and locally. To understand concepts, it requires a lot of analysis as well as memory of past incidents. Machines have memory and not analysis power, humans have analysis power, but less memory. So, teaching concepts is challenging in nature. Modeling thousands of story using a single interface and adding human model in it is a major problem.\\\\ 
Example of the concept teaching case for the famous Lion and Rabbit Story of Panchatantra have been presented below in detail. Input of a user has been presented in Table \ref{Lr}.
Values directly calculated from the user inputs are presented in Table \ref{Sv}.
Algorithm relating library with slope-variation is directly applied to draw conclusion of the story as "Brain power very very good than work power" (Simplified version of "Intelligence is better than gross power" in the language of algorithm).
\begin{table}[]
\centering
\caption{Donkey-Washerman Story}
      \centering
\label{Dh}
\resizebox{\linewidth}{!}{
\begin{tabular}{|l|c|l|c|c|}
\hline
\multicolumn{1}{|c|}{1} & Scenes                                                                                                                                                                                                                                                                                                                                                                                                                                                                       & \multicolumn{1}{c|}{Q1}  & Q2                       & Q3                       \\ \hline
                        & \begin{tabular}[c]{@{}c@{}}S1 (Introduction\\ of characters: the washer man \& rest of the world, The washer-man could not take proper care of his\\ donkey. The surroundings where he lived, lacked grass; and he did not have\\ enough to offer the donkey to eat. As a result, the donkey had grown lean and\\ weak.)\end{tabular}                                                                                                                                        & \multicolumn{1}{c|}{1}   & 1                        & 1                        \\ \hline
                        & \begin{tabular}[c]{@{}c@{}}S2 (One particular day,\\ he was wandering in the jungle, where he came across a dead tiger. He at once\\ struck an idea \& thought, "I will skin the tiger and take the skin\\ home. I will cover the donkey with the tiger's skin and let him graze in the\\ nearby barley fields after sunset. The farmers will not dare to come near him\\ fearing my donkey as a tiger. This way, he will be able to eat as much as he\\ wants)\end{tabular} & \multicolumn{1}{c|}{4.1} & 4.1                      & 4.2                      \\ \hline
                        & \begin{tabular}[c]{@{}c@{}}S3 (The washer-man did so\\ after sunset and the donkey returned unharmed after he had eaten to his heart's\\ content.)\end{tabular}                                                                                                                                                                                                                                                                                                              & \multicolumn{1}{c|}{4.2} & 4.2                      & 4.3                     \\ \hline
                        & \begin{tabular}[c]{@{}c@{}}S4 (From then onwards,\\ the washer-man would cover his donkey with the tiger's skin every night and\\ lead him to the fields. The farmers did spot him, but mistook it for a tiger.\\ They did not even venture out of their homes in fear. All the time, the donkey\\ ate as much as he liked and returned home. In the morning, he would stand in\\ the washer-man's stall without anybody suspecting anything. )\end{tabular}                 & 4.4                      & \multicolumn{1}{l|}{4.3} & \multicolumn{1}{l|}{4.4} \\ \hline
                        & \begin{tabular}[c]{@{}c@{}}S5 (As time passed, the donkey\\ regained his health \& the washer-man did not have to worry about his food.)\end{tabular}                                                                                                                                                                                                                                                                                                                        & \multicolumn{1}{c|}{4.8} & 4.6                      & 4.7                      \\ \hline
                        & \begin{tabular}[c]{@{}c@{}}S6 (One night, as he was\\ feeding on the fresh barley crops in the fields, he heard a sound. It was a\\ female donkey braying from a distance. He was attracted and brayed in return.))\end{tabular}                                                                                                                                                                                                                                             & \multicolumn{1}{c|}{1.2} & 1.3                      & 1.2                      \\ \hline
                        & \begin{tabular}[c]{@{}c@{}}S7 (The farmers, who were\\ watching him from inside for fear of the tiger, heard this and realized that it\\ was a donkey and not a tiger. They came out to observe, it was indeed a donkey\\ dressed in tiger's skin. They chased the donkey with sticks and killed him.)\end{tabular}                                                                                                                                                          & 0                        & \multicolumn{1}{l|}{0}   & \multicolumn{1}{l|}{0}    \\ \hline
\multicolumn{1}{|c|}{2} & 1. What is the moral of the story?                                                                                                                                                                                                                                                                                                                                                                                                                                           &                          & \multicolumn{1}{l|}{}    & \multicolumn{1}{l|}{}    \\ \hline
\multicolumn{1}{|c|}{3} & \multicolumn{1}{l|}{}                                                                                                                                                                                                                                                                                                                                                                                                                                                        & 4.6,2.3                        & \multicolumn{1}{l|}{4.8,2.5} & \multicolumn{1}{l|}{4.9,2.7} \\ \hline
\multicolumn{1}{|c|}{4} & Do, be, actually                                                                                                                                                                                                                                                                                                                                                                                                                                                             & \multicolumn{3}{l|}{}                                                          \\ \hline
\multicolumn{1}{|c|}{5} & Not required                                                                                                                                                                                                                                                                                                                                                                                                                                                                 & \multicolumn{3}{l|}{}                                                          \\ \hline
\multicolumn{1}{|c|}{6} & Not required                                                                                                                                                                                                                                                                                                                                                                                                                                                                 & \multicolumn{3}{l|}{}                                                          \\ \hline
\multicolumn{1}{|c|}{7} & No                                                                                                                                                                                                                                                                                                                                                                                                                                                                 & \multicolumn{3}{l|}{}                                                          \\ \hline
\multicolumn{1}{|c|}{8} & \begin{tabular}[c]{@{}c@{}}Don’t be with not\\ actually\end{tabular}                                                                                                                                                                                                                                                                                                                                                                                                         & \multicolumn{3}{l|}{}                                                          \\ \hline
\end{tabular}
}
\end{table}
\begin{table}[]
\centering
\caption{Heron-Crab Story}
\label{my-label}
\resizebox{\linewidth}{!}{
\begin{tabular}{|c|c|l|c|c|}
\hline
1                      & Scenes                                                                                                                                                                                                                                                                                                                                                                                                                                                                                                                                                                                                                                                                                                                                                                                                                                                                                                                                                                                                                                                                                                                                                                                                                                                                                                                                                                                                                                                                                                                                                                                                                                                                                                                                         & \multicolumn{1}{c|}{Q1}  & Q2                       & Q3                       \\ \hline
\multicolumn{1}{|l|}{} & \begin{tabular}[c]{@{}c@{}}S1 ((Introduction of characters: heron \& crab,)\end{tabular}                                                                                                                                                                                                                                                                                                                                                                                                                                                                                                                                                                                                                                                                                                                                                                                                                                                                                                                                                                                                                                                                                                                                                                                                                                                                                                                                                                                                                                                                                                        & \multicolumn{1}{c|}{1.5} & 1.4                      & 1.4                      \\ \hline                
\multicolumn{1}{|l|}{} & \begin{tabular}[c]{@{}c@{}}S2  The heron had grown so \\old that he could not catch fishes from the lake.\\ Unable to bear hunger, he hit upon a plan, sat at the edge of the lake for everybody to see \& began crying. When\\ Crab asked for the reason as “Why are you crying instead of catching fishes”, he replied, “I would not touch any fish\\ anymore. I have decided to renounce all worldly matters, and vowed to undertake a fast unto death. I hear from a wise \\ astrologer that this lake will dry up as there will be no rains for the next twelve years. The crab told this to other water \\ creatures. On hearing this, everybody started to panic. They believed the heron \& asked "Please guide \& save us from \\ this disaster". The heron said “There is a beautiful lake not far from here \& it would not dry even if it did not rain for \\ twenty four years. I can take you there, if you can ride on my back." He had already gained their confidence. So, they \\ requested to carry them one at a time. The wicked heron succeeded in his plan. Every day, he would carry on of them \\ on his back. After flying a little away from the lake, he would smash them against a rock and eat them up.He would \\ then return after some time to the lake and relate false messages how they are happy in the other lake. One day crab \\ asked, "Uncle, you take others to the lake but it is me who is your first friend. Please take me to the other lake to save\\ my life". The heron started carrying the crab to the same rock. The crab looked down from above, saw the heap of \\ bones, skeletons \& he understood what the heron was up to.) )\end{tabular} & \multicolumn{1}{c|}{1.8} & 1.5                      & 1.4                      \\ \hline
\multicolumn{1}{|l|}{} & \begin{tabular}[c]{@{}c@{}}S3 (He remained calm, and\\ said to the heron, "Uncle, the lake seems far and I am quite heavy. You must be getting tired, let us stop\\ for some rest)\end{tabular}                                                                                                                                                                                                                                                                                                                                                                                                                                                                                                                                                                                                                                                                                                                                                                                                                                                                                                                                                                                                                                                                                                                                                                                                                                                                                                                                                                                                                                                                                                                                                & \multicolumn{1}{c|}{2.2} & 2.1                      & 2                        \\ \hline
\multicolumn{1}{|l|}{} & \begin{tabular}[c]{@{}c@{}}S4 (The confident heron replied,\\ "There is no lake for real. This trip is for my own meal. As I do every\\ day, I will smash you against a rock and make a meal out of you.)\end{tabular}                                                                                                                                                                                                                                                                                                                                                                                                                                                                                                                                                                                                                                                                                                                                                                                                                                                                                                                                                                                                                                                                                                                                                                                                                                                                                                                                                                                                                                                                                                                         & \multicolumn{1}{c|}{2.7} & 2.6                      & 2.5                      \\ \hline
\multicolumn{1}{|l|}{} & \begin{tabular}[c]{@{}c@{}}S5 (The crab got hold of\\ the heron's neck with its strong claws \& strangled him to death.)\end{tabular}                                                                                                                                                                                                                                                                                                                                                                                                                                                                                                                                                                                                                                                                                                                                                                                                                                                                                                                                                                                                                                                                                                                                                                                                                                                                                                                                                                                                                                                                                                                                                                                                          & 0                        & \multicolumn{1}{l|}{0}   & \multicolumn{1}{l|}{0}   \\ \hline
2                      & 1. What is the moral of the story?                                                                                                                                                                                                                                                                                                                                                                                                                                                                                                                                                                                                                                                                                                                                                                                                                                                                                                                                                                                                                                                                                                                                                                                                                                                                                                                                                                                                                                                                                                                                                                                                                                                                                                             &                          & \multicolumn{1}{l|}{}    & \multicolumn{1}{l|}{}    \\ \hline
3                      & \multicolumn{1}{l|}{}                                                                                                                                                                                                                                                                                                                                                                                                                                                                                                                                                                                                                                                                                                                                                                                                                                                                                                                                                                                                                                                                                                                                                                                                                                                                                                                                                                                                                                                                                                                                                                                                                                                                                                                          & 4.6,2.3                       & \multicolumn{1}{l|}{4.8,2.5} & \multicolumn{1}{l|}{4.9,2.7} \\ \hline
4                      & Brain Power, Work Power, Good                                                                                                                                                                                                                                                                                                                                                                                                                                                                                                                                                                                                                                                                                                                                                                                                                                                                                                                                                                                                                                                                                                                                                                                                                                                                                                                                                                                                                                                                                                                                                                                                                                                                                                                  & \multicolumn{3}{l|}{}                                                          \\ \hline
5                      & Not required                                                                                                                                                                                                                                                                                                                                                                                                                                                                                                                                                                                                                                                                                                                                                                                                                                                                                                                                                                                                                                                                                                                                                                                                                                                                                                                                                                                                                                                                                                                                                                                                                                                                                                                                   & \multicolumn{3}{l|}{}                                                          \\ \hline
6                      & Not required                                                                                                                                                                                                                                                                                                                                                                                                                                                                                                                                                                                                                                                                                                                                                                                                                                                                                                                                                                                                                                                                                                                                                                                                                                                                                                                                                                                                                                                                                                                                                                                                                                                                                                                                   & \multicolumn{3}{l|}{}                                                          \\ \hline
7                      & No                                                                                                                                                                                                                                                                                                                                                                                                                                                                                                                                                                                                                                                                                                                                                                                                                                                                                                                                                                                                                                                                                                                                                                                                                                                                                                                                                                                                                                                                                                                                                                                                                                                                                                                                   & \multicolumn{3}{l|}{}                                                          \\ \hline
8                      & \begin{tabular}[c]{@{}c@{}}Brain power very very good than work power\end{tabular}                                                                                                                                                                                                                                                                                                                                                                                                                                                                                                                                                                                                                                                                                                                                                                                                                                                                                                                                                                                                                                                                                                                                                                                                                                                                                                                                                                                                                                                                                                                                                                                                                                                                           & \multicolumn{3}{l|}{}                                                          \\ \hline
\end{tabular}
}
\end{table}
Final results along with user inputs have been shown for 2 different stories of Panchatatantra in Table \ref{Dh}.
\section{User-Friendly Tuning} \label{Sec:User-Friendly Tuning}
\subsection{Types of User-inputs}
\begin{table}[]
\centering
\begin{minipage}{.5\linewidth}
      \caption{User input type-1}
      \centering
\label{U1i}
\begin{tabular}{|c|c|c|c|c|}
\hline
\multirow{7}{*}{1} & Scenes                & Q1  & Q2  & Q3  \\ \cline{2-5} 
                   & S1                    & 3.2 & 3.3 & 3.2 \\ \cline{2-5} 
                   & S2                    & 4.7 & 4.9 & 4.7 \\ \cline{2-5} 
                   & S3                    & 4.1 & 4.2 & 4.1 \\ \cline{2-5} 
                   & S4                    & 3.7 & 3.8 & 3.7 \\ \cline{2-5} 
                   & S5                    & 2.4 & 2.5 & 2.4 \\ \cline{2-5} 
                   & S6                    & 0   & 0   & 0   \\ \hline
3                  & \multicolumn{1}{l|}{} & 4.6 & 4.8 & 4.9 \\ \hline
\end{tabular}
\end{minipage}%
    \begin{minipage}{.5\linewidth}
      \centering
        \caption{User input type-2}
        \label{U6i}
        \begin{tabular}{|c|c|c|c|c|}
\hline
\multirow{7}{*}{1} & Scenes                & Q1  & Q2  & Q3  \\ \cline{2-5} 
                   & S1                    & 3.5 & 4   & 3.9 \\ \cline{2-5} 
                   & S2                    & 3.8 & 4.3 & 4.2 \\ \cline{2-5} 
                   & S3                    & 4   & 4.4 & 4.5 \\ \cline{2-5} 
                   & S4                    & 4.5 & 4.6 & 4.7 \\ \cline{2-5} 
                   & S5                    & 4.8 & 4.9 & 4.8 \\ \cline{2-5} 
                   & S6                    & 0   & 0   & 0   \\ \hline
3                  & \multicolumn{1}{l|}{} & 2.3 & 2.5 & 2.7 \\ \hline
\end{tabular}
 \end{minipage}       
\end{table}
\begin{table}[]
\centering
\begin{minipage}{.5\linewidth}
      \caption{User input type-3}
      \centering
\label{U3i}
\begin{tabular}{|c|c|c|c|c|}
\hline
\multirow{7}{*}{1} & Scenes                & Q1  & Q2  & Q3  \\ \cline{2-5} 
                   & S1                    & 3.2 & 3.3 & 3.2 \\ \cline{2-5} 
                   & S2                    & 4.7 & 4.9 & 4.7 \\ \cline{2-5} 
                   & S3                    & 4.1 & 4.2 & 4.1 \\ \cline{2-5} 
                   & S4                    & 3.7 & 3.8 & 3.7 \\ \cline{2-5} 
                   & S5                    & 2.4 & 2.5 & 2.4 \\ \cline{2-5} 
                   & S6                    & 0   & 0   & 0   \\ \hline
3                  & \multicolumn{1}{l|}{} & 2.3 & 2.5 & 2.7 \\ \hline
\end{tabular}
\end{minipage}%
    \begin{minipage}{.5\linewidth}
      \centering
        \caption{User input type-4}
        \label{U4i}
\begin{tabular}{|c|c|c|c|c|}
\hline
\multirow{7}{*}{1} & Scenes                & Q1  & Q2  & Q3  \\ \cline{2-5} 
                   & S1                    & 3.5 & 4   & 3.9 \\ \cline{2-5} 
                   & S2                    & 3.8 & 4.3 & 4.2 \\ \cline{2-5} 
                   & S3                    & 4   & 4.4 & 4.5 \\ \cline{2-5} 
                   & S4                    & 4.5 & 4.6 & 4.7 \\ \cline{2-5} 
                   & S5                    & 4.8 & 4.9 & 4.8 \\ \cline{2-5} 
                   & S6                    & 0   & 0   & 0   \\ \hline
3                  & \multicolumn{1}{l|}{} & 4.6 & 4.8 & 4.9 \\ \hline
\end{tabular}
 \end{minipage}
\end{table}

\begin{table}[]
\centering
\begin{minipage}{.5\linewidth}
      \caption{Output of User input type-1}
      \centering
\label{U1o}
\begin{tabular}{|c|c|c|}
\hline
Scenes & pos value & $|$slope$|$ \\ \hline
S1     & 6.17      & -       \\ \hline
S2     & 9.09      & 2.92    \\ \hline
S3     & 7.63      & 1.46    \\ \hline
S4     & 6.22      & 1.41    \\ \hline
S5     & 3.28      & 2.93    \\ \hline
S6     & 0         & 3.28    \\ \hline
\end{tabular}
 \end{minipage}%
    \begin{minipage}{.5\linewidth}
      \centering
        \caption{Output of User input type-2}
        \label{U6o}
\begin{tabular}{|c|c|c|}
\hline
Scenes & pos value & $|$slope$|$ \\ \hline
S1     & 3.81      & -       \\ \hline
S2     & 2.31      & 1.50    \\ \hline
S3     & 1.36      & 0.95    \\ \hline
S4     & 0.82      & 0.54    \\ \hline
S5     & 0.48      & 0.33    \\ \hline
S6     & 0         & 0.48    \\ \hline
\end{tabular}
 \end{minipage}
\end{table}
\begin{table}[]
\centering
\begin{minipage}{.5\linewidth}
      \caption{Output of User input type-3}
      \centering
\label{U3o}
\begin{tabular}{|c|c|c|}
\hline
Scenes & pos value & $|$slope$|$ \\ \hline
S1     & 3.24      & -       \\ \hline
S2     & 2.68      & 0.55    \\ \hline
S3     & 1.31      & 1.37    \\ \hline
S4     & 0.66      & 0.64    \\ \hline
S5     & 0.24      & 0.42    \\ \hline
S6     & 0         & 0.24    \\ \hline
\end{tabular}
 \end{minipage}%
   \begin{minipage}{.5\linewidth}
      \centering
        \caption{Output of User input type-4}
        \label{U4o}
\begin{tabular}{|c|c|c|}
\hline
Scenes & pos value & $|$slope$|$ \\ \hline
S1     & 7.25      & -       \\ \hline
S2     & 7.82      & 0.57    \\ \hline
S3     & 8.20      & 0.38    \\ \hline
S4     & 8.77      & 0.57    \\ \hline
S5     & 9.22      & 0.44    \\ \hline
S6     & 0         & 9.22    \\ \hline
\end{tabular}
 \end{minipage}
\end{table}
To demonstrate the user-friendly tuning feature of this network, many user inputs were taken for Lion-Rabbit story and for simplified analysis, those were classified in to 4 types of inputs after averaging. These 4 inputs are presented in Table \ref{U1i}, \ref{U3i}, \ref{U4i} and \ref{U6i} . Network was trained to get standard conclusions as the output. It has been shown how tuning (which has been a tough problem in machine learning algorithms) is very much user-friendly here. Interestingly it can also be shown how by being a better human (with more sensitivity to emotion), one can use scientific literature and intelligence in a better way to solve various problems.

\subsection{Tuning parameters}
Various Network parameters wchich can be used for tuning have been mentioned below.
\begin{enumerate}
\item Number of Sensors:\\
- Add new node representing another sensory organs.\\
- Add a node representing combination of some sensory organs or modified form of a sensory organ.\\
- Add a modified form of node (with focus on different property of node).
- Switching between forms (e.g. varying significance of a node with respect to parameters (e.g. time)\\
\item Connection of Links:\\
- Addition of multiple links between nodes with dynamic variation of its parameters.\\
- Removal of link between nodes.\\
- Changing the shape of links (e.g. link connecting more than 2 nodes).\\
- Varying the non-uniformity in link distribution with respect to some parameter (e.g. time).\\
\item Node division:\\
- Nonlinear relation (with respect link values) in the definition of "contribution factor".\\
- Dynamic variation of weight in the contribution factor definition with respect to different parameters like time, other nodes, any other activity etc.\\
\item Multitasking based adjustments:\\
- Variation of multi-tasking index with respect to any change we like.\\
- Extending multi-tasking to other structure of the network like link, emotion meter etc.\\
-Dependency on past history can be shown by updating parameters based on the relevant tasks completed.\\
\item Network parameters to the user interface relation.\\
- Suitable Control algorithm and other relevant scientific literature can be used.\\
- Connectivity to processes we like can be done simply by a variable weight based formulation.\\
\item Emotion variable definition.\\
- Through suitable control algorithm (by treating this as a simple input-output system) etc., emotion meter can be connected to suitable nodes (with suitable mathematical transformation if needed).\\
- Dynamic (with respect to time and other parameters) definition of emotion and dream states.\\
\item Contribution and confidence value update.\\
- Any short of process can be tracked (e.g. inputs in the IOT section) and update equation can be connected to anything of our interest.\\
- Mathematical algorithm etc. also can be added for more accurate tracking.\\
\item Point of observation selection and slope to library relation.\\
- It also can be expressed as a linear combination of weights multiplied by suitable network parameters (e.g. node value after suitable transformation) and weights can vary with respect to time and other parameters.\\
- Suitable mathematical functions also can be multiplied with network parameters depending on the requirement.\\
- Slope to library relation need to be established based on personal experience about the problem.\\
- Network structure can be vary with respect to time, iteration or any other parameter.
\end{enumerate}

%
%
%
\subsection{Change in output and necessity for tuning}
For different user inputs various output values are shown in Table \ref{U1o}, \ref{U3o},\ref{U4o},and \ref{U6o}.\\

To distiniguish various inputs and analyze better, Algorithm of table \ref{Algo} (for the substep which gives rise to 2nd conclusion) is extended as follows:\\
 If ($ | \text{slope at any point}|$\textgreater3), additional moral (in addition to the main moral of "Brain power very very good than work power") will be, "Have patience!".\\
 If ($\left| highest\quad slope \right| \le 1.5\quad and\quad \left| lowest\quad value \right| \le 0.5$), additional moral (in addition to the main moral of "Brain power very very good than work power") will be, "Use your intelligence instead of simply believing your work to do the action naturally".\\\\
 This extension in algorithm can be justified using the same concept mentioned in \ref{Sec:Network Operation}-C. With this modification, user input type-1 will have an additional moral "Have patience!" and user input type-2, 3 will have "Use your intelligence instead of simply believing your work to do the action naturally" as the additional moral. Main Moral for user 1,2,3 is "Brain power very very good than work power". Table \ref{U4o} Output for user input type-4 is "Do not be with not actually".
\subsection{Analysis and Tuning}
 These two additional morals shouldn't exist (as naturally there is only one conclusion ("Brain power very very good than work power") of the story), but additional morals have been coming in the output because of some problem associated with the user (as reflected in its inputs). In user-input type 1, problem is the mismatch of external and internal inputs (y and x respectively). Based on his experience user should know about his senses (specifically the relilability part) and should have kept confidence values a bit low. But suppose user doesn't want to change his confidence (though this is the most permanent solution), then we should change the threshold to higher value (e.g. 5) and there are many other ways as mentioned in the list of parameters above. Precedence order is from bottom to top. Because top parameters represent characterstics of a person which is very difficult to change. The bottom parameters are easier to change.\\\\ 
 The beauty of this approach is to get personal guidance at each step and rectify it with more and more use by selecting appropriate tuning parameter and choosing a suitable algorithm. This way the user makes the network more and more personalized (tuning is what we mean basically) and can have fun with the diverging application (e.g. generating story case). Similarly for user input type-3 the mismatch is because of the lesser confidence values. It should increase or accordingly parameters in the algorithm need to be changed (i.e. changing the limit from 0.5 and 1.5). Again there are many other ways.\\\\
 In case of user input type-4, the change is so much that conclusion slips to "Don't be with not actually". Here also the problem is confidence values. Based on previous experience, user should know his sensitivity etc.(comparison of this input with inputs of user 3 clearly says the absence of intelligence of the user in understanding the situation in the story) and should have lesser confidence value. A simple alternate option is to divide pos by 2. Similar problem also can happen if the user doesn't believe in past history or somehow missed the 1st scene (As the value will drastically increase from the next scene). Same problem happens with user input type-2, here the confidence value should be a bit higher. Changing algorithm values and other parameters are again more options here. User should do the change to which he agrees to (based on his own experience).\\\\
 System can be made to switch between various schemes (depending on the priority basis for example) using various existing algorithms.E.g. Generalized optimal sensor controller (problem-solver in a more general way) connection design algorithm for a Cyber physical system can be used by proper selection of constraints based on user's preference to change various parameters.
 \subsection{Other possible variations}
 If a user considers the opposite ratio (rabbit:lion) while answering questions, then the conclusions will be just opposite. For the same 6 user inputs, instead of 1st and 2nd moral of table \ref{Algo}, 3rd and 4th moral will come as output and will have different sub-morals. From the rabbit frame, 3rd moral can be a valid moral similar to the 1st natural moral which comes in the lion's frame. So, accordingly standard stories should have atleast two types of conclusion considering most active and most inactive character. This needs to be taken into account while tuning.\\\\
 In the story, we saw all 3 values are very near to each other. If one value is very much different than the other and it is causing the moral to be different or produce a sub-moral. In that case though decreasing confidence value of that node is a good option, still best one is to divide pos by the degree which distinguishes that sensory organ from others.
 \subsection{Tuning illustration}
This section is to justify the feature of customized and user-friendly tuning in our network. Similar to Section \ref{Sec:User-Friendly Tuning}-A, here also 4 types of user-inputs are used . Lion-Rabbit Story has been used for tuning and other stories Heron-crab, Donkey-washerman, Dove-hunter have been used for testing. For better analysis,  
algorithm of table \ref{Algo} is extended by replicating the 2nd conclusion based modified algorithm of Section \ref{Sec:User-Friendly Tuning}-C. 
If ($ | \text{slope at any point}|$\textgreater5.6), additional moral (in addition to the 1st moral "Do not be with not actually"(Dont pretend)) will be, "Behave like a human!".\\
 If ($\left| highest\quad slope \right| \le 4.55\quad and\quad \left| lowest\quad value \right| \le 3.85$), additional moral (in addition to the 1st moral "Do not be with not actually" (Dont pretend)) will be, Don't change yourself infront of others.\\\\
 If ($ | \text{slope at any point}|$\textgreater5.6), additional moral (in addition to the 1st moral "Not natural work is not good") will be, "Calm down!".\\
 If ($\left| highest\quad slope \right| \le 4.55\quad and\quad \left| lowest\quad value \right| \le 3.85$), additional moral (in addition to the 1st moral "Not natural work is not good") will be, Don't disbelieve!.\\\\
 If ($ | \text{slope at any point}|$\textgreater5.6), additional moral (in addition to the 1st moral "Very good natural work with no other thing has very much power") will be, "Control urself with patience!".\\
 If ($\left| highest\quad slope \right| \le 4.55\quad and\quad \left| lowest\quad value \right| \le 3.85$), additional moral (in addition to the 1st moral "Very good natural work with no other thing has very much power") will be, Don't get attached to any situation.\\\\
 This extension in algorithm again can be justified using the same concept mentioned in \ref{Sec:Network Operation}-C.\\ 
 Confidence values are taken as tuning parameters as these are the permanent solution to tune our observations. This has been already discussed in \ref{Sec:Network Operation}-B.
Our input (which gives correct moral with out any additional moral) has been used as the standard reference. Other user inputs have used the same confidence values to tune the network. Results show consistency in behaviour for the testing stories after tuning it based on the Lion-rabbit story. By taking any other user-input as reference, network can be tuned to make it behave like that user for all stories. The same can be observed from the results below.\\
User input type-1 values have been shown in table \ref{21}. Its output for Donkey-washerman story is "Do not be with not actually". But it has an additional output "Behave like a human!". So this would be the output of User input type-1 without tuning  and this is not desired. User input type-3 values have been shown in table \ref{23}. Its output for Donkey-washerman story is "Brain power very very good than work power". So this would be the output of User input type-3 without tuning  and this is not desired.\\
Tuned version of user input type-1 and tuned user input type-3 values have been shown in table \ref{22}. Its output for Donkey-washerman story is "Do not be with not actually".\\
\begin{table}[]
\centering
\begin{minipage}{.5\linewidth}
\caption{Story-1 Output for user-1}
\centering
\label{21}

\begin{tabular}{|c|c|c|}

\hline

Scenes & pos value & $|$slope$|$                   \\ \hline

S1     & 2.16      & 2.16                      \\ \hline

S2     & 7.88      & 5.72                      \\ \hline

S3     & 8.07      & 0.19                      \\ \hline

S4     & 8.33      & 0.25                      \\ \hline

S5     & 8.96      & 0.64                      \\ \hline

S6     & 2.28      & 6.68                      \\ \hline

S7     & 0         & \multicolumn{1}{l|}{2.28} \\ \hline

\end{tabular}

\end{minipage}%
 \begin{minipage}{.5\linewidth}

\centering

\caption{Story-1 Output for user-3}

\label{23}

\begin{tabular}{|c|c|c|}

\hline

Scenes & pos value & $|$slope$|$                   \\ \hline

S1     & 1.13      & 1.13                      \\ \hline

S2     & 3.35      & 2.22                      \\ \hline

S3     & 1.93      & 1.43                      \\ \hline

S4     & 1.12      & 0.82                      \\ \hline

S5     & .0.68     & 0.44                      \\ \hline

S6     & 0.10      & 0.58                      \\ \hline

S7     & 0         & \multicolumn{1}{l|}{0.10} \\ \hline

\end{tabular}
\end{minipage}
\end{table}

User input type-4 values have been shown in table \ref{21n}. Its output for Donkey-washerman story is "Do not be with not actually". But it has an additional output "Don't change yourself infront of others". So this would be the output of user input type-4 without tuning  and this is not desired. User input type-2 values have been shown in table \ref{23n}. Its output for Donkey-washerman story is "Brain power very very good than work power". So this would be the output of user input type-2 without tuning  and this is not desired.\\
Tuned version of User input type-4 and tuned User input type-2 values) have been shown in table \ref{22n}. Its output for Donkey-washerman story is "Do not be with not actually".\\

\begin{table}[]

\centering
\begin{minipage}{.5\linewidth}
\caption{Story-1 Output for user-4}
\centering
\label{21n}

\begin{tabular}{|c|c|c|}

\hline

Scenes & pos value & $|$slope$|$ \\ \hline

S1     & 2.16      & 2.16    \\ \hline

S2     & 7.95      & 5.78    \\ \hline

S3     & 5.53      & 2.42    \\ \hline

S4     & 4.45      & 1.08    \\ \hline

S5     & 3.56      & 0.89    \\ \hline

S6     & 0.70      & 2.86    \\ \hline

S7     & 0         & 0.70    \\ \hline

\end{tabular}

\end{minipage}%
 \begin{minipage}{.5\linewidth}

\centering

\caption{Story-1 Output for user-2}

\label{23n}

\begin{tabular}{|c|c|c|}

\hline

Scenes & pos value & $|$slope$|$ \\ \hline

S1     & 1.13      & 1.13    \\ \hline

S2     & 3.38      & 2.24    \\ \hline

S3     & 1.32      & 2.06    \\ \hline

S4     & 0.60      & 0.72    \\ \hline

S5     & 0.27      & 0.33    \\ \hline

S6     & 0.03      & 0.24    \\ \hline

S7     & 0         & 0.03    \\ \hline

\end{tabular}
\end{minipage}
\end{table}

\begin{table}[]

\centering
\begin{minipage}{.5\linewidth}
\caption{Story-1 O/p for tuned user 1,3}
\centering
\label{22}

\begin{tabular}{|c|c|c|}

\hline

Scenes & pos value & $|$slope$|$                   \\ \hline

S1     & 1.74      & 1.74                      \\ \hline

S2     & 5.73      & 3.99                      \\ \hline

S3     & 3.46      & 2.27                      \\ \hline

S4     & 2.01      & 1.45                      \\ \hline

S5     & 1.21     & 0.79                      \\ \hline

S6     & 0.18      & 1.03                      \\ \hline

S7     & 0         & \multicolumn{1}{l|}{0.18} \\ \hline

\end{tabular}

\end{minipage}%
 \begin{minipage}{.5\linewidth}

\centering

\caption{Story-1 O/p for tuned user 2,4}

\label{22n}

\begin{tabular}{|c|c|c|}

\hline

Scenes & pos value & $|$slope$|$ \\ \hline

S1     & 1.74      & 1.74    \\ \hline

S2     & 5.77      & 4.03    \\ \hline

S3     & 2.26      & 3.51    \\ \hline

S4     & 1.02      & 1.24    \\ \hline

S5     & 0.46      & 0.56    \\ \hline

S6     & 0.05      & 0.41    \\ \hline

S7     & 0         & 0.05    \\ \hline

\end{tabular}
\end{minipage}
\end{table}
User input type-1 values have been shown in table \ref{31}. Its output for Heron-Crab story is "Do not be with not actually". So this would be the output of User input type-1 without tuning  and this is not desired.
User input type-3 values have been shown in table \ref{33}. Its output for Heron-Crab story is "Brain power very very good than work power".  But it has an additional output "Use your intelligence instead of simply believing your work to do the action naturally". So this would be the output of User input type-3 without tuning  and this is not desired.\\
Tuned version of User input type-1 and tuned User input type-3 values have been shown in table \ref{32}. Its output for Heron-Crab story is "Brain power very very good than work power".\\
\begin{table}[]

\centering
\begin{minipage}{.5\linewidth}
\caption{Story-2 Output for user-1}
\centering
\label{31}

\begin{tabular}{|c|c|c|}

\hline

Scenes & pos value & $|$slope$|$ \\ \hline

S1     & 2.73      & 2.73    \\ \hline

S2     & 2.98      & 0.25    \\ \hline

S3     & 4.00      & 1.02    \\ \hline

S4     & 4.96      & 0.95    \\ \hline

S5     & 0         & 4.96    \\ \hline

\end{tabular}

\end{minipage}%
 \begin{minipage}{.5\linewidth}

\centering

\caption{Story-2 Output for user-3}

\label{33}

\begin{tabular}{|c|c|c|}

\hline

Scenes & pos value & $|$slope$|$ \\ \hline

S1     & 1.43      & 1.43    \\ \hline

S2     & 1.22      & 0.21    \\ \hline

S3     & 1.14      & 0.08    \\ \hline

S4     & 0.80      & 0.35    \\ \hline

S5     & 0         & 0.80    \\ \hline

\end{tabular}
\end{minipage}
\end{table}

User input type-4 values have been shown in table \ref{31n}. Its output for Heron-Crab story is "Brain power very very good than work power". But it has an additional output Have patience!. So this would be the output of User input type-4 without tuning  and this is not desired.User input type-6 values have been shown in table \ref{33n}. Its output for Heron-Crab story is "Brain power very very good than work power".  But it is in the verge of having an additional output "Use your intelligence instead of simply believing your work to do the action naturally". So this would be the output of User input type-6 without tuning  and this is not desired.\\
Tuned version of User input type-4 and User input type-2 values have been shown in table \ref{32n}. Its output for Heron-Crab story is "Brain power very very good than work power".\\
\begin{table}[]

\centering
 \begin{minipage}{.5\linewidth}
\caption{Story-2 Output for user-4}
\centering
\label{31n}

\begin{tabular}{|c|c|c|}

\hline

Scenes & pos value & $|$slope$|$ \\ \hline

S1     & 3.05      & 3.05    \\ \hline

S2     & 4.32      & 1.27    \\ \hline

S3     & 2.16      & 2.16    \\ \hline

S4     & 1.08      & 1.08    \\ \hline

S5     & 0         & 1.08    \\ \hline

\end{tabular}

\end{minipage}%
 \begin{minipage}{.5\linewidth}

\centering

\caption{Story-2 Output for user-2}

\label{33n}

\begin{tabular}{|c|c|c|}

\hline

Scenes & pos value & $|$slope$|$ \\ \hline

S1     & 1.60      & 1.60    \\ \hline

S2     & 1.70      & 0.10    \\ \hline

S3     & 0.48      & 1.22    \\ \hline

S4     & 0.17      & 0.31    \\ \hline

S5     & 0         & 0.17    \\ \hline

\end{tabular}
\end{minipage}
\end{table}
\begin{table}[]

\centering
\begin{minipage}{.5\linewidth}

\caption{Story-2 O/p for tuned user 1,3}
\centering
\label{32}

\begin{tabular}{|c|c|c|}

\hline

Scenes & pos value & $|$slope$|$ \\ \hline

S1     & 2.20      & 2.20    \\ \hline

S2     & 2.17      & 0.03    \\ \hline

S3     & 2.54      & 0.37    \\ \hline

S4     & 2.27      & 0.27    \\ \hline

S5     & 0         & 2.27    \\ \hline

\end{tabular}

\end{minipage}%
 \begin{minipage}{.5\linewidth}

\centering

\caption{Story-2 O/p for tuned user 2,4}

\label{32n}

\begin{tabular}{|c|c|c|}

\hline

Scenes & pos value & $|$slope$|$ \\ \hline

S1     & 2.45      & 2.45    \\ \hline

S2     & 3.14      & 0.69    \\ \hline

S3     & 1.27      & 1.87    \\ \hline

S4     & 0.55      & 0.72    \\ \hline

S5     & 0         & 0.55    \\ \hline

\end{tabular}
\end{minipage}%
\end{table}
\begin{table}[]
\centering
\caption{Dove hunter story User 1,3 input..}
\label{my-label}
\resizebox{\linewidth}{!}{
\begin{tabular}{|l|c|l|l|l|l|l|}
\hline
1 & \multicolumn{3}{c|}{Scenes}                                                                                                                                                                                                                                                                                                                                                                                                                                                                                                                                                                                                                        & \multicolumn{1}{c|}{Q1} & \multicolumn{1}{c|}{Q2} & \multicolumn{1}{c|}{Q3}  \\ \hline
  & \multicolumn{3}{c|}{\begin{tabular}[c]{@{}c@{}}S1(Introduction of Characters dove and the world: There was a mean\\ hunter, who used to roam in the jungle in search of birds \& other\\ small animals. He was deserted by all his friends and relatives for\\ his cruel deeds. In the same jungle, there lived a happy couple of\\ doves. They had built a beautiful nest in the top of a big tree)\end{tabular}}                                                                                                                                                                                                                                 & 2.4                     & 2                       & \multicolumn{1}{c|}{4.6} \\ \hline
  & \multicolumn{3}{c|}{\begin{tabular}[c]{@{}c@{}}S2(One evening, the hunter caught hold of the female dove when she\\ was alone, and trapped her into a cage. He was looking for more\\ hunting, when a sudden storm broke in. It was accompanied by heavy\\ rains. Shivering in rain and cold, he took shelter under the tree.\\ This happened to be the very tree where the female dove he had caught\\ lived. )\end{tabular}}                                                                                                                                                                                                                     & 0.3                     & 0.2                     & 0.4                      \\ \hline
  & \multicolumn{3}{c|}{\begin{tabular}[c]{@{}c@{}}S3(After some time, the rain started to cease and the sky started\\ to clear. But it was late into the night, so the hunter decided to\\ spend the night under the tree. Meanwhile, in the nest above, the\\ male dove was very worried as his wife had not returned. The fact\\ that there was a storm even compounded his worries. The female dove\\ could hear her husband worrying from above, and she called out to him\\ \& said, "I am being held by the hunter who has taken shelter\\ under the very tree. But I will tell you something that will be for\\ your own good.")\end{tabular}} & 0.5                     & 0.4                     & 0.9                      \\ \hline
  & \multicolumn{3}{c|}{\begin{tabular}[c]{@{}c@{}}S4(She continued, "The guest is always God. If someone ever\\ comes to your house for shelter, one must do his best to him, even\\ risking own life. This hunter is in cold and is hungry. He is not to\\ be blamed that I have been caged by him. But this must be the result\\ of my past deeds.,Please welcome him according to our traditions,\\ and not hate him for me." )\end{tabular}}                                                                                                                                                                                                      & 1.1                     & 0.9                     & 1.9                      \\ \hline
  & \multicolumn{3}{c|}{\begin{tabular}[c]{@{}c@{}}S5(The male dove was touched by his wife's virtuous guidance. The\\ dove at once flew to a distant place and brought back a piece of\\ burning coal. He then climbed up the tree and dropped some dry\\ leaves. The leaves caught fire. He said to the hunter, "Please\\ warm yourself from this fire. I am already unfortunate for not being\\ able to provide food. As you are my guest, I offer myself. Please\\ accept my sacrifice and make a meal out of me." Saying so, he\\ flew into the fire, which killed him.)\end{tabular}}                                                            & 1.7                     & 1.5                     & 2.6                      \\ \hline
\end{tabular}
}
\end{table}
\begin{table}[]
\centering
\caption{Dove hunter story User 1,3 input..continued}
\label{my-label}
\resizebox{\linewidth}{!}{
\begin{tabular}{|l|c|l|l|l|l|l|}
\hline
1                       & \multicolumn{3}{c|}{Scenes}                                                                                                                                                                                                                                                                                                                                                                                                                                                                                                                                                         & \multicolumn{1}{c|}{Q1} & \multicolumn{1}{c|}{Q2} & \multicolumn{1}{c|}{Q3} \\ \hline
                        & \multicolumn{3}{c|}{\begin{tabular}[c]{@{}c@{}}S6(The hunter was very hungry, and could not refuse to accept his\\ offer. At the same time, he was moved by such warmth. His heart was\\ filled with pity. He said, "I am certain to go to hell, for the\\ cruel misdeeds I have done for so long. But this dove has set a\\ virtuous example, I will lead a life of discipline and well-being\\ from today. I promise to sacrifice all my lavish unwanted pleasures."\\ With this, he threw away his cage, which broke and released the\\ unfortunate female dove. )\end{tabular}} & 2.2                     & 2.1                     & 3.1                     \\ \hline
                        & \multicolumn{3}{c|}{\begin{tabular}[c]{@{}c@{}}S7(When the female dove realized that her husband was already dead\\ and was burning in the fire, she began to wail \& flew into the\\ flames, which got her killed too. After her death, she met her\\ husband in heaven. He was transformed into a divine creature, riding\\ a chariot in costly ornaments. The female dove realized, she had\\ assumed a divine form, too.)\end{tabular}}                                                                                                                                         & 4.8                     & 4.8                     & 4.8                     \\ \hline
\multicolumn{1}{|c|}{2} & \multicolumn{3}{c|}{1. What is the moral of the story?}                                                                                                                                                                                                                                                                                                                                                                                                                                                                                                                             &                         &                         &                         \\ \hline
\multicolumn{1}{|c|}{3} & \multicolumn{3}{l|}{}                                                                                                                                                                                                                                                                                                                                                                                                                                                                                                                                                               & 4.6,2.3                 & 4.8,2.5                 & 4.9,2.7                 \\ \hline
\multicolumn{1}{|c|}{4} & \multicolumn{3}{c|}{Work, not, is, good, natural}                                                                                                                                                                                                                                                                                                                                                                                                                                                                                                                                   & \multicolumn{3}{l|}{}                                                       \\ \hline
\multicolumn{1}{|c|}{5} & \multicolumn{3}{c|}{Not Required}                                                                                                                                                                                                                                                                                                                                                                                                                                                                                                                                                   & \multicolumn{3}{l|}{}                                                       \\ \hline
\multicolumn{1}{|c|}{6} & \multicolumn{3}{c|}{Not Required}                                                                                                                                                                                                                                                                                                                                                                                                                                                                                                                                                   & \multicolumn{3}{c|}{}                                                       \\ \hline
7                       & \multicolumn{3}{c|}{No}                                                                                                                                                                                                                                                                                                                                                                                                                                                                                                                                                             & \multicolumn{3}{l|}{}                                                       \\ \hline
8                       & \multicolumn{3}{c|}{Not natural work is not good}                                                                                                                                                                                                                                                                                                                                                                                                                                                                                                                                   & \multicolumn{3}{l|}{}                                                       \\ \hline
\end{tabular}
}
\end{table}
\begin{table}[]
\centering
 \begin{minipage}{.5\linewidth}

\caption{Dove hunter story User 2,4 i/p}
\centering
\label{my-label}
\begin{tabular}{|c|c|c|c|c|}
\hline
\multirow{8}{*}{1} & Scenes & Q1      & Q2      & Q3      \\ \cline{2-5} 
                   &        & 2.3     & 2.0     & 4.6     \\ \cline{2-5} 
                   &        & 0.3     & 0.2     & 0.4     \\ \cline{2-5} 
                   &        & 0.4     & 0.3     & 0.5     \\ \cline{2-5} 
                   &        & 0.5     & 0.4     & 0.7     \\ \cline{2-5} 
                   &        & 0.6     & 0.4     & 0.8     \\ \cline{2-5} 
                   &        & 0.8     & 0.7     & 1.0     \\ \cline{2-5} 
                   &        & 4.9     & 4.9     & 5.0     \\ \hline
3                  &        & 4.6,2.3 & 4.8,2.5 & 4.9,2.7 \\ \hline
\end{tabular}
\end{minipage}%
 \begin{minipage}{.5\linewidth}
\centering
\caption{Donkey-Washerman User 2,4 i/p}
\label{my-label}
\begin{tabular}{|c|c|c|c|c|}
\hline
\multirow{8}{*}{1} & Scenes & Q1      & Q2      & Q3      \\ \cline{2-5} 
                   & S1     & 1.1     & 1.1     & 1.2     \\ \cline{2-5} 
                   & S2     & 4.1     & 4.1     & 4.3     \\ \cline{2-5} 
                   & S3     & 2.9     & 2.8     & 3.0     \\ \cline{2-5} 
                   & S4     & 2.3     & 2.3     & 2.4     \\ \cline{2-5} 
                   & S5     & 1.9     & 1.8     & 1.9     \\ \cline{2-5} 
                   & S6     & 0.4     & 0.3     & 0.4     \\ \cline{2-5} 
                   & S7     & 0       & 0       & 0       \\ \hline
3                  &        & 4.6,2.3 & 4.8,2.5 & 4.9,2.7 \\ \hline
\end{tabular}
\end{minipage}%
\end{table}

\begin{table}
\centering
\caption{Heron-Crab Story User 2,4 input}
\label{my-label}
\begin{tabular}{|l|c|c|c|c|}
\hline
\multicolumn{1}{|c|}{1} & Scenes & Q1      & Q2      & Q3      \\ \hline
                        & S1     & 1.7     & 1.5     & 1.6     \\ \hline
                        & S2     & 2.3     & 2.2     & 2.3     \\ \hline
                        & S3     & 1.1     & 1.2     & 1.1     \\ \hline
                        & S4     & 0.5     & 0.6     & 0.6     \\ \hline
                        & S5     & 0       & 0       & 0       \\ \hline
\multicolumn{1}{|c|}{3} &        & 4.6,2.3 & 4.8,2.5 & 4.9,2.7 \\ \hline
\end{tabular}
\end{table}
User input type-1 values have been shown in table \ref{41}. Its output for Dove-hunter story is "Not natural work is not good". It also has an additional output as "Calm down!". So this would be the output of User input type-1 without tuning  and this is not desired.User input type-3 values have been shown in table \ref{43}. Its output for Dove-hunter story is "Not natural work is not good". It also has an additional output as "Don't disbelieve!". So this would be the output of User input type-3 without tuning  and this is not desired.\\
Tuned version of User input type-1 and User input type-3 values have been shown in table \ref{42}. Its output for Dove-hunter story is "Not natural work is not good".\\
\begin{table}[]

\centering
 \begin{minipage}{.5\linewidth}
\caption{Story-3 Output for user-1}
\centering
\label{41}

\begin{tabular}{|c|c|c|}

\hline

Scenes & pos value & $|$slope$|$ \\ \hline

S1     & 5.73      & 5.73    \\ \hline

S2     & 0.58      & 5.16    \\ \hline

S3     & 1.15      & 0.57    \\ \hline

S4     & 2.48      & 1.34    \\ \hline

S5     & 3.69      & 1.21    \\ \hline

S6     & 4.71      & 1.02    \\ \hline

S7     & 9.15      & 4.45    \\ \hline

\end{tabular}

\end{minipage}%
 \begin{minipage}{.5\linewidth}

\centering

\caption{Story-3 Output for user-3}

\label{43}

\begin{tabular}{|c|c|c|}

\hline

Scenes & pos value & $|$slope$|$ \\ \hline

S1     & 3.02      & 3.02    \\ \hline

S2     & 0.20      & 2.83    \\ \hline

S3     & 0.37      & 0.18    \\ \hline

S4     & 0.70      & 0.33    \\ \hline

S5     & 0.72      & 0.03    \\ \hline

S6     & 0.55      & 0.17    \\ \hline

S7     & 0.61      & 0.06    \\ \hline

\end{tabular}
\end{minipage}%

\end{table}

User input type-4 values have been shown in table \ref{41n}. Its output for Dove-hunter story is "Not natural work is not good". It also has an additional output as "Calm down!". So this would be the output of User input type-4 without tuning  and this is not desired.User input type-2 values have been shown in table \ref{43n}. Its output for Dove-hunter story is "Not natural work is not good". It also has an additional output as "Don't disbelieve!". So this would be the output of User input type-2 without tuning  and this is not desired.\\
Tuned version of User input type-4 and User input type-2 values have been shown in table \ref{42n}. Its output for Dove-hunter story is "Not natural work is not good".
\begin{table}[]

\centering
 \begin{minipage}{.5\linewidth}

\caption{Story-3 Output for user-4}
\centering
\label{41n}

\begin{tabular}{|c|c|c|}

\hline

Scenes & pos value & $|$slope$|$ \\ \hline

S1     & 5.67      & 5.67    \\ \hline

S2     & 0.57      & 5.09    \\ \hline

S3     & 0.76      & 0.19    \\ \hline

S4     & 1.02      & 0.25    \\ \hline

S5     & 1.14      & 0.13    \\ \hline

S6     & 1.59      & 0.45    \\ \hline

S7     & 9.41      & 7.82    \\ \hline

\end{tabular}

\end{minipage}%
 \begin{minipage}{.5\linewidth}

\centering

\caption{Story-3 Output for user-2}

\label{43n}

\begin{tabular}{|c|c|c|}

\hline

Scenes & pos value & $|$slope$|$ \\ \hline

S1     & 2.98      & 2.98    \\ \hline

S2     & 0.19      & 2.79    \\ \hline

S3     & 0.25      & 0.06    \\ \hline

S4     & 0.31      & 0.06    \\ \hline

S5     & 0.30      & 0.01    \\ \hline

S6     & 0.37      & 0.07    \\ \hline

S7     & 1.72      & 1.35    \\ \hline

\end{tabular}
\end{minipage}%
\end{table}
\begin{table}[]

\centering
 \begin{minipage}{.5\linewidth}

\caption{Story-3 O/p for tuned user 1,3}
\centering
\label{42}

\begin{tabular}{|c|c|c|}

\hline

Scenes & pos value & $|$slope$|$ \\ \hline

S1     & 4.63      & 4.63    \\ \hline

S2     & 0.37      & 4.26    \\ \hline

S3     & 0.74      & 0.37    \\ \hline

S4     & 1.55      & 0.81    \\ \hline

S5     & 1.94      & 0.39    \\ \hline

S6     & 1.74      & 0.20    \\ \hline

S7     & 1.94      & 0.20    \\ \hline

\end{tabular}

\end{minipage}%
 \begin{minipage}{.5\linewidth}

\centering

\caption{Story-3 O/p for tuned user 2,4}

\label{42n}

\begin{tabular}{|c|c|c|}

\hline

Scenes & pos value & $|$slope$|$ \\ \hline

S1     & 4.58      & 4.58    \\ \hline

S2     & 0.37      & 4.21    \\ \hline

S3     & 0.50      & 0.12    \\ \hline

S4     & 0.66      & 0.17    \\ \hline

S5     & 0.72      & 0.06    \\ \hline

S6     & 0.96      & 0.25    \\ \hline

S7     & 5.18      & 4.21    \\ \hline

\end{tabular}
\end{minipage}%

\end{table}
 \subsection{User attribute as significance of different outputs}
 This way all such combination which are giving the same moral represent similar person (though their characteristics are completely different e.g. confidence value etc.). This is the beauty and intelligence of this network. User 2 and User 5 are similar here.
 As various parts of the algorithm signify different emotional activities, different user attributes can be tracked based on user inputs. User is made to tune the network by selecting appropriate tuning parameter from the tuning parameters section  of \ref{Sec:User-Friendly Tuning}-B. Based on his/her tuning response, conclusion drawn about the user attributes are mentioned below in the table. For fair understanding, user's input need to be taken for multiple stories and also at different times and average of results need to be taken after analyzing in the above approach. An alternate approach could be to change 1 of the 8 tuning parameters and work on the user inputs to get the desired moral and whichever is allowing us  to get the moral, corresponding element represents the user attribute (in case of multi options, user need to be asked to choose).
 Based on the options selected by users, similar users can be found out (user 2 and user 5  similarity is a special case of this analysis). Note that similarity can only be confirmed after analyzing a lot of cases of same users (Further it needs to be taken for a range of time before declaring similar else it will signify similar mood) .\\\\
Because of easy-understandable significance of our proposed network, user can easily vary various tuning parameters based on his/her understanding about the error between desired and actual outuput. Reverse application of this helps us to track user-attribute based on selection of tuning options. Thus any new environment can be simulated by expressing our perception about the user and a common user can implement his thoughts and can experiment. But this feature has been completely missing in the literature. E.g. without knowing proper significance of nodes, change in their numbers are made based on mathematical sensitivity function of output to number of nodes. To visualize the sensor based significance and implement our common-sense based idea, understanding and decoding mathematical sensitivity is the issue. Conversion of our idea to mathematical literautre is the central issue in those. 
\section{Uniqueness of this network}\label{Sec:unique}
As already discussed in the literature survey section thinking process of a technical innovator gets automatically biased by the development process of science. Discretness can be traced back to the mathematical number system which holds the base of science. So computational techniques (like Neural Networks) have been developed again in a dry discrete sense without the presence of any intelligence (like a intelligent logic which is used to relate many physics process or some law of nature which is recursive for a set of processes under certain condition). Because we learn very quickly and never take so much time unlike the way training phase of conventional NNs have been designed. As humans are the source of intelligence, we tried to understand human processes and try to put it in a network. So, this is our attempt to add water in the form of intelligence to the dry algorithms developed so far. Conventional techniques have been based on computational power as the logic was superposition of huge  number of things can bring intelligence. As humans understand very fast, we tried to replicate the same in a network without going for much computation. We have used just one data to train which is hardly possible for any existing intelligent techniques. Basically through this network we extracted a lot other information based on the one data which is more holistic way of doing things rather than discrete step based work of NNs and etc. As model exists in our proposed network, there is no point in comparing with other techniques where model doesn't exist (black box model) and it is almost impossible to train based on a single data. An motivation of this work was also been to show difference between actual humans (unbiased) and machines (most of the humans). How unbiased humans can use their sense of humour to develop things in a holistic way which is almost impossible for biased humans to collect all discret parts and complet the same work. E.g. in the development of this work, emotion experts, intelligent technique devlopers and many other experts of disciplines could have added different parts and make it work. Help should be taken for better information (based one experience), but my point is the final designer should be one. Through this we also wanted to establish how holistic emotion can be basic unit of analysis unlike the discrete number system. Adding manual inputs and putting user at the center by adding the feature of user-friendly tuning is interesting in the sense that it allows users to analyze and be a better personality.\\\\
Because of biasing effect, putting our logic in existing intelligent techniques like NN is quite cumbersome task. To justify the same, we have tried to put one of our concept "contribution factor" in NNs. Similarly emotion based concept can be depicted by rederiving update equations after taking into account the structure. \\\\
The second part of this algorithm i.e. forming library and correlating it with algorithm is obvious and every body does it in its life style at-least the case of generating story.(other family members definitely give this instruction to kid once he is sad because some family member scolded him). The ”concept teaching”, ”multi-tasking” cases can also be made related to fuzzy based applications. Even research on areas like self-analysis, personalized automation system, education system etc. are going on. But people are doing research in a discrete manner. The need of a holistic approach instead of so much reduction based approaches is the important target. So, automating innovation through a common platform was the main aim and the major part of the contribution is development of the human-model in the network (and making it user-friendly, emotion based), selection of cases and applications. Basically the important thing is about relation based things and co-operation.\\
This gives motivation to develop emotional world of peace and love where every user-interface starting from conversation language to scientific terms is replaced by emotional variables. Because everything in this world affects us and best way to keep track of maximum things is the language of emotion. This is the major thing we attempted to prove in this paper. We already have so much data because of Big data etc., here it is just shown how handling those will be easier using this language of emotion (Also how co-operation can be better maintained using emotion as well).
\subsection{Contribution factor concept in a MLN}\label{Sec:CF}
\begin{figure}[h!]
\centering
\includegraphics[width=0.75\textwidth, height=3in]{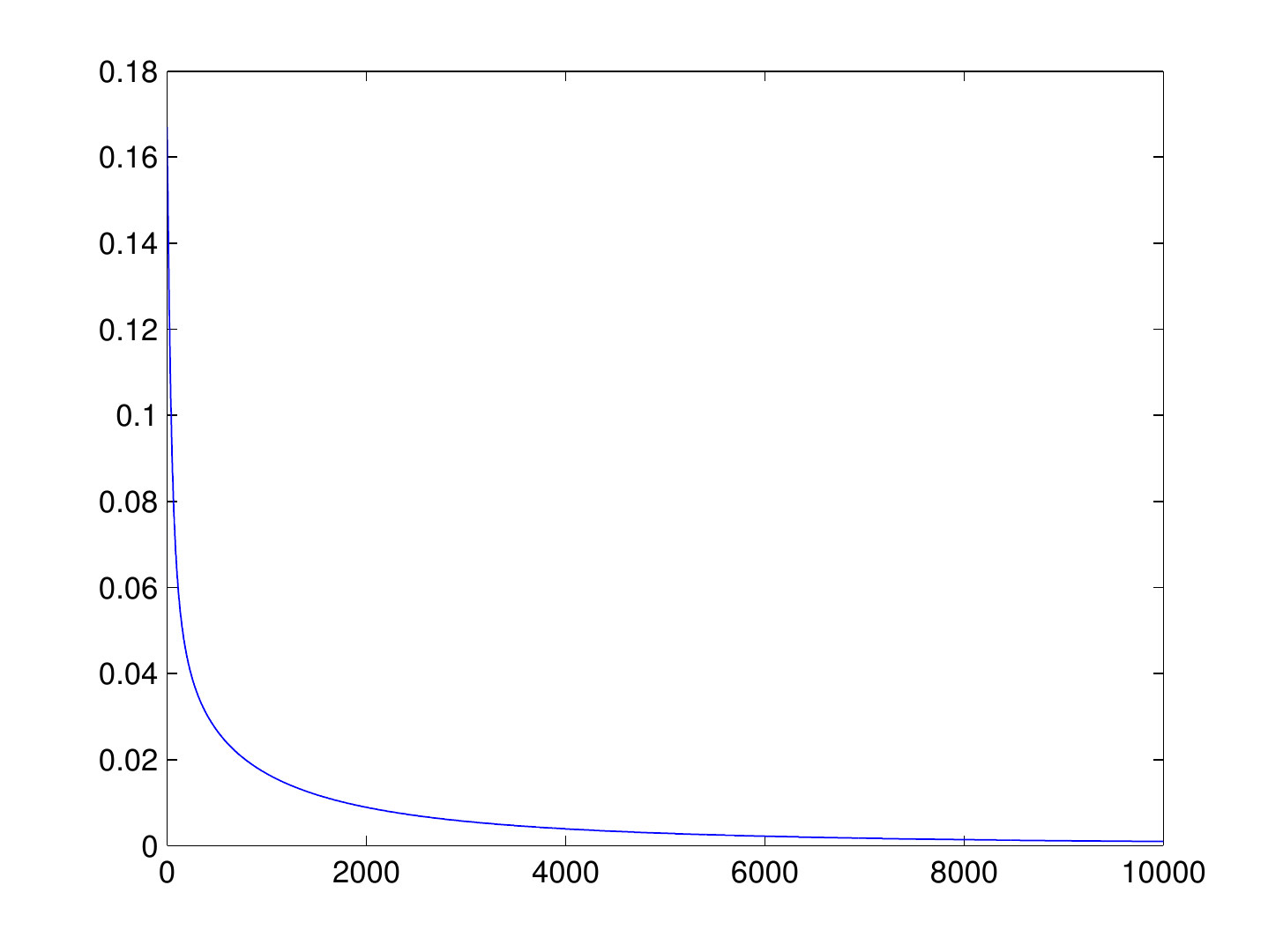}
\caption{Contribution factor result}
\label{figure84}
\end{figure}
\begin{figure}[h!]
\centering
\includegraphics[width=0.75\textwidth, height=3in]{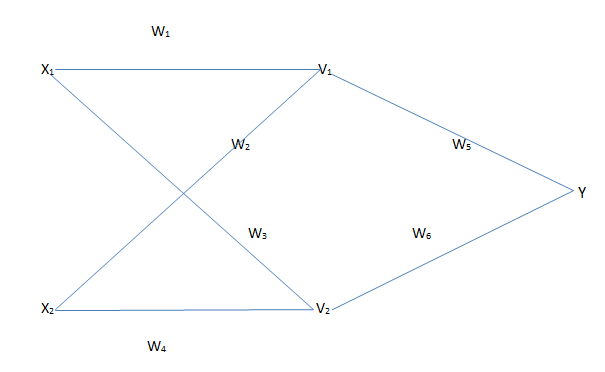}
\caption{Conventional MLN structure}
\label{figure85}
\end{figure}
\begin{figure}[h!]
\centering
\includegraphics[width=0.75\textwidth, height=3in]{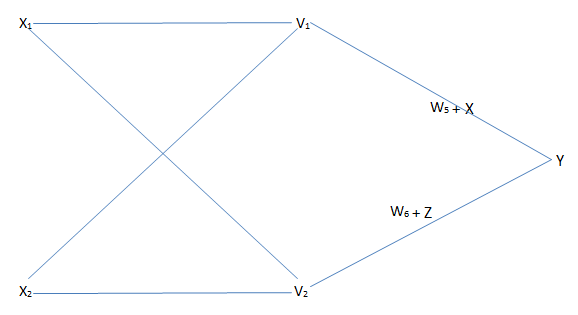}
\caption{MLN structure with contribution factor concept}
\label{figure86}
\end{figure}
To illustrate the difficulty in incorporating features of our proposed network in conventional MLNs, one important feature of our network "contribution factor concept" has been added in conventional MLN architecture. This concept is against the backpropagation algorithm as we only have to update final layer weights without bothering about all other weights. Prefinal layer responses are updated and the concept of any other weights except final layer weights gets eliminated. This can be scalable and generalized for standard MLN architecture.\\ 
\begin{eqnarray*}
\Delta v_1 &=& \frac{1}{4}\Delta h_1- \frac{1}{48} \Delta {h_1}^3 \\
&=& \frac{1}{4}{\eta e y (1-y)v_1(1-v_1)w_5({x_1}^2+{x_2}^2) -\frac{1}{48}({\eta e y}^3 (1-y)v_1(1-v_1)w_5({x_1}^2+{x_2}^2)})^3\\
\Delta v_2 &=& \frac{1}{4}{\eta e y (1-y)v_2(1-v_2)w_6({x_1}^2+{x_2}^2) -\frac{1}{48}({\eta e y}^3 (1-y)v_2(1-v_2)w_6({x_1}^2+{x_2}^2)})^3\\
\end{eqnarray*}
 For output y to remain same, change in $v_1$ and $v_2$ has to be compensated by adding extra terms in $w_5$ and $w_6$ respectively.\\
 Compensation term $x$ for $w_5$ becomes:
 \begin{eqnarray*}
x &=& \frac{\Delta v_1.w_5}{v_1}\\
x &=& \frac{1}{4}{\eta e y (1-y)(1-v_1){w_5}^2({x_1}^2+{x_2}^2) -\frac{1}{48}({\eta e y (1-y)(1-v_1)(x_1}^2+{x_2}^2)})^3{v_1}^2{w_5}^4 \\
\end{eqnarray*}
With the same error function, for gradient descent rule to be appplicable,
\begin{eqnarray*}
w_5 &=& w_5+ \frac{\int\frac{1}{4}{\eta e y (1-y)(1-v_1){w_5}^2({x_1}^2+{x_2}^2) -\frac{1}{48}({\eta e y (1-y)(1-v_1)(x_1}^2+{x_2}^2)})^3{v_1}^2{w_5}^4 dw_5 }{\eta e y (1-y)v_1}\\
w_5 &=& w_5+ \frac{1}{12v_1}{(1-v_1){w_5}^3({x_1}^2+{x_2}^2) -\frac{1}{240}(({\eta e y (1-y))^2(1-v_1)(x_1}^2+{x_2}^2)})^3{v_1}{w_5}^5 \\
\end{eqnarray*}
Similarly
\begin{eqnarray*}
 w_6 &=& w_6+ \frac{1}{12v_2}{(1-v_2){w_6}^3({x_1}^2+{x_2}^2) -\frac{1}{240}(({\eta e y (1-y))^2(1-v_2)(x_1}^2+{x_2}^2)})^3{v_2}{w_6}^5 \\
\end{eqnarray*}
v will be updated as:
\begin{eqnarray*}
v_1 &=& v_1 + \Delta v_1\\
v_2 &=& v_2 + \Delta v_2\\
\end{eqnarray*}
Proof:
\begin{eqnarray*}
\Delta w_5 &=& \eta e y (1-y) v_1 \\
h_1 &=& w_1x_1 + w_2x_2 \\
\Delta w_1 &=& \eta e y(1-y)v_1(1-v_1)w_5x_1 \\
\Delta w_2 &=& \eta e y(1-y)v_1(1-v_1)w_5x_2 \\
\Delta h_1 &=& \Delta w_1x_1 + \Delta w_2x_2 \\
           &=& \eta e y (1-y)v_1(1-v_1)w_5{x_1}^2 + \eta e y(1-y)v_1(1-v_1)w_5{x_2}^2 \\ 
\Delta v_1 &=& f(h^1)-f(h) \\
\end{eqnarray*}
Where f(x) is the sigmoid activation function.\\
As per Maclaurin series, 
\begin{equation*}
f(x)= \frac{1}{2} + \frac{1}{4}x - \frac{1}{48}x^3
\end{equation*}
Note that series expansion is truncated after the 3rd term because range of sigmoid function f(x) is (0,1), so higher order terms has been ignored.\\
\section{Summary}
Our world is simply combination of various inputs from our sensory organs and naturally we 
are more sensitive to emotions than any other thing. Humans are treated as intelligent compared to other living beings. So, both these concepts are added to design this user-friendly cognitive network. It has been applied to write a story based on an incident. It should be able to do a thing that normal AI technique can't. So, detailed application of this network to find moral from a set of morals (as a part of concept teaching to machine) has been shown. These applications can be extended to develop an educational platform to help students better learn tools, concepts.\\ 
Presently we are working to write a story based on an incident using this network. Because of user-friendly interface and human-like structure, this can have very interesting applications. Few examples mentioned in the paper are: helping people use daily life correlations to develop concepts/applications, improving social relations as well as individual personality, acting as a personal guide like teacher/mother/wife, allowing to have control over public for reducing negative thinking based activities like crime, showing human-like behavior through modifications in the basic algorithm, compelling students to focus more on thinking and innovating than gaining skills, helping students specially those of villages to enjoy learning and to improve personally, changing our problem solving style, developing personalized entertainment system and designing thousands of applications by relating science and arts using this algorithm. Additionally we should further develop a platform to extend it and enable it to help solve real-world problems.

 \chapter{Conclusions and Future Work\label{Chap7}}
 \section{Conclusion}
In this thesis, we have developed a user-friendly cyber physical smart-grid architecture. First we focused on achieving stability of the cyber-physical smart grid system by designing a proper cyber system under various constraints present in both physical and cyber system. Then we developed communication hardware modules and tested our proposed architecture  there. We focused on the cyber-aspect more in the hardware development so that this can be useful for other Cyber physical systems. While testing we used virtual grid modelled in the form of state space equations. Then we added robust and adaptive formulations in our architecture to deal with various critical conditions like fault, communication lode failure, load variation. Further we made our architecture user-centric by addition of real time peak load shaving concept in it. After realizing the importance of intelligent algorithms in every sector (including daily life activities) and the amount complexity involved in their use (tuning for example), we used our experience in user-friendly architecture design to develop a cognitive network. We used our network for text summarization application because of its huge research interest due to the natural presence of time and space constraints in various domains.
Some directions of future research are as follows
\subsection{Future Work on Cyber Physical Smart Grid}
\begin{itemize}
\item Instead of smart grid, other interesting similar areas like commerce, entertainment can be targeted using our proposed cyber physical architecture. Heuristic model can be used for the physical system in case actual model is not available.
\item Non linear system model can be used along with our robust and adaptive framework. This will help us to represent physical system in a better manner. 
\item Stochastic formulation need to be added with our robust and adaptive framework to handle critical conditions like fault, theft in a more effective way.
\item Our prosed sensor-controller connection design algorithm can be mathematically modeled and an integrated optimization problem can be formulated to automate our process of smart grid architecture development. For this, we have to compromise the user-friendly aspect.
\item Decentralized routing can also be added and relay dynamics and constraints can also be taken in to picture for better visualization of practical problems associated with Cyber-Physical system.
\item A priority based algorithm can be coupled with our sensor-controller design algorithm to achieve additional objectives like performance, security, safety.
\item Along with delay packet loss, computation time, control effort equalization need to considered in our architecture.
\end{itemize}
\subsection{Future Work on Emotion Recognition and Cognitive Network}
\begin{itemize}
\item Story generation is another interesting and fast-growing area along with text summarization because of the natural presence of time and space constraints in various domains. Our user-friendly cognitive network can be applied to this story generation problem. 
\item User attribute tracking can be correlated with various human activities along with the tuning parameter selected by the user to produce interesting personal applications. 
\item This concept also can be used for developing intersting products like The next generation learning equipment for education sector, public assistance device in problem-solving, personal guidance device. 
\item The develop cognitive network based interface will be applied for complex smart grid issues for assisting users and utilities intelligently.

\end{itemize}

 \newpage
\vspace {1.5in}
\begin{center}
\begin{large}
{\bf \LARGE {List of Publications}}
\end{large}
\end{center}

\begin{itemize}
\item[1.] Swaroop Ranjan Mishra, N Venkata Srinath, K Meher Preetam and Laxmidhar Behera, ``\textit{A generalized novel framework for optimal sensor-controller connection design to guarantee a stable cyber physical smart grid }", 2015 IEEE International Conference on Industrial Informatics (INDIN 2015), Cambridge, UK,pp. 424 - 429, July 22-24, 2015. 
\end{itemize}
%

\bibliography{References}{}
\bibliographystyle{plain}

\end{document}